# MIXED STATE HALL EFFECT IN A TWINNED YBa$_2$Cu$_3$O$_{7-\delta}$ SINGLE CRYSTAL

PAR

## Vincent BERSETH

Ingénieur physicien diplômé EPF
de nationalité suisse et originaire de Saint-George (VD)

# *VERSION ABRÉGÉE*

La dynamique des vortex est étudiée dans un échantillon monocristallin maclé d'YBa$_2$Cu$_3$O$_{7-\delta}$ de haute qualité à l'aide de mesures de résistivité. Neuf contacts en or sont déposés sur l'une des surfaces de l'échantillon, permettant ainsi d'effectuer différentes sortes d'études.

En premier lieu, la résistivité longitudinale révèle la transition de phase de vortex. Lorsque le champ magnétique est légèrement incliné au dehors des plans de macles, la transition est vraisemblablement du premier ordre, séparant le liquide de vortex d'un réseau de vortex ou un verre de Bragg. Si par contre le champ magnétique est parallèle à l'axe cristallographique *c*, c'est-à-dire le long des plans de macles, la phase solide est alors apparemment un verre de Bose.

D'autre part, des mesures de l'orientation et de l'amplitude du champ électrique en fonction de la direction du courant dans le plan *ab* fourni un grand nombre d'informations au sujet de l'influence des macles sur la dynamique des vortex. On observe un mouvement partiellement dirigé des vortex le long de la famille de plans de macles dominante, débutant déjà dans le liquide de vortex et devenant de plus en plus marqué lorsque la température ou le champ magnétique sont réduits. Il n'y a pas de changement soudain de l'influence des macles à la transition de phase de vortex. On montre aussi que les mesures habituelles de résistivité dans des échantillons maclés avec un courant appliqué le long des axes *a* ou *b* doivent être interprétées avec prudence, précisément en raison du fait que suite à ce mouvement dirigé des vortex le champ électrique n'est plus parallèle au courant.

Finalement, l'effet Hall de l'état mixte est étudié, avec une attention particulière pour la transition de phase de vortex. On observe l'habituelle anomalie Hall, c'est-à-dire un changement de signe de l'effet Hall dans l'état mixte, et on montre que la loi d'échelle pour la résistivité Hall $\rho_{xy} \propto \rho_{xx}^{\beta}$ reste inchangée dans le solide de vortex ($\rho_{xy}$ est la résistivité Hall et $\rho_{xx}$ la résistivité longitudinale). Lorsque le champ magnétique est parallèle aux plans de macles, on obtient un exposant critique de $\beta = 2$, correspondant naturellement à une conductivité Hall constante au-dessous et légèrement au-dessus de la transition de phase de vortex. En présence d'un champ magnétique incliné, l'exposant devient $\beta = 1.4$. Dans ce cas, on observe un soudain changement de comportement de la conductivité Hall précisément à la fusion du réseau de vortex. Cet effet étant fortement dépendant de la densité de courant, on l'interprète comme une conséquence de l'ancrage des vortex. Ainsi, on démontre que la conductivité Hall est bien *dépendante* de l'ancrage, résolvant en principe une traditionnelle controverse sur cette question. A la lumière des présents résultats, on passe en revue différents modèles théoriques concernant l'anomalie Hall aussi bien que la loi d'échelle. Un nouveau modèle phénoménologique pour la loi d'échelle de la résistivité Hall est aussi proposé, s'appuyant sur une analogie directe avec la théorie de la percolation dans les conducteurs métalliques.

# *THESIS ABSTRACT*

Resistivity measurements are used to study vortex dynamics in a high quality superconducting YBa$_2$Cu$_3$O$_{7-\delta}$ twinned single crystal. Nine gold contacts have been deposited on one sample surface, allowing us to perform different kinds of investigations.

Firstly, simple longitudinal resistivity data reveal the vortex phase transition. When the magnetic field is slightly tilted away from the twin planes, this transition is presumably of first order, separating a vortex liquid phase from a solid vortex lattice or Bragg glass. If on the other hand the magnetic field is parallel to the crystallographic *c*-axis, that is along the twin boundaries, the solid phase then apparently becomes a Bose glass.

Secondly, measurements of the orientation and magnitude of the electric field as a function of the current direction in the *ab* plane yield much information about the influence of the twins on the vortex dynamics. We observe a partially guided vortex motion along the dominating twin plane family, already apparent in the vortex liquid, and becoming progressively more and more pronounced as the temperature or the magnetic field is reduced. There is no sharp change of the twin influence at the vortex phase transition. We also show that standard resistivity measurements in twinned samples, in which the current is applied along the *a* or *b*-axis, should be interpreted carefully, precisely because as a consequence of this guided motion the electric field is no longer parallel to the current.

Finally, the mixed state Hall effect is studied, with particular focus given to the vortex phase transition. We observe the usual Hall anomaly, that is, a sign reversal of the Hall effect in the mixed state, and show that the often reported Hall resistivity scaling law $\rho_{xy} \propto \rho_{xx}^\beta$ remains unchanged in the vortex solid phase ($\rho_{xy}$ is the Hall resistivity and $\rho_{xx}$ the longitudinal resistivity). When the magnetic field is parallel to the twin boundaries, we obtain the critical exponent $\beta = 2$, naturally corresponding to a constant Hall conductivity below and slightly above the vortex phase transition. In the presence of a tilted magnetic field the exponent is then $\beta = 1.4$. In this case, we observe a sharp change of behavior in the Hall conductivity right at the vortex lattice melting point, its slope becoming much larger in the vortex solid phase. This effect being strongly dependent on the current density, we interpret it as a result of vortex pinning. Hence this demonstrates that the Hall conductivity *is* pinning dependent, hopefully solving a long term controversy with regards to this topic. We review some theoretical models concerning the Hall anomaly as well as the Hall scaling law in light of our data. A novel phenomenological model for the Hall resistivity scaling law is also given, directly inspired from the theory of percolation in metallic conductors.

# *CONTENT*

CHAPTER I    *VORTEX MATTER*

## 1. Introduction

The basic magnetic properties of type II superconductors are well represented in the mean-field *H-T* phase diagram (Fig. I-1). Besides the normal metallic phase at $H > H_{c2}(T)$, the superconducting domain is divided into two phases :

- the low field Meissner-Ochsenfeld phase, where the external magnetic field $H < H_{c1}(T)$ is completely expelled from the bulk material, such that the induction vanishes inside it, *i.e.* $B = 0$,
- the intermediate field regime $H_{c1}(T) < H < H_{c2}(T)$, called the Shubnikov phase or mixed state, in which the magnetic field enters the superconductor in the form of quantized flux lines, the *vortices*.

The vortices are essential in the determination of the mixed state properties : they are indeed responsible for both the magnetization and the resistivity in the superconducting state. On the theoretical level, most of the investigation of vortex physics can, in principle, be performed in the framework of the Ginzburg-Landau theory [1]. This theory is based on the expansion of the free energy in powers of a complex order parameter $\Psi$, the squared modulus of which represents the density of superconducting electrons. The mean field solution for $\Psi$ is then obtained by minimizing the free energy $F(\Psi)$. This approach leads to a second order superconducting to normal phase transition at $H_{c2}(T)$.

One of the important results of the Ginzburg-Landau theory is a spatially periodic solution, obtained by Abrikosov [2], predicting the formation of a triangular lattice of straight vortices parallel to the applied field, subsisting practically up to the phase transition at $H_{c2}(T)$. This picture is experimentally confirmed in most of the conventional superconductors.

However, since the mean field solution of this model does not take into account critical fluctuations, it is not valid close to the phase transition at $H_{c2}(T)$. Fortunately, the fluctuation region surrounding the superconducting-normal transition is negligible in conventional superconductors.

The situation has dramatically changed with the discovery of high temperature superconductors in 1986 [3]. These copper-oxide based materials have a critical temperature, $T_c$, of the order of 100 K, that is almost an order of magnitude larger than the previous "conventional" materials. Also their critical fields are quite extreme : $H_{c1}(T = 0)$ is of the order of $10^{-2}$ Tesla, whereas the upper critical field $H_{c2}(T = 0)$ is estimated to be as high as 100 T. Actually, with their very large anisotropy, due to their layered crystallographic structure, and their very short coherence length and large penetration depth, these materials are subject to significant fluctuations, such that the superconducting transition is smoothed out over a few degrees. Below this broad "crossover", the mixed state covers most of the phase diagram below $T_c$, as we can guess from the typical critical fields mentioned above.

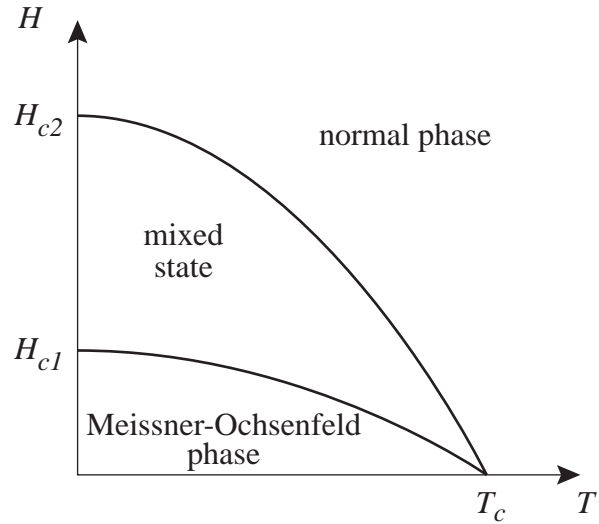

**Figure I-1 :** Mean field phase diagram of a conventional type II superconductor.

Moreover, due to the importance of fluctuations, the simple view of a triangular lattice of straight vortices emerging from the mean field theory is far from reality. The mixed state is a system of flexible interacting lines in the presence of disorder, generating a nontrivial statistical mechanics and leading to a very complex collection of various vortex phases [4,5,6]. The richness of this new vortex physics leads to consider the mixed state like a new state of matter, the *vortex matter*, emerging as an independent subject of investigation (see *e.g.* the review from Blatter *et al.* [6]).

The vortex matter is all the more interesting since its relevant parameters can very easily be varied over wide ranges : the vortex density, and thus the vortex repulsive interaction energy, can be changed over many orders of magnitude through the magnetic field, the thermal energy through the temperature, the pinning energy for example by introducing defects in the material through irradiation, and finally the anisotropy, by selecting the coupling energy between the crystallographic layers among various materials.

It is the competition between all of these energies, which are often of the same order of magnitude, that leads to the large variety of vortex phases. A generic picture of the resulting new phase diagram is represented in Fig. I-2. Note that, as we have mentioned above, the superconducting-normal transition has broaden into a wide zone of critical fluctuations (shaded area). For the sake of simplicity, this diagram focuses on the case of isotropic superconductors (or slightly anisotropic, like the sample used in this study), and thus does not exhaustively represent the richness of all the possible phases and transitions. However, it still allows us to discuss a very general feature of the vortex matter : the existence of a vortex liquid. Due to temperature induced fluctuations in the vortex assembly, the usual Abrikosov vortex lattice indeed melts into a liquid (paragraph 2.3), similarly to the melting of conventional matter.

The solid phase can be ordered and form a lattice (or an ordered Bragg glass, see paragraph 2.1), and in this case the corresponding phase transition has been proved to be of first order. But if the interaction with the defects present in the material is too strong, vortices are influenced by the pinning forces, and the disorder induced fluctuations leads to the formation of a disordered solid with a glassy structure (see paragraph 2.2). Therefore, a transition from a periodic vortex structure to an amorphous vortex solid can happen while the magnetic field is increased, since the latter accentuates disorder effects. Obviously, the position of this transition line, and thus the extent of the ordered phase, is significantly affected by the sample dependent amount and strength of pinning. The solid-liquid transition is of second order, whereas the lattice-glass transition is rather a continuous crossover.

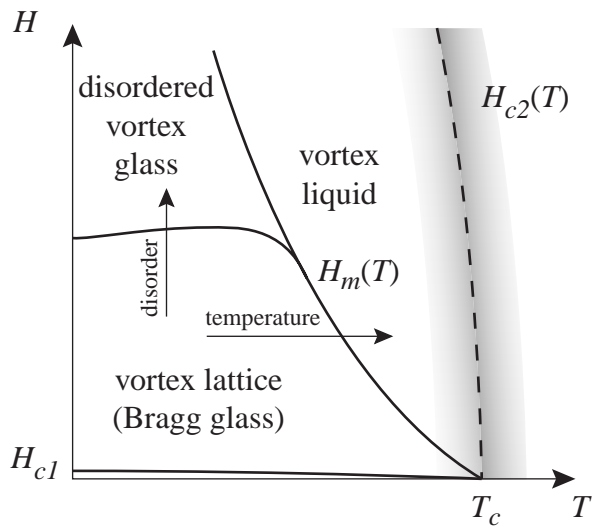

**Figure I-2 :** Phase diagram of an isotropic high temperature superconductor in the presence of point disorder.

Before we give a more detailed description of these vortex phases, we first have to mention that the above picture is considerably modified at least in two cases. First, when the defects in the material are spatially correlated (in contrast to the randomly distributed pointlike disorder we have implicitly assumed so far), the previously ordered crystalline solid phase then becomes a Bose glass, and the liquid-glass transition is again of second order. This specific vortex phase will be further described in paragraph 2.2.

Second, in more anisotropic materials, the coupling between the superconducting $CuO_2$ planes, separated by blocking layer, can weaken to the point at which the vortices can only form inside the bidimensional parallel layers, forming what is usually called *pancakes*. Depending on the structure in each of these layers, we can then again have a pancake liquid or solid. Since the decoupling transition is only reached at high magnetic fields, the phase diagram of strongly anisotropic compounds is therefore similar to what is shown in Fig. I-2 with an additional transition line at high fields separating the linear vortex phases from the pancake phases [5]. However, since the superconducting material used in this study is sufficiently weakly anisotropic to be presumably free of pancake phases, we will not discuss this concept in more details.

Finally, we also have to point out that, whereas we have only treated of the different vortex phases and the corresponding transitions from the static point of view, the dynamics of these phases are also extremely rich and fundamentally important. In a few words, vortices can be set into motion with the help of an electric current, provided the resulting driving force exceeds the pinning interaction of the vortices with material defects. Since a moving vortex generates an electric field, this motion is associated with the appearance of non-zero resistivity and energy dissipation. We can

# I

now understand how resistivity measurements in superconductors can be a very appropriate tool to investigate the vortex dynamics on a practical as well as on a fundamental level (see section 3). Note that it is also extremely important to understand and control pinning and vortex dynamics with the view of technical applications, in order to fully take advantage of the most appealing property of superconductors, namely its very low (ideally vanishing) resistivity.

In the next section, we provide more detailed information on the different vortex phases relevant for the present work, the aims and achievements of which is given in the last section of the chapter.

## 2. Vortex phases

The sample used in this study is a superconducting $YBa_2Cu_3O_{7-\delta}$ crystal, with $\delta$ of the order of 0.05. It has an orthorhombic structure, the *c*-axis being orthogonal to the $CuO_2$ layers. The anisotropy $\Gamma = (m_{ab} / m_c)^{1/2}$ is of the order of 5 to 7, where $m_{ab}$ and $m_c$ are the effective masses of the charge carriers in the *ab* plane and along the *c*-axis respectively [7]. In this compound, the oxygen vacancies provide a source of weak pointlike pinning centers for vortices. Oxygen atoms are also anisotropically distributed along chains, generating a small in-plane anisotropy, the *b* axis being very slightly greater than *a*. The consequence of this is the formation of twins, that are adjacent domains with successively exchanging *a* and *b* axes. The twin boundaries are oriented along the (110) directions, and are an important second source of vortex pinning (see section IV.2), this time generated by correlated disorder. Therefore, beside the discussion of the different phases represented in Fig. I-2, the Bose glass mentioned above will also be presented here.

### 2.1 The ordered vortex "lattice"

In the case of a perfect sample (in absence of pinning forces), the ideal triangular Abrikosov vortex lattice is formed at low temperature and low magnetic field. But, already in the presence of weak disorder (like we find in any real sample), the question of the effect of the small vortex pinning on the vortex assembly is not trivial. Of course, we can imagine quite intuitively that if we have only sparsely distributed weak pointlike pinning centers, vortices will only be slightly distorted locally without any impact on the overall vortex structure. However, since point defects in $YBa_2Cu_3O_{7-\delta}$, namely oxygen vacancies, have a large density, the collective effect of all of these pinning centers might be significant.

A formalization of this problem has been done by Larkin and Ovchinikov in their collective pinning theory [8]. The main idea is that the periodic order of the lattice is conserved in a characteristic volume $V_c$ given by a collective pinning length $L_c$ along the vortices direction, and a corresponding radius $R_c$ perpendicularly to the flux lines. Beyond these length scales, we find distortions of the order of the pinning potential range $\xi$. As a consequence, $V_c$ corresponds to the size of ordered vortex bundles, or Larkin domains, that can be considered to be pinned as a whole, and such that the vortices in these domains move collectively, by successive jumps of the order of $\xi$. In this model, it was also considered that for larger distances, the cumulated distortions might become of the order of the vortex lattice parameter, destroying the crystalline order through topological defects. The main consequence of this description is that, due to the accumulated distortions between all the domains over long distances, even weak pinning can suppress the long range order.

On the other hand, a new theoretical approach to the same problem has recently shown that the vortex lattice can take another form in the presence of weak disorder. Based on the early work of Natterman [9], Giamarchi and Le Doussal have introduced the notion of *Bragg glass* [10]. They have indeed shown that, due to the periodicity of the vortex lattice, the weak random disorder introduces displacements in the crystalline structure that only grow logarithmically with the distance, such that long range order is maintained in the resulting distorted lattice, which does not contain any topological defects such as dislocations. As a consequence, the structure function of the vortex phase shows Bragg peaks, hence the name of this glass. Note that we still have to speak about a glass – and not simply a lattice – in the sense that the system can occupy different equivalent metastable states in the random disorder potential. These different configurations are separated by energy barriers diverging at low currents, in contrast to the original Anderson–Kim theory [11] of flux creep for vortex bundles, which predicted a linear voltage-current characteristic at any nonzero temperature. As a consequence, this solid phase is truly superconducting, whereas the traditional vortex creep always leads to dissipations.

The Bragg glass have another fundamentally different dynamic behavior from the collectively pinned vortex solid introduced above. When it is set into motion, the ordered glass indeed flows through well defined static channels [12]. An important prediction, still without clear experimental support, is the existence of a transverse critical current : once the Bragg glass is moving in a given direction, the corresponding channel pattern can be modified to take another orientation only provided the transverse driving force is exceeding a certain threshold.

On the experimental point of view, the solid phase and the transition separating it from the vortex liquid have been characterized by many different methods [13]. For example, the spatial distribution of the vortices can be directly observed after decoration of the sample surface by small ferromagnetic particles [14] (Bitter decoration). Subsequent observation of the revealed vortex pattern under a microscope have confirmed the existence of a triangular lattice with or without dislocations. The structure function can also be measured directly by small angle neutron scattering [15]. A hexagonal symmetry pattern is observed in the vortex solid phase, with well defined peaks. These peaks then decrease as the vortex phase transition is approached, and the symmetry pattern is lost at the vortex lattice melting.

What probably represents the largest experimental effort at the present time is the investigation of the nature of the transition from the vortex solid to the liquid phase [16,17], instead of the characterization of the solid phase itself. Recent magnetic measurements have indeed shown that, in some regime at least, this transition is associated with a step in the magnetization [18]. More precisely, the vortex matter is more dense in the liquid phase (the magnetization is larger), just like the water-ice system. Moreover, sensitive calorimetric measurements have also revealed that a peak in the specific heat happens simultaneously, and is quantitatively consistent with the magnetization jump via the Clausius-Clapeyron relation for first order phase transitions [19]. Therefore, we have now strong evidence that the vortex liquid to solid phase transition can be of first order. However, the presence of critical points on the phase transition line limiting the first order regime have been reported on several occasions [17]. Although the question is still open, these changes of regimes seem to be related to the sample disorder, and will be discussed in paragraph 2.2 of chapter IV.

Note that, although it is not a thermodynamical investigation tool, the phase transition is also visible in resistivity measurements : at low currents, the resistivity suddenly drops to zero at the transition from the vortex liquid to the solid [20,21]. Even if neither the width of this step nor its height can bring information on the thermodynamic character of the transition, since the measurements

# I

are clearly performed out of equilibrium (the resistivity step simply corresponds to a sharp increase in the critical current when going from the vortex liquid into the vortex solid), the existence of hysteresis at the transition observed in Ref. [20] provides more evidence for a first order transition.

## 2.2 The vortex glasses

When more disorder is present in the sample, the vortex phase transition is not of first order anymore, as we have mentioned above. This means that the solid phase is altered by the pinning : the long range translational order is lost. We therefore speak about a glass instead of a lattice.

For the present work, we consider two different approaches to vortex glasses. The first one, due to Fischer *et al.* [22], which refers to a classical vortex glass, has some similarities with the theory of spin glasses. The central idea is that the system is qualified by a characteristic length $\xi_g$ and a relaxation time $\tau_g$, both diverging as the glassy transition at the temperature $T_g$ is approached :

$$\xi_g \propto |T - T_g|^{-\nu} \qquad \text{and} \qquad \tau_g \propto |T - T_g|^{-\nu z},$$

where $\nu$ and $z$ are universal critical exponents. This scaling behavior has repercussions on the transport properties, which in turn obeys some universal scaling relation. The non-linear voltage-current characteristic resulting from a diverging barrier at low currents makes this phase a true superconducting phase, similarly to the Bragg glass discussed above. The possible consequences of the scaling relations for the Hall effect will be given in the next chapter.

The second type of vortex glass, more relevant for the present work, appears in the presence of correlated strong disorder, such as columnar defects induced by irradiating the sample or twin planes. Its properties were studied mainly by Nelson and Vinokur [23], using a two-dimensional boson localization analogy. Let us write the $i^{\text{th}}$ vortex position as $\mathbf{r}_i(z)$, where $\mathbf{r}_i$ is a 2d vector localizing a position in the *ab* plane, and $z$ is the position along the *c*-axis (which is also the orientation of the correlated disorder). The energy of the system of many vortices with pair interactions and interaction with a random $z$-independent potential (the $z$-correlated defects) is then similar to the evolution of a system of interacting bosons in two dimensions with the same 2d potential. The vector $\mathbf{r}_i$ is the position of the $i^{\text{th}}$ boson, whereas $z$ becomes a time variable : the vortices propagate across the sample just as bosons move in two dimensions with time.

The result is again a glass (with diverging barriers at low currents, and a scaling behavior, though with a different universality class than the vortex glass discussed above), in which the vortices are pinned on the defects. Actually, its dynamic properties are very similar to those of the above vortex glass corresponding to point disorder. The Bose glass also has a phase transition to a vortex liquid, in which vortices are no longer localized on defects, and where kinks allow them to jump from one defect to the other. The main difference is that the transition temperature is shifted to larger values with respect to the phase transition observable in the presence of point defects : the stronger the disorder, the higher the transition temperature. As a consequence, the transition temperature will depend on the orientation of the magnetic field, since the pinning efficiency is maximum when the field is exactly aligned with the correlated disorder. Actually, there is even a threshold angle above which the Bose glass disappears, since localization on defects no longer occurs. An expected angular dependence is sketched in Fig. I-3. On the same plot is represented the transition temperature for pointlike defects, which is slightly angle dependent due to the material anisotropy. It is very interesting to point out that such a behavior closely corresponds to experimental data, as the cusp at low angles of Fig. I-3 is indeed observed in twinned $YBa_2Cu_3O_{7-\delta}$ crystals [24,25]. Our sample also follows the same angular dependence (chapter IV).

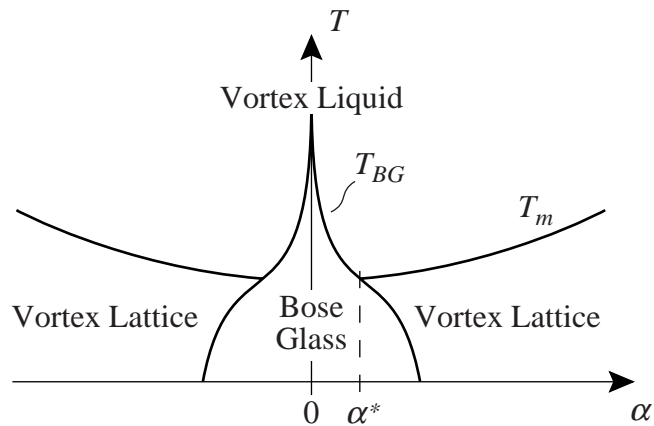

**Figure I-3 :** Relative dependences of the vortex phase transitions as a function of the angle $\alpha$ between the applied magnetic field and the *c*-axis. $T_m$ corresponds to the vortex lattice melting in the presence of point disorder, and $T_{BG}$ is the Bose glass transition for correlated defects. When $\alpha < \alpha^*$, the liquid freezes into a Bose glass, whereas at larger angles the correlation of pinning is no longer dominant, and the liquid freezes into a usual vortex lattice.

## 2.3 The vortex liquid

Above the melting line presented in Fig. I-2, vortex interactions and disorder become less important than fluctuations, and we have to consider a liquid of weakly confined lines. These lines are usually entangled, in the sense that their wandering away from the straight vortex configuration can be larger than the intervortex spacing.

Close above the melting line, the vortex liquid is still viscous with an energy barrier for vortex motion $U$ independent of the current, such that the resistivity takes the thermally activated flux flow (TAFF) form $\rho \propto e^{-U/kT}$ [26]. On the other hand, when the temperature is further increased, as we go deep into the liquid phase (farther from the vortex phase transition line), the energy barrier vanishes, and we reach the flux flow regime, where $\rho \approx \rho_n B / B_{c2}$, in which $\rho_n$ is the normal state resistivity. Note that this important relation is the result of the Bardeen–Stephen [27] and the Nozière–Vinen [28] theories, which will be discussed in details in the next chapter.

Finally, we have to say that the above picture of a liquid of line, even though it is the most widely accepted description of the phase found above the melting line, is not the only possible model. If the Abrikosov lattice is indeed observed at low temperatures by decoration experiments, and its existence can be tracked up to the melting transition by neutron diffraction [29], there is no direct evidence of the existence of vortices above this transition. Therefore, an alternative idea has emerged, considering that the so called liquid phase is actually a sort of normal state dominated by critical fluctuations of the order parameter [30]. The melting transition would then correspond to the actual creation of vortices (when reducing the temperature or magnetic field).

However, note that, although this description might seem to be reasonable as a substitute for a second order or continuous transition from a vortex liquid to a vortex glass, it is more difficult to see how it could lead to a first order transition at the end of the critical fluctuation region. Moreover, it is now not only established that the vortex melting is of first order in the most clean samples, but it has also been shown that both the magnitude of the corresponding thermodynamic properties (such as the entropy jump) and their temperature dependences are in very good agreement with a configurational melting of a vortex lattice within the London model [31].

## 3. The mixed state Hall effect

In the preceding section, we have given a short overview of the main vortex phases, with a few indications on their dynamic properties. However, as far as the Hall effect is concerned, almost all of the existing models consider only the dynamics of a single vortex, without taking into account vortex-vortex interactions, elasticity of the vortex lattice, vortex phases, *etc*.

It is the aim of the present section to present the open questions and challenges related to the mixed state Hall effect in type II superconductors, as well as to introduce the achievements of the present work in this context.

### 3.1 Single vortex equation of motion

We shall derive formally in the next chapter the equation of motion of a single vortex in the absence of pinning according to the Bardeen–Stephen [27] and the Nozière–Vinen [28] models. Here, we merely use the result to provide a general introduction to the subject. When mass terms are neglected, this equations of motion can be written as

$$(k_1 \mathbf{v}_T - k_2 \mathbf{v}_L) \times \mathbf{e}_z - \eta \mathbf{v}_L = 0, \tag{I.1}$$

where $\mathbf{v}_T$ is related to the applied transport current (providing the driving force) by $\mathbf{j}_T = n_s e \mathbf{v}_T$ (with $n_s$ the density of superconducting carriers, $e$ their charge, and $\mathbf{v}_T$ their velocity), $\mathbf{v}_L$ is the velocity of the straight vortex, and $\mathbf{e}_z$ a unit vector oriented along the vortex. With a very intuitive view emerging from, for example, the Bardeen–Stephen theory, the different terms can be justified as follows (see Fig. I-4) :

- $k_1 \mathbf{v}_T \times \mathbf{e}_z$ is the Lorentz force resulting from the interaction between the transport current, $\mathbf{j}_T = n_s e \mathbf{v}_T$, and the vortex self magnetic field,
- $-k_2 \mathbf{v}_L \times \mathbf{e}_z$ can be seen as the consequence of the Lorentz force acting on the charges localized in the vortex core, moving at the vortex velocity $\mathbf{v}_L$,
- $-\eta \mathbf{v}_L$ is simply a viscous drag force for the vortex motion, due to the scattering of these localized charges.

Note that $\eta > 0$, whereas both $k_1$ and $k_2$ have the same sign as the charge $e$ in this simple model. More precisely, the damping coefficient $\eta$ comes from the interactions between the excited, non-superconducting states localized in the vortex core and the imperfections of the background material made of positive ions, leading to a finite characteristic collision time for the excited states. The coefficients $k_1$ and $k_2$ have multiple and more complex origins. They are in general determined by hydrodynamic and core forces, arising from the interactions between the superflow around the vortex and the excitations of the superconducting condensate (both localized in the vortex core and delocalized in the whole charged fluid). The interface between the vortex core and the bulk super-fluid also plays an major but non-trivial role in the determination of the different factors in the equation of motion.

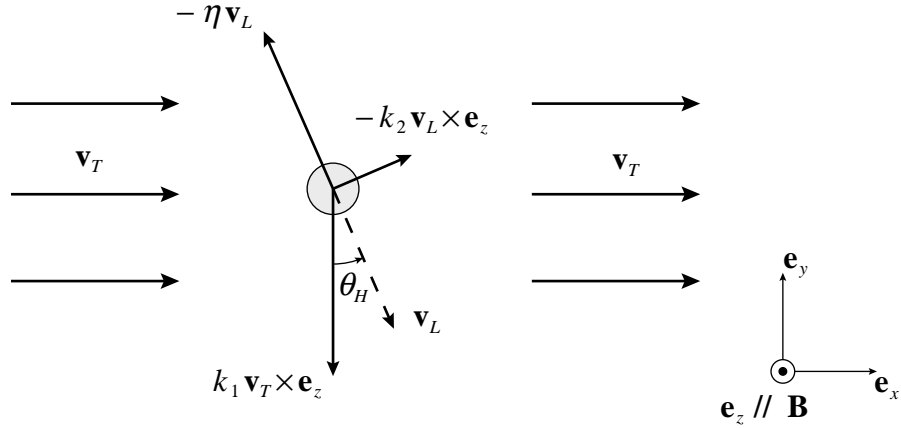

**Figure I-4 :** Representation of the different forces appearing in the vortex equation of motion (I.1).

From this equation, we can then get the vortex velocity from the applied current. The electric field response is then determined through Josephson's relation [32]

$$\mathbf{E} = -\mathbf{v}_L \times \mathbf{B}.$$

We can actually use this relation to directly relate the equation of motion to the different components of the resistivity tensor. For the sake of simplicity, we consider here a conductor which is invariant under a $\pi/2$ rotation around the $z$-axis, such that the properties along the $x$ and $y$ directions are equivalent. Hence the resistivity tensor (defined by the relation $\mathbf{E} = \rho \mathbf{j}$) has the form

$$\rho = \begin{pmatrix} \rho_{xx} & -\rho_{xy} \\ \rho_{xy} & \rho_{xx} \end{pmatrix},$$

where $\rho_{xx}$ is the longitudinal resistivity and $\rho_{xy}$ is the transverse resistivity. Note that in the present discussion, we assume that $\rho_{xx}$ is symmetric in the magnetic field, namely $\rho_{xx}(\mathbf{H}) = \rho_{xx}(-\mathbf{H})$, whereas $\rho_{xy}$ is antisymmetric, that is $\rho_{xy}(\mathbf{H}) = -\rho_{xy}(-\mathbf{H})$, such that $\rho_{xy}$ is in fact strictly identified as the Hall resistivity. The case of a symmetric part in the transverse resistivity $\rho_{xy}$ will be discussed in chapter V.

From the above definitions and the Josephson relation, it follows straightforwardly that

$$\mathbf{v}_L = \frac{n_s e}{B} \left( \rho_{xy} \mathbf{v}_T + \rho_{xx} \mathbf{v}_T \times \mathbf{e}_z \right).$$

Moreover, if we define the conductivity tensor $\sigma$ as the inverse of $\rho$,

$$\sigma = 1/\rho = \begin{pmatrix} \sigma_{xx} & -\sigma_{xy} \\ \sigma_{xy} & \sigma_{xx} \end{pmatrix},$$

we have

$$\sigma_{xx} = \frac{\rho_{xx}}{\rho_{xx}^2 + \rho_{xy}^2} \quad \text{and} \quad \sigma_{xy} = \frac{\rho_{xy}}{\rho_{xx}^2 + \rho_{xy}^2},$$

and the equation of motion becomes

$$\mathbf{v}_T = \frac{B}{n_s e}\left(\sigma_{xy}\mathbf{v}_L - \sigma_{xx}\mathbf{v}_L \times \mathbf{e}_z\right).$$

In any case, we see that, even in absence of complex situations such as pinning and vortex-vortex interactions, the simplest equation of motion (I.1) of a single vortex has terms that will contribute to the Hall effect : we get indeed

$$\sigma_{xy}^V = \frac{n_s e}{B}\frac{k_2}{k_1} \tag{I.2}$$

for the vortex contribution to the Hall conductivity. *The Hall component of the electric response of a type II superconductor is therefore an integral part of the vortex dynamics, and should not be neglected in the analysis of resistivity measurements.* A complete picture of the vortex response can indeed only be obtained from *both* longitudinal and transverse (Hall) components of the resistivity. Of course, the longitudinal resistivity $\rho_{xy}$ represents by far the main part of the vortex motion, since the Hall angle, defined by

$$\tan\theta_H = \frac{\rho_{xy}}{\rho_{xx}}$$

is experimentally usually very small (of the order of $10^{-1}$ to $10^{-2}$). However, as we will see in paragraph 3.3, there are some features in the vortex dynamics that are only observable in the Hall response.

Note that the above Hall conductivity is strictly related to vortex dynamics, hence the notation $\sigma_{xy}^v$. Although the equation of motion (I.1) corresponds to a hydrodynamic, mean-field description of the vortex contribution, the fluctuations might also be taken into account in this term, for example through a Ginzburg-Landau derivation. Aside from this vortex term, the total Hall conductivity also contains contributions from the delocalized excitations, expressed as the product of the normal state Hall conductivity $\sigma_{xy}^n$ times the normal-like excited states fraction $(1-g)$ (representing the normal fluid density). Finally, we can write

$$\sigma_{xy} = (1-g)\,\sigma_{xy}^n + \sigma_{xy}^v(g), \tag{I.3}$$

where we have stressed the fact that $\sigma_{xy}^v$ also depends in general on the superconducting density $g$.

### 3.2 Pairing symmetry and vortex structure

As we shall see in the next chapter, the vortex flow Hall contribution $\sigma_{xy}^v$ can be related to the microscopic properties of the vortex, its core structure, *etc*. It is therefore useful to first briefly present a few of these aspects. We start by recalling what the structure of a vortex is in the standard Ginzburg–Landau theory.

The vortex is associated with a cylindric distribution of supercurrents, trapping a quantized magnetic flux $\phi_0 = h/2e$. Both the current and the magnetic induction are distributed around the core axis over a distance $\lambda$, called the penetration depth. Inside the vortex core, the superconducting order parameter $\Psi$ is reduced to zero over a distance $\xi$, the coherence length (Fig. I-5). In high temperature superconductors, $\xi \ll \lambda$. The fact that $|\Psi|$ is suppressed in the core means that no superconducting state is found on the vortex axis.

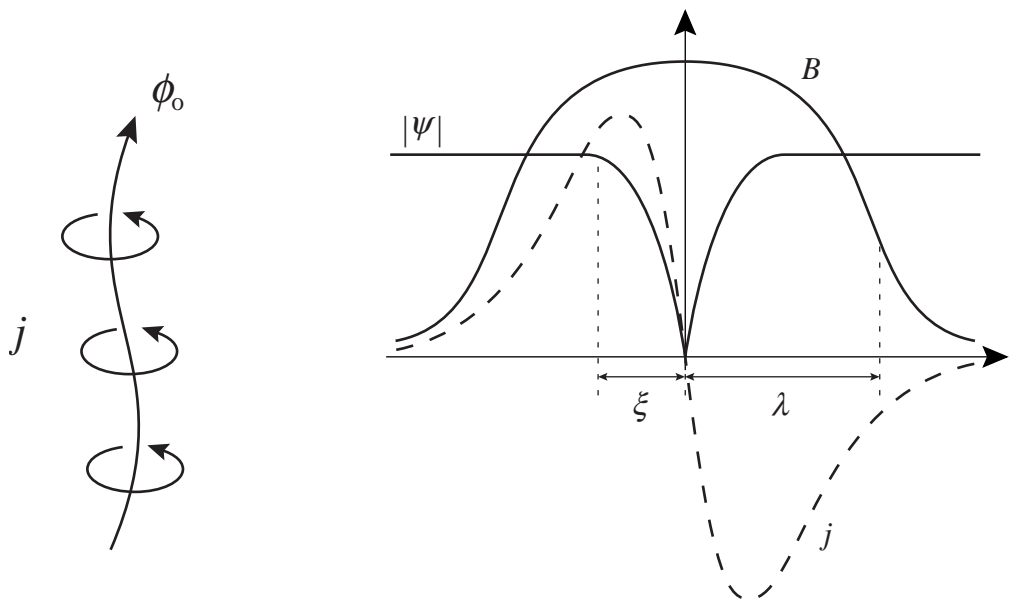

**Figure I-5 :** Left : schematic view of the supercurrents surrounding a flux line in a type II superconductor. Right : appearance of the resulting magnetic field *B*, current density *j*, and superconducting order parameter modulus $|\Psi|$ across such a vortex.

More precisely, the zone in which $|\Psi|$ is reduced corresponds to a zone of reduced superconducting gap $\Delta$, forming a potential well in which localized states can exist. In conventional BCS superconductors, Caroli, de Gennes and Matricon [33] have calculated the energy levels of these localized quasiparticles. It has also been shown that the density of these localized states is the same as in the normal state, so that the picture of a normal core of radius $\xi$ is often used for the vortex. The interlevel spacing, also called the minigap, which is of the order of $\Delta^2/\varepsilon_F$, where $\Delta$ is the BCS gap and $\varepsilon_F$ is the Fermi energy, is unfortunately too small to be observed directly in conventional superconductors (and moreover a level broadening due to scattering by impurities can easily become comparable to the minigap). Nevertheless, Hess *et al.* have shown the existence of (unresolved) low-lying states in the vortex core of $NbSe_2$ with the help of scanning tunneling microscope (STM) spectroscopy [34].

However, recent experiments in high temperature superconductors have shown a different behavior. For $YBa_2Cu_3O_{7-\delta}$, Maggio-Aprile *et al.* [35] have shown that only two bound states were observable in the vortex cores by STM spectroscopy. This result is actually consistent with the Caroli–de Gennes–Matricon picture, since the ratio $\Delta/\varepsilon_F$ in this material is much larger than in conventional superconductors. As a consequence, only a few levels separated by the minigap can exist in the well of depth $\Delta$.

The same type of investigation on $Bi_2Sr_2CaCu_2O_{8+\delta}$ at different oxygen concentrations $\delta$ leads to the conclusion that there is no bound state at all in the vortex cores [36]. The larger value of the gap $\Delta$ (and thus of the ratio $\Delta/\varepsilon_F$) is a possible explanation. However, other surprising features are observed in the density of states (DOS) of this compound. For example, the superconducting gap, which is supposed to close at the superconducting to normal transition, is still clearly visible as a large dip in the DOS. This dip is called the pseudogap, has the same width as $\Delta$, and remains visible up to temperatures much higher than $T_c$ [37]. Note that this pseudogap is generally only visible in underdoped materials (that is materials with a charge carrier content lower than the doping providing the highest $T_c$, see Fig. I-6) in which it can even be measured at room temperature [38], al-though it is also reported to exist in overdoped $Bi_2Sr_2CaCu_2O_{8+\delta}$ in Ref. [38]. This reveals the efficiency of the pairing interaction far above $T_c$, a mechanism quite different from the usual BCS picture.

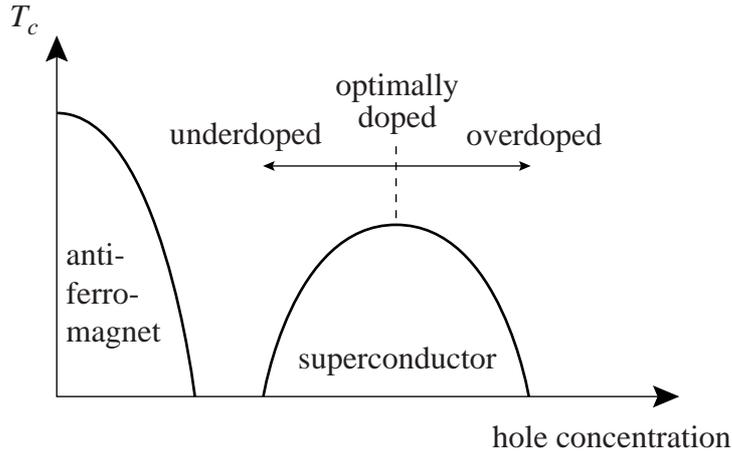

**Figure I-6 :** Typical dependences of the metallic-superconducting and the insulating-antiferromagnetic transition temperatures on the charge carriers concentration in most of the cuprates.

More precisely, this pseudogap in the density of states first opens in some directions of the Brillouin zone at high temperature, whereas the rest of the Fermi surface still has a normal Fermi liquid behavior, as noted from measurements by angle resolved photoemission spectroscopy [37,39]. As the temperature is reduced, this gap spread along the Fermi surface. When the superconducting transition is reached, the gap is then open for all the directions of the Brillouin zone, except for four nodes, at the corners of the Fermi surface, where the spectrum remains gapless.

It is now widely accepted that this striking momentum dependence of the electronic properties corresponds to the *d*-wave symmetry of the order parameter : in simple terms, the wave function of each pair changes sign four times around the Fermi surface. The *d*-wave symmetry of the superconducting state has been verified experimentally on many occasions for many cuprate compounds, for example by the observation of a fractional flux quantum in a superconducting ring laying at a tricrystal junction [40], or the diffraction pattern of the Josephson tunneling in corners of crystals [41]. This *d*-wave nature of the pairing mechanism also has important consequences on the localized vortex core states, since then the wave function has nodes in four different directions. As a consequence, it was argued that no bound state can exist in a *d*-wave superconductor vortex, because the potential well depth related to the gap $\Delta$ is also anisotropic, and vanishes in four directions, such that the particles can in some way "escape" from the well in these directions (they are in other terms infinitely extended in these directions). However, as we saw above, localized states can be found in vortex cores, at least in some materials. From recent rigorous calculations [42], the ideas of *d*-wave symmetry and vortex bound states have been reconciled by assuming a higher symmetry, with a superimposition of two *d*-wave components (actually a $d_{x^2-y^2} + i\, d_{xy}$ symmetry).

Although the existence of *d*-wave components in high temperature superconductors is now well accepted, much more has to be done to understand on the one hand the origin of this symmetry (related to the pairing mechanism), and on the other hand its consequences for the vortex structure and the vortex dynamics.

Based on the above experimental evidences on the Fermi surface and symmetry of the order parameter, Geshkenbein *et al.* have developed a new theory [43], presumably only valid for under-doped high temperature superconductors, or more precisely for superconductors showing a pseudogap above the critical temperature $T_c$. Actually, the question of whether this pseudogap is a general feature in cuprates or is specific to underdoped systems is still open : it is not clear yet if it is really absent from overdoped materials, or if it is only less pronounced and more difficult to observe experimentally, though still present [38].

The idea of this model is that the formation of the pseudogap corresponds to the creation of non-conducting pairs in the corners of the Brillouin zone. When $T_c$ is reached, these preformed bosons then form a superconducting condensate. Note that the idea of a Bose condensation of preformed pairs had been proposed earlier (see for example Ref. [44]). However, these previous models lead to some predictions that do not agree with experimental facts like the width of the fluctuation region and the Hall effect. The crucial difference in this latest model [43] is that the Bose condensation happens against the background of the Fermi liquid, the (unpaired) fermions playing an active role in the process, as they interact with the bosons. This promising model also deduces expressions for the Hall conductivity, as will be discussed in the next chapter.

### 3.3 Hall anomaly and scaling

*a) Hall anomaly*

We now turn back to the Hall effect in the mixed state. As we have seen in paragraph 3.1, the standard flux flow models predict that the vortex Hall conductivity (I.2) has the same sign as in the normal state, and can be seen as a consequence of the Hall effect in the normal cores of the vortices.

However, in high temperature superconductors (and in some conventional materials too), the Hall effect sign changes in the superconducting state (see Fig. I-7). This *Hall anomaly* was first observed in $YBa_2Cu_3O_{7-\delta}$ by Galffy *et al.* [45]. Thereafter, the same kind of behavior was reported for $Bi_2Sr_2CaCu_2O_{8+\delta}$ [46]. In this latter compound, as well as in other very anisotropic cuprates, a second sign change, back to the normal state one, is often seen (note that a detailed literature review is given in chapter III). As it is now widely accepted that this sign change is a genuine effect of vortex dynamics, and not a spurious Hall effect resulting from extrinsic effects, such as materials defects, it has become a challenge from the theoretical point of view to explain why sign changes appear in the combination of the different terms of the total Hall conductivity (I.3). Some of the tentative theories for the Hall anomaly are presented in the next chapter.

We can see from Fig. I-7 that the sign change of the Hall effect is not associated with any feature in the longitudinal resistivity. More precisely, the longitudinal resistivity is known to be quite consistent with the standard flux flow models (Bardeen-Stephen and Nozières-Vinen) when pinning can be neglected (and with the collective pinning picture in the other case), but these standard models fail to explain the Hall behavior. This demonstrates that, as we have mentioned in paragraph 3.1, the Hall effect is a necessary component for a complete picture of the vortex dynamics, as it reveals some features that could not be guessed from the longitudinal resistivity only.

One of the principal experimental controversies for the flux flow Hall effect that is still not completely resolved is the pinning dependence of the Hall conductivity $\sigma_{xy}$. Whereas both resistivities $\rho_{xx}$ and $\rho_{xy}$ are obviously disorder dependent, there is no consensus on the resulting effects on the Hall conductivity, as contradicting results are reported (see chapter III for a review).

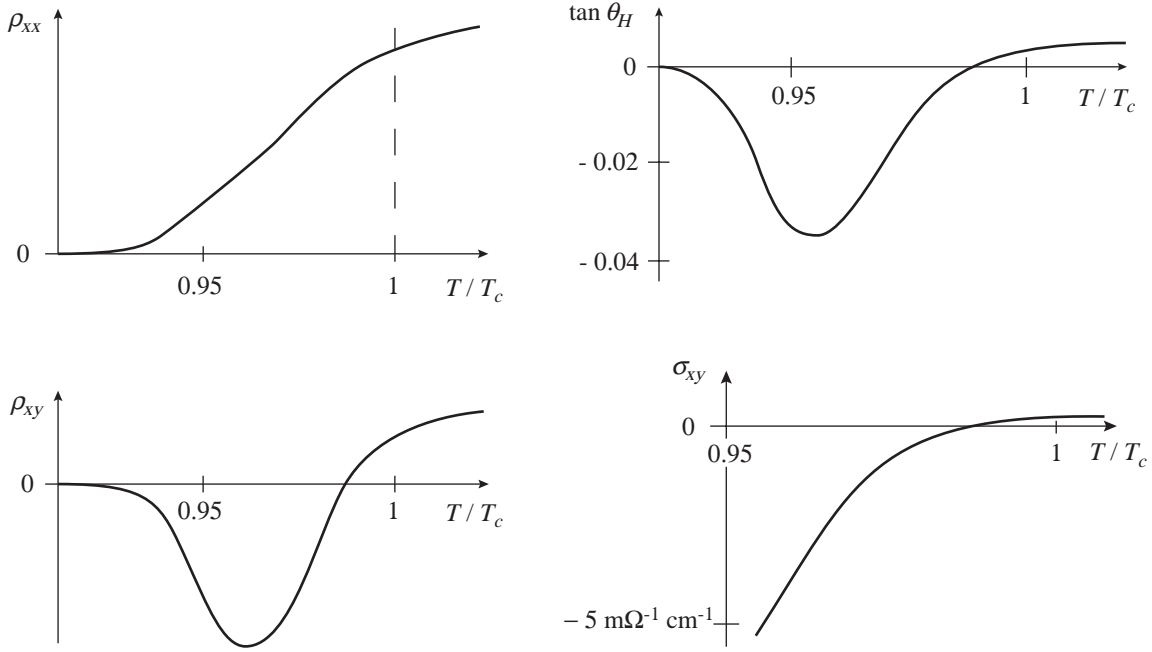

**Figure I-7 :** Schematic temperature dependences of the longitudinal resistivity $\rho_{xx}$, Hall resistivity $\rho_{xy}$, Hall angle $\tan\theta_H$ and Hall conductivity $\sigma_{xy}$ for $YBa_2Cu_3O_{7-\delta}$ (at a moderate magnetic field $B = 2$ T), showing the Hall sign reversal below $T_c$.

The influence of pinning can also become relevant in the presence of correlated and anisotropic pinning, just like in the case of twin boundaries. When vortices tend to move along such extended defects, the resistive response will have a transverse component that is even with respect to the magnetic field (thus the name *even Hall effect*), in contrast to the Hall effect which is by definition odd in the magnetic field. The question is then to know to what extent this guided motion also influences the true (odd) Hall effect, or if the latter can be considered as independent of this guided motion. This question of guided motion along twin boundaries is addressed in chapter V.

*b) Hall scaling law*

Another striking feature is also observed in the Hall behavior of all the high temperature superconductors : the vanishing part of the data, at low temperature or low field, follows the scaling relation

$$\rho_{xy} = A\,\rho_{xx}^{\beta}$$

as first noted by Luo *et al.* [47]. This behavior is easily seen on a log-log plot of $\rho_{xx}$ versus $\rho_{xy}$, as shown in Fig. I-8.

Although this scaling seems to be very general (it is for example observed for any sign of the Hall effect, after one or two sign changes, or even in the absence of Hall sign change), the reported values of $\beta$ are quite scattered (see chapter III for a review of the literature). The highest reported exponent is $\beta = 2$, whereas values as low as 0.8 can be found. The upper limit $\beta = 2$ can be justified by noting that $\beta > 2$ would imply a vanishing Hall conductivity as the resistivities vanish, since $\sigma_{xy} = \rho_{xy}/(\rho_{xx}^2+\rho_{xy}^2) \approx \rho_{xy}/\rho_{xx}^2 \propto \rho_{xx}^{\beta-2}$, knowing that $\rho_{xx} \gg \rho_{xy}$. Therefore, if we assume that both components of the conductivity $\sigma_{xx}$ and $\sigma_{xy}$ should diverge simultaneously, the condition $\beta < 2$ has to be satisfied.

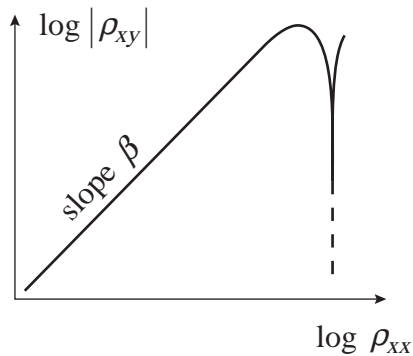

**Figure I-8 :** Log-log plot of the Hall resistivity absolute value $|\rho_{xy}|$ as a function of the longitudinal resistivity $\rho_{xx}$ in the low temperature range, bringing to the fore the Hall scaling relation $\rho_{xy} \propto \rho_{xx}^{\beta}$. The dashed line represents the divergence corresponding to the Hall resistivity sign change (see Fig. I-7).

To our knowledge, four theories have been proposed in an attempt to explain this scaling relation, and will be presented in the next chapter. Unfortunately, two of these theories predict a constant universal exponent, in contradiction to the strongly sample dependent experimental data. The other two are more compatible with the experimental diversity, but they are related to corresponding scenarios for the Hall anomaly, which are in turn difficult to accept : one is based on the influence of pinning, the other on the existence of vacancies in the vortex lattice, both incompatible with the existing observations of a Hall sign change in the critical fluctuation region, very close to or even above $T_c$ (see chapters II and III for details).

## 4. Summary of the present work

As we see, the Hall effect in high temperature superconductors is far from being fully understood. Among the open questions is the effect of vortex-vortex and pinning interactions on the Hall behavior. We have seen that different solid phases can form thanks to the repulsive interaction between the vortices and the influence of disorder (pinning). It would be interesting to know what are the Hall properties of these vortex phases, to help establish what are the important factors entering the mixed state Hall effect.

Conversely, we have also noted at the beginning of this section (page 10) that the Hall effect is necessary to investigate all the aspects of the vortex dynamics, to get a complete picture of all the forces acting on the vortices. As a consequence, a study of the Hall effect in different well identified vortex phases can obviously bring crucial information on the dynamics of these phases.

In summary, the central question of this work is : what are the correlations between the different vortex phases and the Hall effect ? Very few similar investigations have been performed up to the present time. Very recently, Morgoon *et al.* have studied both vortex guided motion [48] and Hall effect [49] in unidirectionally twinned single crystals. Although one of these samples might possibly show a sharp vortex lattice melting (as is apparent by comparing their data with the results of the present work – but the vortex phase transition is not discussed in the above papers), all the others appear to be rather disordered. It is then possible that the irreversibility line, at which the vortices become pinned, lies above the phase transition, such that it prevents the formation of an ordered phase (no step in the longitudinal resistivity, nor any other signatures of a phase transition, are shown nor reported for these samples). Moreover, the *odd* Hall effect shows surprising features, not quite reproducible from one sample to the other and apparently related to twin effects.

Actually, the only work reporting Hall measurements across a vortex phase transition is made by Wöltgens *et al.* [50], using $YBa_2Cu_3O_{7-\delta}$ films. However, the glassy phase transition, identified through vortex glass scaling, reveals the presence of significant disorder in the system. Only the Hall scaling is shown in this paper : no data for the Hall conductivity are presented. Moreover, the effect of the glassy transition on the Hall behavior is barely touched on. Note that Harris *et al*. [51] also measured the Hall effect in untwinned $YBa_2Cu_3O_{7-\delta}$ crystals showing a very clear resistive vortex phase transition, but the measurements were done at too low current to set the vortex solid into motion, therefore remaining limited to the vortex liquid phase.

Although we shall further discuss these pioneering works in the light of the present new results, it already appears clearly that a detailed study of the Hall effect behavior at the vortex phase transition is lacking. The plan for the present work can be sketched as follows :

- choose a very clean, twinned crystal
- identify a true vortex *phase* transition
- measure both components of the resistivity, $\rho_{xx}$ and $\rho_{xy}$, as a function of the applied magnetic field and the temperature
- analyze the even part of the transverse voltage in terms of vortex guided motion
- deduce from the odd transverse voltage the corresponding Hall conductivity $\sigma_{xy}$, as well as study the Hall power law scaling behavior for $\rho_{xy}(\rho_{xx})$, and correlate the data with the vortex phase transition
- check for disorder (pinning) influence by changing the current density and the magnetic field orientation with respect to the twin planes.

To conclude this chapter, we briefly summarize the obtained results that will be presented in chapters V and VI. We first discuss in a few words the results for the even Hall effect. We observe indeed a preferential motion of the vortices along the twin planes. This guided motion occurs already in the vortex liquid, and becomes more and more marked as the vortex phase transition is approached. Note that there is no sharp change in this guided motion as the vortex solid is reached : the vortex motion is simply progressively more and more influenced by the twins direction, just as in the smooth continuation of the vortex liquid trend. By measurements performed at different current directions with respect to the twin planes, we conclude that, whereas the direction of vortex motion is strongly influenced by the twin planes, the overall vortex mobility is itself not much affected, much less than what might be guessed from usual measurements at as single current orientation (which is often at 45 ° from the twin planes). The detailed results for the even Hall effect are shown and discussed in chapter V.

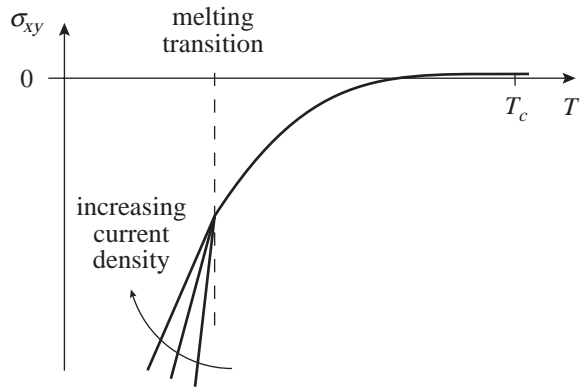

**Figure I-9 :** Change of behavior of the Hall conductivity $\sigma_{xy}$ at the vortex phase transition, for a magnetic field tilted away from the twin planes. The same effect is observed for measurements as a function of the magnetic field.

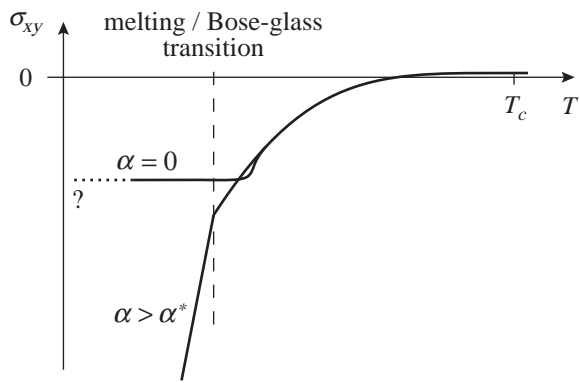

**Figure I-10 :** Comparison of the behavior of the Hall conductivity $\sigma_{xy}$ for a magnetic field tilted away from the twin planes ($\alpha > \alpha^*$) and parallel to the planes ($\alpha = 0$). In the latter case, the conductivity seem to remain constant around the vortex phase transition. At lower temperature, the noise prevents the observation of the trend for $\sigma_{xy}$ (dotted line).

Finally, we come to the discussion of the (odd) Hall effect results. Of course, consistently with the preceding works on YBa$_2$Cu$_3$O$_{7-\delta}$ samples, we observe a Hall anomaly, just as is represented in Fig. I-7. However, when the angle between the magnetic field and the twin boundaries is large enough ($\alpha > \alpha^*$, see Fig. I-3), the Hall conductivity changes abruptly at the (presumably) first order phase transition (see Fig. I-9). The Hall scaling law is verified for $\beta = 1.4$. It is completely independent of magnetic field, temperature, current density and field orientation with respect to the

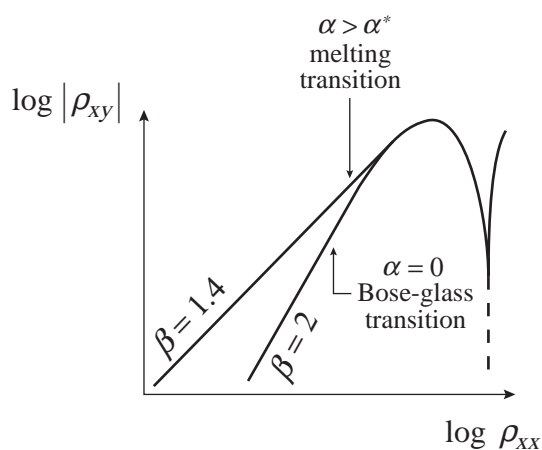

**Figure I-11 :** Log-log plot of the Hall resistivity absolute value $|\rho_{xy}|$ as a function of the longitudinal resistivity $\rho_{xx}$ (see Fig. I-8), revealing that the scaling exponent is $\beta = 1.4$ for a magnetic field tilted away from the twin planes ($\alpha > \alpha^*$) and $\beta = 2$ for a field parallel to the planes ($\alpha = 0$). It also appears that the scaling law is unaffected by the respective vortex phase transitions.

# I

twin boundaries (as long as $\alpha > \alpha^*$). The most interesting observation is that the scaling relation is absolutely unaffected by the vortex phase transition.

The current dependence and the exponent value clearly show that the Hall conductivity *is pinning dependent*. This pinning dependence of the new Hall behavior observed in the solid phase also means that it is probably not directly related to effects of the vortex bound states or analogous microscopic and electronic processes that were discussed in paragraph 3.2.

When the twins induce a Bose-glass phase (see Fig. I-3), the Hall behavior becomes very different. First, the Hall conductivity is apparently constant in the vortex solid phase (Fig. I-10). It is then not clear whether $\sigma_{xy}$ finally diverges, vanishes, or remains constant as $\sigma_{xx}$ diverges, since the observation in the low temperature limit is too sensitive to the noise. Secondly, the Hall scaling exponent is precisely $\beta = 2$ (see Fig. I-11). Again, the scaling relation is unaffected by the vortex liquid to solid phase transition. These results are further discussed in chapter VI.

# CHAPTER II   *HALL EFFECT REVIEW : THEORY*

## 1. Overview

In this chapter, we discuss the pertinent theoretical models concerning the Hall effect in the mixed state of type II superconductors. We start with the standard Bardeen-Stephen and Nozières-Vinen models for the flux flow. These theories are based on a hydrodynamic approach within a two fluid description, in which the charged fluid is split into a superfluid part and a dissipative normal-like component. The vortex is considered as a core of normal fluid. The asymptotic flow is supposed to be purely superfluid. The interface between the superfluid and the normal core is assumed to be of negligible width in the Nozière-Vinen model, whereas a smooth transition region is introduced in the Bardeen-Stephen theory. In both cases, only the interaction between the superfluid and the vortex core, as well as the usual scattering of the normal component (which would lead to the normal resistivity in a non-superconducting system) are considered; no pinning force on the vortex is introduced. As we shall see, both lead to the same expression for the flux flow, which is in good agreement with the experimental findings for the free flux flow longitudinal resistivity, *e.g.* at high enough temperature, such that pinning is not relevant. They also both predict a finite Hall resistivity, however with a slight difference in the field dependence. Unfortunately, the Hall effect in these models is expected to have the same sign as in the normal state. Therefore, these two theories cannot satisfyingly explain the data for the Hall anomaly, namely a Hall sign change (see Fig. I-7 in the preceding chapter) in the mixed state of type II superconductors (and some conventional superconductors as well).

As a consequence, it is necessary to look for other descriptions of the mixed state Hall effect. In section 3, we present what we think are the most relevant theories intended to explain this Hall anomaly. Some of them are still based on a hydrodynamic approach. For example, the Nozière-Vinen model is modified in order to take into account the effects of the pinning force (paragraph 3.1). Pinning can also be considered to be so strong that it completely and rigidly immobilizes the whole vortex lattice, except for a few lattice defects, *e.g.* vacancies, which then control the overall dynamics of the vortex assembly (paragraph 3.2). On the other hand, two other theories, actually in relation with each other, consider microscopic processes concerning the pairing interaction and the electronic states forming the superconducting phase (paragraphs 3.3 and 3.4). One is based on the phenomenological Ginzburg-Landau theory, in which a particle-hole asymmetry has to be introduced to allow for a non-vanishing Hall effect. This asymmetry is represented by the imaginary part of the relaxation time for the superconducting order parameter $\Psi$. This imaginary relaxation time can be obtained from microscopic considerations either in the frame of the usual weak coupling BCS theory, or for an alternative interaction leading to a non-BCS superconductivity, more consistent with recent experiments on some high temperature superconductors. The second model actually treats the entire problem on a microscopic level, and shows the equivalence with the phenomenological time dependent Ginzburg-Landau and imaginary relaxation time picture.

Finally, the scaling relation between the longitudinal and the Hall resistivities (see Fig. I-8 in the preceding chapter) also deserves an explanation. The theoretical predictions for this scaling law are grouped in section 4, even though two of these theories are in fact direct consequences of models for the Hall anomaly summarized above and presented in section 3 (namely the two "hydrodynamic" approaches including pinning, paragraphs 3.1 and 3.2). In these two cases, the exponent $\beta$ apparently can take different values in the range $1 \leq \beta \leq 2$, consistent with the rather scattered experimental data. The other two are specifically developed to explain this Hall scaling : one again introduces pinning, in the form of a very simplified average pinning force, again from a hydrodynamic point of view, and leads strictly to $\beta \equiv 2$, but does not predict the sign of the Hall effect; the other is more a consideration of universal scaling around critical phenomena, in this case the vortex liquid to solid phase transition, where scaling exponents are used to describe a power law divergence of the quantities which are assumed to diverge at the transition. A universal critical exponent is then introduced for the Hall conductivity, and is related to the other known critical exponents. The Hall resistivity scaling is then a consequence of this universal scaling at the transition, such that the scaling exponent $\beta$ can be expressed as a function of the other exponents as well. It is then expected to be between 1 and 2, but should be universal (not sample dependent).

## 2. The standard Bardeen-Stephen and Nozières-Vinen models

In this section, we review the two original models [27,28] phenomenologically describing the motion of vortices in a charged superfluid. We first derive the usual Magnus force in the case of an ideal superfluid with non-dissipative vortex motion. Then we introduce the dissipative force, to reach the single vortex equation of motion already discussed in the preceding chapter, paragraph 3.1.

We consider a single, straight vortex line, oriented along the *z*-axis, and immersed in an applied superfluid flow $\mathbf{v}_T(\mathbf{r})$ in the laboratory frame of reference (which is also the referential of the superconducting material, the lattice of positive ions). We choose S.I. units, thus $\mathbf{B} = \mu_o \mathbf{H}$. In the following we write $\mathbf{b} = \mu_o \mathbf{h}$, the microscopic (local) induction, and $\mathbf{B} = \langle \mathbf{b} \rangle$, the macroscopic (averaged) induction. The following discussions are given within the framework of a two fluid model [52]. The total supercurrent is $\mathbf{j}_s = n_s e \mathbf{v}_s$, where $n_s$ and $\mathbf{v}_s$ are respectively the density and the velocity of the superfluid component and the carriers charge is $e < 0$ for electrons. For the sake of simplicity, we make the following assumptions :

  i) An extreme type-II superconductor, namely $\xi \ll \lambda$.
  ii) A two-dimensional geometry, the vortex remaining parallel to the *z*-axis, and the fluid flow lying in the (*xy*)-plane.
  iii) A very low temperature, such that there are no normal electrons outside the vortex core.
  iv) No pinning.

In this case the transport flow $\mathbf{v}_T$ can be considered to be constant over the vortex core.

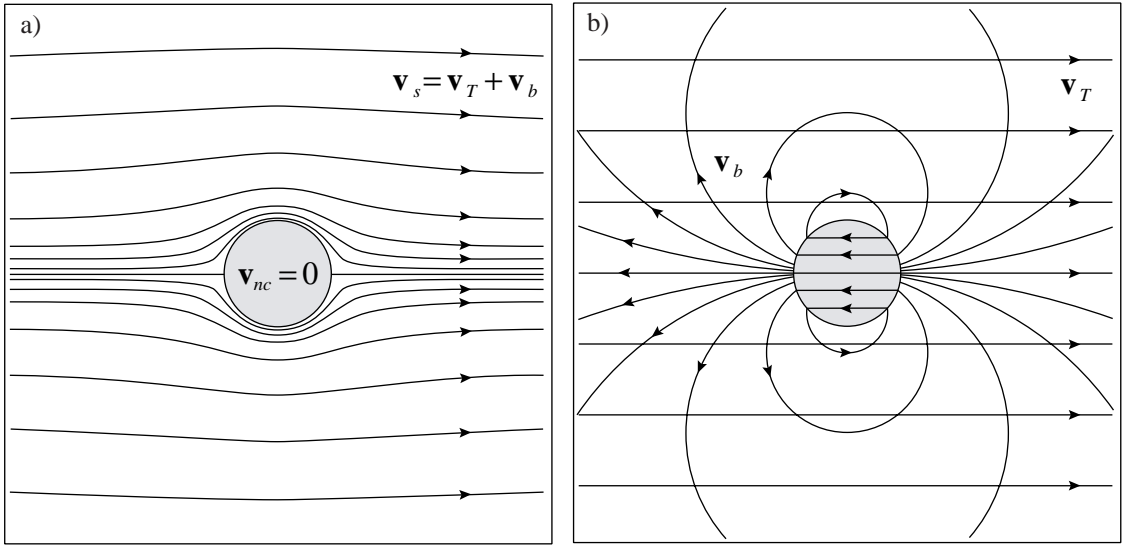

**Figure II-1 :** a) Flow lines of a transport current flowing around a rigid cylindric core. b) Decomposition of the same pattern into the homogeneous transport flow and a dipolar backflow given by Eq. (II.1) with $\mathbf{v}_{nc} = 0$ (the vorticity has been omitted for more clarity).

We also assume that the vortex line moves at a constant velocity $\mathbf{v}_L$. This motion induces a current in the core $\mathbf{v}_{nc}$, the exact distribution of which will depend on the core structure and on the vortex velocity itself. For example, in an extreme case corresponding to a vortex at rest ($\mathbf{v}_L = 0$), like in the presence of strong pinning, the supercurrent would be flowing *around* the vortex (see Fig. II-1), such that no fluid flows across the core ($\mathbf{v}_{nc} = 0$). This requires the existence of a dipolar backflow $\mathbf{v}_b(\mathbf{r})$ to "screen" the applied transport flow $\mathbf{v}_T$, as shown in Fig. II-1.

On the other hand, if the vortex follows the applied flow ($\mathbf{v}_L = \mathbf{v}_T$), this means that it is actually at rest in the *transport flow* frame of reference, thus, from the same argument as above, the core velocity is now zero in this reference frame. Returning to the laboratory reference, we find $\mathbf{v}_{nc} = \mathbf{v}_T$ : the applied flow is not screened at all in the core, the backflow $\mathbf{v}_b(\mathbf{r})$ is zero. These extreme examples show that in general circumstances, the total supercurrent has to be decomposed as

$$\mathbf{v}_s(\mathbf{r}) = \mathbf{v}_T + \mathbf{v}_v(\mathbf{r}) + \mathbf{v}_b(\mathbf{r}),$$

where $\mathbf{v}_v$ is the circular velocity distribution of the vortex itself :

$$\mathbf{v}_v(\mathbf{r}) = \frac{m}{e} \frac{\phi_o}{2\pi r} \mathbf{e}_\theta,$$

and $\mathbf{v}_b$ depends on $\mathbf{v}_{nc}$ and the exact core structure.

However, in both models considered here, the backflow $\mathbf{v}_b$ is neglected, though for different reasons. Nozière and Vinen (NV) [28] consider the vortex core as a sharply bounded cylinder of radius $a$, consistent with microscopic calculations from Caroli *et al.* [33]. In this case, for a given $\mathbf{v}_{nc}$, the backflow is given by the standard result of hydrodynamics :

$$\mathbf{v}_b(\mathbf{r}) = \begin{cases} \mathbf{v}_{nc} - \mathbf{v}_T & \text{for } r < a \\ \nabla\left[ (\mathbf{v}_T - \mathbf{v}_{nc}) \cdot \mathbf{r} \frac{a^2}{r^2} \right] & \text{for } r > a \end{cases} \quad \text{(II.1)}$$

which is negligible outside the core. Bardeen and Stephen (BS) [27] consider a more detailed local model with a normal core of radius *a* surrounded by a normal to superconducting transition region, which allows them to carefully investigate the continuity of potential and velocity components at the core boundary. They show from energy conservation that $\mathbf{v}_b = 0$, and thus that $\mathbf{v}_{nc} = \mathbf{v}_T$. In the NV model, consistently with the calculations of BS (and even if their core modelization is different), they also assume $\mathbf{v}_{nc} = \mathbf{v}_T$.

In summary, both models come to the same picture : a core of radius $a \approx \xi$ is composed of normal electrons, backflow is neglected, and the velocities are

$$\mathbf{v}_s(\mathbf{r}) = \mathbf{v}_T + \mathbf{v}_v(\mathbf{r}) \tag{II.2}$$

for the superfluid component outside the vortex core and

$$\mathbf{v}_{nc} = \mathbf{v}_T \tag{II.3}$$

in the normal core.

## 2.1 The Magnus force

As a first step, we want to calculate the driving force on the vortex exerted by the superfluid. To separate this force from the other contributions like the dissipative scattering of normal charges by impurities, we first make an additional assumption :

   v)   A clean material, to avoid frictional forces on the vortex.

The dissipative force will be reintroduced once we have determined what the driving force is.

Well outside the core ($r \gg \xi$), the superfluid then obeys London's and Maxwell's equations :

$$\nabla \times \mathbf{v}_s + \frac{e}{m}\mathbf{b} = 0 \tag{II.4}$$

$$\nabla \times \mathbf{b} = \mu_0 N e \mathbf{v}_s \tag{II.5}$$

$$\nabla \times \mathbf{E} = (\mathbf{v}_L \cdot \nabla)\mathbf{b}. \tag{II.6}$$

Note that the last equation derives from Maxwell's equation $\nabla \times \mathbf{E} = -\partial \mathbf{b}/\partial t$ knowing that the magnetic field is constant in the vortex frame :

$$\frac{d\mathbf{b}}{dt} = \frac{\partial \mathbf{b}}{\partial t} + (\mathbf{v}_L \cdot \nabla)\mathbf{b} = 0.$$

In (II.5), we have used the total electron density $N$ instead of $n_s$, since from the assumption (iii) we know that the normal fluid density $n_n = 0$. From (II.6) we see that

$$\mathbf{E}(\mathbf{r}) = -\mathbf{v}_L \times \mathbf{b}(\mathbf{r}) - \nabla V(\mathbf{r}) \tag{II.7}$$

with an arbitrary potential *V*.

The superfluid flow is governed by Euler's equation

$$\frac{d\mathbf{v}_s}{dt} = \frac{\partial \mathbf{v}_s}{\partial t} + (\mathbf{v}_s \cdot \nabla)\mathbf{v}_s = \frac{e}{m}(\mathbf{E} + \mathbf{v}_s \times \mathbf{b}) - \frac{1}{m}\nabla\mu, \tag{II.8}$$

where $\mu$ is the chemical potential. Note that, following our initial assumptions, the total energy in the vortex frame is constant :

$$\mu_{tot} = \mu + eV + \frac{m}{2}(\mathbf{v}_s - \mathbf{v}_L)^2 = \text{const}, \tag{II.9}$$

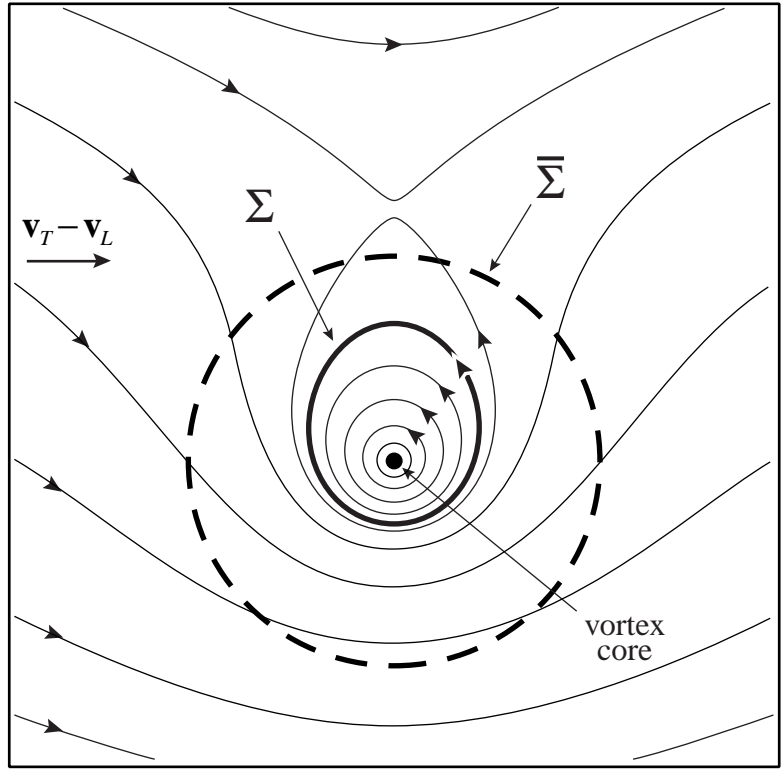

**Figure II-2 :** Top view of the flow pattern around the vortex, in the vortex frame. The surface $\Sigma$ is chosen on a closed flow line, whereas $\overline{\Sigma}$ is a circular cylinder centered on the vortex.

since in this frame the superflow is stationary[1].

In order to calculate the force acting on the vortex, we now construct a unit length cylinder $\Sigma$ parallel to the $z$-axis, on one closed flow line in the vortex frame, far enough from the core, as illustrated in Fig. II-2. There is no fluid transfer across $\Sigma$, and the total force exerted on the fluid inside the surface derives directly from (II.8) :

$$\mathbf{F}_\Sigma = \iiint_\Sigma Ne\left(\mathbf{E} + \mathbf{v}_s \times \mathbf{b}\right) d^3r - \iint_\Sigma N\mu\, d^2\mathbf{r}$$
$$= -\iint_\Sigma N(\mu + eV)\, d^2\mathbf{r} + Ne \iiint_\Sigma (\mathbf{v}_s - \mathbf{v}_L) \times \mathbf{b}\, d^3r,$$

where we have used Eq. (II.7). With the help of Eq. (II.9) and the cylindric symmetry of $\mathbf{v}_v$ and $\mathbf{b}$, it becomes

$$\mathbf{F}_\Sigma = \frac{Nm}{2} \iint_\Sigma (\mathbf{v}_s - \mathbf{v}_L)^2 d^2\mathbf{r} + Ne \iiint_\Sigma (\mathbf{v}_T - \mathbf{v}_L) \times \mathbf{b}\, d^3r \,. \qquad (\text{II.10})$$

---

1. It can be shown that the electric potential in (II.9) is indeed the same as the potential $V$ used in (II.7) by noting that, in Eq. (II.8), $d\mathbf{v}_s/dt = 1/2\, \nabla (\mathbf{v}_s - \mathbf{v}_L)^2 - (\mathbf{v}_s - \mathbf{v}_L) \times (\nabla \times \mathbf{v}_s)$. This derives from the fact that the superflow is stationary in the vortex frame and $\mathbf{v}_L$ is constant. We can then see, using (II.4) and (II.8) with the above relation, that the gradient of (II.9) vanishes precisely if $V$ is the same as in (II.7).

If we instead take a surface $\overline{\Sigma}$ which is not on a flow line, we must also include the net momentum flowing through the surface per unit time :

$$\mathbf{F}_{conv} = \iint_{\overline{\Sigma}} N m \, \mathbf{v}_s \, (\mathbf{v}_L - \mathbf{v}_s) \cdot d^2\mathbf{r} \,. \tag{II.11}$$

To allow us to easily perform the integration, we chose $\overline{\Sigma}$ to be circular and centered on the vortex core. In this case, $\mathbf{v}_v \cdot d^2\mathbf{r} = 0$, and $\mathbf{F}_{conv}$ becomes

$$\mathbf{F}_{conv} = Nm \iint_{\overline{\Sigma}} [(\mathbf{v}_L - \mathbf{v}_T) \times \mathbf{v}_v] \times d^2\mathbf{r}$$
$$= Nm \iiint_{\overline{\Sigma}} \nabla \times [(\mathbf{v}_L - \mathbf{v}_T) \times \mathbf{v}_v] d^3r.$$

With the help of Eq. (II.2) and (II.4) and the standard vector identities for $\nabla \times (\mathbf{a} \times \mathbf{b})$ and $\frac{1}{2}\nabla \mathbf{a}^2$, one easily gets, for constant $\mathbf{v}_L$ and $\mathbf{v}_T$,

$$\mathbf{F}_{conv} = \frac{Nm}{2} \iint_{\overline{\Sigma}} (\mathbf{v}_s - \mathbf{v}_L)^2 d^2\mathbf{r} + Ne \iiint_{\overline{\Sigma}} (\mathbf{v}_T - \mathbf{v}_L) \times \mathbf{b} \, d^3r \,,$$

which is the same expression as in Eq. (II.10). The convective term (II.11) resulting from the momentum flow through the surface is thus equal to the force acting on the core of the vortex :

$$\mathbf{F}_{conv} = \mathbf{F}_{\overline{\Sigma}} = \frac{\mathbf{F}_{Mag}}{2} \,. \tag{II.12}$$

We can then finally calculate the total driving force, neglecting the second integral, which vanishes as $\xi^2/\lambda^2$ :

$$\mathbf{F}_{tot} = 2 \, \mathbf{F}_{\overline{\Sigma}} = Nm \iint_{\overline{\Sigma}} (\mathbf{v}_s - \mathbf{v}_L)^2 d^2\mathbf{r}$$
$$= Nm \iint_{\overline{\Sigma}} (\mathbf{v}_T^2 + \mathbf{v}_v^2 + \mathbf{v}_L^2 - 2 \, \mathbf{v}_T \cdot \mathbf{v}_L - 2 \, \mathbf{v}_v \cdot \mathbf{v}_L + 2 \, \mathbf{v}_T \cdot \mathbf{v}_v) d^2\mathbf{r} \,.$$

Since $\mathbf{v}_T^2 + \mathbf{v}_v^2 + \mathbf{v}_L^2 - 2 \, \mathbf{v}_T \cdot \mathbf{v}_L$ is constant over $\overline{\Sigma}$, we finally have

$$\mathbf{F}_{tot} = 2 \, Nm \iint_{\overline{\Sigma}} (\mathbf{v}_T - \mathbf{v}_L) \cdot \mathbf{v}_v \, d^2\mathbf{r}$$

or, after integration,

$$\mathbf{F}_{tot} = Ne \, (\mathbf{v}_T - \mathbf{v}_L) \times \boldsymbol{\phi}_o = \mathbf{F}_{Mag} \,. \tag{II.13}$$

Finally, we can see that *the force exerted by the superfluid on the vortex is the Magnus force.*

Now that we have the driving force acting on the vortex, we can drop the last assumption (v), and introduce the effect of friction between the vortex and the background superconducting material made of positive ions. By balancing the driving Magnus force with the drag force $\mathbf{f}_{drag}$, we then find the vortex equation of motion $\mathbf{F}_{Mag} + \mathbf{f}_{drag} = 0$, from which we calculate the vortex velocity $\mathbf{v}_L$ as a function of the transport current $\mathbf{j}_T = n_s e \, \mathbf{v}_T$. In the two next paragraphs, we summarize the respective calculations of both models.

## 2.2 The Bardeen-Stephen model

In their paper [27], Bardeen and Stephen consider, as already mentioned at the beginning of this chapter, a normal core of radius $a$ which is surrounded by a transition region, in which the order parameter $|\Psi|$ goes smoothly from zero to its asymptotic equilibrium value $|\Psi_\infty|$. In this case, friction occurs both in the normal core and in the transition layer. To avoid the difficult calculations necessary for the transition layer, we can restrict the equilibrium condition between the driving force and the drag force to the normal core only. However, this requires us first to determine what part of the total Magnus force (concerning the whole vortex) is acting over the core, the rest being applied to the interface. Once we know this driving force $\mathbf{F}_{nc}$ acting only on the normal core charges, we can then balance it against the core friction

$$\mathbf{f}_{drag}^{nc} = -\frac{Nm}{\tau}\pi a^2 \mathbf{v}_{nc}, \tag{II.14}$$

where $\tau$ is the electron-lattice collision time.

Considering that the normal electrons are in local equilibrium with the lattice, BS determine $\mathbf{F}_{nc}$ by writing

$$\mathbf{F}_{nc} = -N\pi a^2 \nabla \tilde{\mu}_{tot} + \mathbf{F}_{Lorentz},$$

where $\tilde{\mu}_{tot}$ is the total energy in the lattice frame, which is related to the total energy $\mu_{tot}$ in the vortex frame written in Eq. (II.9) by $\tilde{\mu}_{tot} = \mu_{tot} + \mathbf{p}_s \cdot \mathbf{v}_L$ with $\mathbf{p}_s = m\mathbf{v}_s - e\mathbf{A}$, $\mathbf{A}$ being the vector potential. $\mathbf{F}_{Lorentz}$ is the Lorentz force on the charges in the normal core $\mathbf{F}_{Lorentz} = N\pi a^2 e\, \mathbf{v}_{nc} \times \mathbf{b}$ (**b** is considered constant over the core).

In this model, the vortex in motion is treated as a perturbation of the stationary vortex by the following development : in the vortex frame, the total superfluid momentum $\mathbf{p}_s$ is written

$$\mathbf{p}_s(\mathbf{r} - \mathbf{v}_L t) = \mathbf{p}_o(\mathbf{r} - \mathbf{v}_L t) + \mathbf{p}_1(\mathbf{r} - \mathbf{v}_L t),$$

where $\mathbf{p}_o$ is the momentum for a vortex at rest, and $\mathbf{p}_1 \ll \mathbf{p}_o$ is the modification of the superfluid flow arising from the motion, which is then neglected. With the appropriate gauge and a constant **b** in the core, we get in cylindric coordinates

$$\mathbf{p}_s = \mathbf{p}_o = \left(\frac{e\phi_o}{2\pi r} - \frac{e}{2} rb\right) \mathbf{e}_\theta.$$

Outside the normal core $\nabla \tilde{\mu}_{tot} = (\mathbf{v}_L \cdot \nabla)\mathbf{p}_o$. The corresponding force field in the core is then obtained by an appropriate continuation of the above field at the core boundary. Finally, we get the total driving force for the normal electrons in the core :

$$\mathbf{F}_{nc} = N\pi a^2 (\mathbf{v}_L \times \mathbf{e}_z)\left(\frac{\partial p_o}{\partial r}\right)_{r=a} + \mathbf{F}_{Lorentz}$$

$$= -\frac{Ne}{2}\mathbf{v}_L \times (\phi_o + \pi a^2 \mathbf{b}) + N\pi a^2 e\, \mathbf{v}_{nc} \times \mathbf{b}.$$

By stating that the vortex circular velocity distribution $\mathbf{v}_v$ reaches the depairing current at the normal core boundary ($r = a$), and interpolating linearly between the low and high field limits of the $\phi_o(\mathbf{B}_{c2})$ relation, BS reach the following equation for $a$ as a function of the applied magnetic field :

$$\frac{\phi_o}{2\pi a^2} = \mathbf{B}_{c2} - \frac{\mathbf{B}}{2}.$$

By balancing $\mathbf{F}_{nc}$ against $\mathbf{f}_{drag}^{nc}$ (II.14), and for $\mathbf{v}_{nc} = \mathbf{v}_T$ (see the discussion on page 21), we get the vortex equation of motion which takes the form

$$e\mathbf{v}_T \times \mathbf{B} - \frac{m}{\tau}\mathbf{v}_T - e\mathbf{v}_L \times \mathbf{B}_{c2} = 0,  \tag{II.15}$$

since for $\xi \ll \lambda$ we can approximate $\mathbf{b} \approx \mathbf{B}$ (see Ref. [53] and references therein). From Eq. (II.15) we calculate the electric field given by Josephson's relation

$$\mathbf{E} = -\mathbf{v}_L \times \mathbf{B} = -\frac{B}{B_{c2}}\left(\mathbf{v}_T \times \mathbf{B} - \frac{m}{e\tau}\mathbf{v}_T\right)  \tag{II.16}$$

with $B = \|\mathbf{B}\| > 0$. Since the longitudinal and transverse components of $\mathbf{E}$ with respect to the transport current $\mathbf{j}_T = Ne\mathbf{v}_T$ are, respectively,

$$\rho_{xx} = \mathbf{E} \cdot \frac{\mathbf{v}_T}{Nev_T^2} \qquad \text{and} \qquad \rho_{xy} = -\mathbf{E} \cdot \frac{\mathbf{v}_T \times \mathbf{e}_z}{Nev_T^2},  \tag{II.17}$$

we get for the longitudinal and transverse resistivities:

$$\rho_{xx} = \rho_n \frac{B}{B_{c2}} \qquad \text{and} \qquad \rho_{xy} = \frac{B^2}{eNB_{c2}},  \tag{II.18}$$

where we have introduced the normal (Drude) resistivity

$$\rho_n = \frac{m}{e^2 N \tau}.$$

Therefore, the longitudinal resistivity can be seen as the contribution of normal electrons of the core, occupying a volume ratio of $B/B_{c2}$. From Eq. (II.16) and (II.17) we note that $\rho_{xx}$ is symmetric in the magnetic field ($\rho_{xx}(-\mathbf{B}) = \rho_{xx}(\mathbf{B})$), whereas $\rho_{xy}$ is antisymmetric ($\rho_{xy}(-\mathbf{B}) = -\rho_{xy}(\mathbf{B})$), such that it is just the Hall resistivity. Then from Eq. (II.18) we extract the Hall angle

$$\tan\theta_H := \frac{\rho_{xy}}{\rho_{xx}} = \omega_c \tau$$

with $\omega_c = eB/m$. This expression is exactly the same as for a normal metal.

### 2.3 The Nozières-Vinen model

On the other hand, Nozières and Vinen [28] consider a normal core of radius $a$, with a discontinuous normal superconducting boundary, in contrast to the continuous transition region of the BS model. Again, since we only know the friction force for the core electrons $\mathbf{f}_{drag}^{nc}$ (II.14), but not for the vortex-superconductor interface, the equilibrium between friction in the normal core and the driving force can only be explicitly written provided we know which part of the total Magnus force applies on the bulk of the core, the remaining part being localized on the sharp interface. To answer this question, we first recall that the convective part (II.11) of the Magnus force arising from the flow of the momentum inside the core represents half the Magnus force, as written in Eq. (II.12). NV then assume that this convective force is localized right at the core boundary, a statement that can be justified by noting that the momentum crossing this interface should be dissipated over a distance $(\mathbf{v}_{nc} - \mathbf{v}_L)\tau$ that can be considered to be much smaller than $\xi$.

Therefore, the remaining part, $\mathbf{F}_{Mag}/2$, is the driving force acting on the normal core[1], which is balanced against the drag force (II.14). Assuming, as already mentioned, $\mathbf{v}_{nc} = \mathbf{v}_T$, the equation of motion is then :

$$\frac{Ne}{2}(\mathbf{v}_T - \mathbf{v}_L) \times \boldsymbol{\phi}_o - N\pi a^2 \frac{m}{\tau} \mathbf{v}_T = 0 \,. \tag{II.19}$$

For high fields, NV choose the following relation for $a$ :

$$\frac{\phi_o}{2\pi a^2} = \mathbf{B}_{c2} \,.$$

With this, we can again deduce the electric field from Eq. (II.19) :

$$\mathbf{E} = -\mathbf{v}_L \times \mathbf{B} = \frac{B}{B_{c2}} \frac{m}{e\tau} \mathbf{v}_T - \mathbf{v}_T \times \mathbf{B} \,.$$

It follows from the definition of the resistivities (II.17) :

$$\rho_{xx} = \rho_n \frac{B}{B_{c2}} \qquad \text{and} \qquad \rho_{xy} = \frac{B}{eN} \,. \tag{II.20}$$

The Hall angle is given by

$$\tan\theta_H = \omega_c \tau \frac{B_{c2}}{B} = \omega_{c2}\tau$$

with $\omega_{c2} = \omega_c B_{c2}/B = eB_{c2}/m$. Note that we can also calculate the resulting mean field vortex flow Hall conductivity

$$\sigma_{xy}^v := \frac{\rho_{xy}}{\rho_{xx}^2 + \rho_{xy}^2} = \frac{eN}{B} \cdot \frac{\omega_{c2}^2 \tau^2}{1 + \omega_{c2}^2 \tau^2} \,. \tag{II.21}$$

In summary, NV reach the same longitudinal resistivity as BS, and have a Hall effect $B/B_{c2}$ times smaller. Therefore, even though both models do not absolutely agree with each other, they bring very similar results, and yet are still both unable to give an explanation for the Hall anomaly. The Hall effect in these models can indeed be seen as the normal Hall effect on the normal charge carriers present in the core, and, hence, always has the same sign as the normal state Hall effect. As we have mentioned in the preceding chapter, many experiments contradict this prediction.

Before discussing other models providing possible explanations for the Hall anomaly, we should mention that the disagreement between the two models presented here can apparently be solved by carefully considering Andreev scattering at the core boundary [54]. The convective force of the NV model, absent in the BS model and responsible for the discrepancy, is then found to be microscopically justified through this process.

---

1. Provided the electric potential $V$ is continuous at the core interface, which means that there is no charge layer around the vortex. If on the other hand a contact potential $\Delta V$ is introduced in order to keep the total energy $\tilde{\mu}_{tot}$ continuous at the interface, we recover the result of the BS model.

## 3. The Hall anomaly

In this section, we present the main recent theories that are intended to provide an explanation for the experimentally observed Hall anomaly, namely a change in the sign of the Hall effect in the mixed state. This overview is not exhaustive, but focuses on the most widely accepted or discussed models.

### 3.1 Pinning and backflow

As we have seen above, both BS and NV models neglect the backflow current reducing the flow in the vortex core, and thus consider that $\mathbf{v}_{nc} = \mathbf{v}_T$. But as we have discussed at the beginning of this chapter, a completely pinned vortex will have a maximal backflow $\mathbf{v}_b = \mathbf{v}_{nc} - \mathbf{v}_T$ such that there is no current inside the core $\mathbf{v}_{nc} = 0$. Therefore, Wang and Ting [55,56] have reconsidered the NV model by including the pinning force and its effect on the backflow current, leading generally to $\mathbf{v}_{nc} \neq \mathbf{v}_T$ (see Eq. (II.1)).

To keep the discussion of this model as simple as possible, we choose the high-$\kappa$ ($\xi \ll \lambda$) approximation in all that follows : we will consider that, since $B \gg B_{c1}$, $\mathbf{b} = \mathbf{B}$ represents the microscopic local field, the averaged induction, as well as the applied external field [53].

The new equation of motion for the normal core is now :

$$\mathbf{F}_{nc} + \mathbf{F}_p^{nc} = \frac{N m}{\tau} \pi a^2 \mathbf{v}_{nc}, \tag{II.22}$$

where $\mathbf{F}_p^{nc}$ is the pinning force acting on the normal core. This force is related to the backflow in the core $\mathbf{v}_b = \mathbf{v}_{nc} - \mathbf{v}_T$ by

$$\mathbf{F}_p^{nc} = k N e \left( \mathbf{v}_{nc} - \mathbf{v}_T \right) \times \boldsymbol{\phi}_o, \tag{II.23}$$

since the backflow should be oriented perpendicularly to the pinning direction ($\mathbf{v}_b \cdot \mathbf{F}_p^{nc} = 0$). Energy conservation in the system leads to $k \approx 1$, and the total pinning force acting on the whole vortex (not only on the normal core) is $\mathbf{F}_p \approx \mathbf{F}_p^{nc}$.

Aside from the introduction of pinning effects, Wang *et al.* make more exact derivations than NV, although exactly within the same normal core model and with the same basic assumptions. For example, unlike NV, they keep all the terms in the calculation of the integral which led to the Magnus force in the NV model (see page 24). Therefore, all terms proportional to $B / B_{c2}$, which were neglected in the NV theory, are kept here. Consistent with this generalization, no particular approximation for the relation $a(B_{c2})$ is made, but a parameter $\chi = a^2 / \xi^2$ is introduced to account for the exact value of $a$. However, most of the time $1/2 \leq \chi \leq 1$, thus this parameter will not bring any noteworthy contribution.

The most important generalization in this model, compared with NV or BS theories, concerns the contact force applied on the superconducting-normal core interface. Wang *et al.* write this force as

$$\mathbf{F}_{cont} = f_\ell \frac{N e}{2} \left( 1 - \frac{B}{2 B_{c2}} \right) \left( 2 \mathbf{v}_T - \mathbf{v}_{nc} \right) \times \boldsymbol{\phi}_o,$$

where $0 \leq f_\ell \leq 1$ depends on the charge carrier mean free path $\ell$. Asymptotically, for $\ell \leq \xi$ (moderately dirty limit), the fluid at the interface is in equilibrium with the lattice, the chemical potential is continuous at the interface, and the contact force is such that $f_\ell = 1$, a situation which corresponds to the BS model (see the footnote on page 23). On the other hand, if $\ell \gg \xi$ (clean limit), the *electric potential* is now continuous, and the contact force vanishes, hence $f_\ell=0$ (NV model). The parameter $f_\ell$ depends on the temperature through $\ell(T)$, and is basically expected to increase with increasing $T$, going rapidly from 0 to 1 around $T_o$ given by $\ell(T_o)=\xi(T_o)$.

With all of these detailed contributions, the force acting on the normal core becomes

$$\mathbf{F}_{nc} = \frac{Ne}{2}\left(1-\frac{B}{2B_{c2}}\right)\left(2\mathbf{v}_T - \mathbf{v}_{nc} - \mathbf{v}_L\right)\times\boldsymbol{\phi}_o + \frac{Ne}{2}\frac{B}{B_{c2}}\left(1-\frac{B}{2B_{c2}}\right)\left(\mathbf{v}_{nc}-\mathbf{v}_L\right)\times\boldsymbol{\phi}_o - \mathbf{F}_{cont}.$$

We can then calculate the resistivities $\rho_{xx}$ and $\rho_{xy}$ from the equilibrium condition Eq. (II.22), much as in the BS or NV models. Assuming that $\mathbf{F}_p \propto -\mathbf{v}_L$, we get

$$\rho_{xx} = \rho_n \frac{B}{B_{c2}} \frac{1}{\chi + B/2B_{c2}}\left(1-\frac{F_p}{j_T\phi_o}\right)$$

and

$$\frac{\rho_{xy}}{\rho_{xx}\omega_{c2}\tau} = \left[(1-f_\ell)\chi + \frac{(1+f_\ell)B}{2B_{c2}}\right] - \frac{F_p}{j_T\phi_o}\left[(1+f_\ell)\chi + \frac{(3-f_\ell)B}{2B_{c2}}\right]. \tag{II.24}$$

Both brackets above are always positive, and their ratio depends on $f_\ell$. Since flux flow only happens if the Lorentz force overrides the pinning ($j_T\phi_o > F_p$), the pinning-related factor satisfies

$$0 \leq \frac{F_p}{j_T\phi_o} = \frac{\tilde{F}_p}{j_T B} < 1$$

with $\tilde{F}_p = F_p B/\phi_o$ the pinning force density, which we consider as independent of $B$ around the Hall anomaly to keep the discussion as simple as possible. It can be easily seen from Eq. (II.24) that the Hall resistivity undergoes a sign change close to the depinning threshold. When $f_\ell=0$ (for low temperature in clean materials), depinning occurs at $B \ll B_{c2}$, and the negative Hall resistivity occurs only in a narrow field region, with very small absolute value. When the field is further increased, $\rho_{xy}$ changes sign and $\rho_{xy} \propto B$. On the other hand, larger negative values of $\rho_{xy}$ happen for $f_\ell=1$ (at higher temperatures), where

$$\rho_{xy}(B) = \rho_{xx}(B)\,\omega_{c2}\tau\left[\frac{B}{B_{c2}} - \frac{\tilde{F}_p}{j_T B}\left(2\chi + \frac{B}{B_{c2}}\right)\right]$$

is negative just above depinning, reaches a negative maximum (of amplitude comparable to the normal state Hall effect), and approaches the (positive) normal state behavior as $B \to B_{c2}$.

Even though this formalism is more adapted to an analysis of the $\rho_{xy}(B)$ dependence at fixed $T$, this model also gives qualitatively correct results for the temperature dependence of the Hall effect, provided the temperature dependences of $\tilde{F}_p$ and $f_\ell$ are taken into account.

However, we see that in no way can this theory predict the *second* sign change observed in some materials at very low temperature. Another limitation of this model is the absence of thermal fluctuation processes, ruling out crucial effects of vortex dynamics as, for example, thermally activated flux flow (TAFF). Therefore, this theory was later completed by Wang, Dong and Ting (WDT) [57] to explicitly include thermal fluctuations. The vortex equation of motion is then

$$\mathbf{F}_{Mag} + \mathbf{F}_p + \mathbf{F}_T + \mathbf{f}_{drag} = 0, \tag{II.25}$$

where $\mathbf{F}_T$ is the thermal noise force and $\mathbf{f}_{drag}$ is the total drag force acting on the vortex (whereas $\mathbf{f}_{drag}^{nc}$ was only exerted on the normal core). A complete discussion of the different contributions to this drag force is given in Ref. [56]. We just note here that it can be shown that the drag force not only depends on the vortex or the superfluid velocities, $\mathbf{v}_L$ and $\mathbf{v}_T$, respectively, but also has components proportional to $\mathbf{F}_p$ and $\mathbf{F}_p \times \mathbf{e}_z$, as a consequence of the pinning-induced backflow (since the core charges velocity $\mathbf{v}_{nc}$ is directly influenced by the pinning, see Eq. (II.23)).

The resistivities $\rho_{xx}$ and $\rho_{xy}$ are then obtained by solving the time-averaged equation of motion. Even though the thermal force itself obviously vanishes in average ($\langle \mathbf{F}_T \rangle = 0$), a difference with the preceding model subsists through the average pinning force $\langle \mathbf{F}_p \rangle$, which appears directly in the equation of motion (II.25), as well as indirectly in the drag force $\mathbf{f}_{drag}$, as mentioned above. In the TAFF and creep regimes, $\langle \mathbf{F}_p \rangle$ is very different from $\mathbf{F}_p$ of Wang and Ting's model above.

Assuming that $\langle \mathbf{F}_p \rangle = -\Gamma_p(v_L) \mathbf{v}_L$ (where the coefficient $\Gamma_p(v_L)$ includes dependence on the temperature $T$ and the pinning energy $U_p$), taking into account the experimental evidence that $\tan(\theta_H) \ll 1$, and with $\tilde{f}_\ell = f_\ell(1 - B/B_{c2})$, we finally have

$$\rho_{xx} = \frac{\phi_\circ B}{\Gamma_p + \eta} \tag{II.26}$$

and

$$\rho_{xy} = \rho_{xx} \frac{\omega_{c2} \tau}{\Gamma_p + \eta} \left[ \eta(1 - \tilde{f}_\ell) - 2 \tilde{f}_\ell \Gamma_p \right] \tag{II.27}$$

($\eta = \phi_\circ B_{c2}/\rho_n$ is the usual viscosity coefficient, namely the factor of the $\mathbf{f}_{drag}$ term which is proportional to $\mathbf{v}_L$, see Ref. [56]).

Clearly, if $f_\ell$ is close to 1 (high enough temperature), $B$ not too large, and for strong enough pinning ($\Gamma_p > \eta$), $\rho_{xy}$ is negative (provided the normal state Hall effect $\omega_{c2} \tau$ is positive, *i.e.* for positive charge carriers). More generally, it can be seen that this version of the model very well describes the observed field as well as temperature dependences of the Hall resistivity in various materials. This theory also features a *second* sign change of $\rho_{xy}$ as a function of temperature : at low temperature, $f_\ell = \tilde{f}_\ell = 0$, and $\rho_{xy}$ is positive. When the temperature is increased, $f_\ell$ increases, and $\rho_{xy}$ becomes negative as stated above (provided the field is low enough). Of course, when the temperature reaches $T_{c2}(B)$, pinning vanishes ($\Gamma_p \to 0$) and $\rho_{xy}$ switches to positive sign again.

Unfortunately, a complete analysis of the above relations in order to compare them to the experimental data is made difficult by the non-trivial dependence on a number of "free" parameters, or more precisely on parameters that are not accurately known experimentally (such as the factors $\Gamma_p$ and $f_\ell$). Therefore, a much more efficient way to investigate the predictions of this pinning induced backflow model is that of numerical simulations [58-60]. Note that the simulations are actually based directly on the single vortex equation of motion (II.25), by explicitly introducing thermal noise in a "molecular dynamics" system, and not on the basis of the final expressions Eq. (II.26) and (II.27) which are the consequence of an assumption on the average pinning force $\langle \mathbf{F}_p \rangle$ (see above). Again, the results are quite consistent with experimental data. These simulations can also be used to *determine* the average pinning force and the coefficient $\Gamma_p(v_L)$ [61].

However, there is in the experimental literature a wide controversy about the pinning effects in this model. In many experiments, the Hall anomaly is indeed said to show a behavior that is opposite to that of WDT's predictions. More precisely, the negative maximum of the Hall resistivity $|\rho_{xy}^{max}|$ often *decreases* as the disorder is *increased* (see next chapter for a review of references). This is usually taken as a proof that the disorder is not the cause of the Hall anomaly. Unfortunately, we can see from Eq. (II.26) and (II.27) that the interpretation within this model is not so simple. We first have to introduce explicitly $\rho_{xx}$ from Eq. (II.26) into $\rho_{xy}$ of Eq. (II.27) :

$$\rho_{xy} = \frac{B}{Ne} \frac{\eta\left[\eta\left(1-\tilde{f}_\ell\right) - 2\tilde{f}_\ell \Gamma_P\right]}{\left(\Gamma_P + \eta\right)^2} \ .$$

We then see that in the thermally activated flux flow (TAFF) pinning dominated regime, where the vortex dynamics is essentially controlled by the pinning ($\Gamma_p \gg \eta$), the above expression for the Hall resistivity leads to $\rho_{xx} \propto -1/\Gamma_p$ : the pinning indeed *reduces* the negative Hall conductivity. It is nevertheless the pinning itself which is responsible for the negative sign (from the term between brackets). On the other hand, closer to the free flux flow region, where the pinning is reduced (and where the negative Hall effect usually occurs), the sign of the above expression and its pinning dependence is far from trivial : once the Hall effect has changed its sign, the influence of an increase of pinning depends on the competition between the decrease of $\rho_{xx}$ and the increase of the (negative) factor in brackets. The discussion on this pinning dependence of the magnitude of the Hall resistivity is hardly straightforward, and is often oversimplified in the literature.

Very recently, this WDT model has also been completed by Zhu *et al.* by adding vortex-vortex interactions in the equation of motion (II.25), which was then solved numerically [62] similarly to the simulations above. In a few words, the result is that the intervortex interactions can apparently play the same role as the thermal fluctuations : this additional parameter, together with the pinning interactions, can also induce a Hall sign change.

Finally, from Eq. (II.26) and (II.27) we see that the WDT model naturally predicts some sort of scaling between the longitudinal and Hall resistivities, since we can also write

$$\rho_{xy} = \frac{\omega_{c2} \tau}{\phi_o B} \left[\eta\left(1-\tilde{f}_\ell\right) - 2\tilde{f}_\ell \Gamma_P\right] \rho_{xx}^2 \ , \tag{II.28}$$

even though the *a priori* field and temperature dependence of the $\rho_{xx}^2$ prefactor should first be discussed in detail. This discussion will be given in the next section (see paragraph 4.2, page 38).

Globally, the only weak point of this model is that is relies strictly on a hydrodynamic approach of the vortex motion. Therefore, it cannot take into account the critical fluctuations of the order parameter near the superconducting-normal transition, which might be important in the Hall anomaly process. The Hall sign change has indeed been reported to be extremely close to $T_c$ (or even above $T_c$ [63]) in some materials.

### 3.2 Vortex lattice defects

Another model that considers pinning on vortices is proposed by Ao [64,65]. Actually, the pinning effects are quite extreme here, since the vortex lattice is considered as completely pinned, the dynamics of the assembly being only controlled by the mobility of lattice defects.

Ao's starting point is the vortex equation of motion

$$Ne\left(\mathbf{v}_T - \mathbf{v}_L\right) \times \boldsymbol{\phi}_o - \eta\, \mathbf{v}_L = 0 \ , \tag{II.29}$$

namely the balance between the Magnus force and a viscous drag parallel to the vortex motion. Note that this equation of motion, although there is some similarity, is not absolutely the same as in the standard BS or NV models of section 2. We recall that in the NV model, the Magnus force is balanced by a drag force proportional to $\mathbf{v}_T$, whereas in the BS model the *Lorentz* force is op-posed to the above drag force $-\eta \mathbf{v}_L$. The viscous coefficient $\eta$ here, nevertheless, takes the BS form $\eta = \phi_o B_{c2}/\rho_n = N e \phi_o \omega_{c2} \tau$. In the usual case of independent free vortices, it is straightforward to see that the resulting longitudinal and Hall resistivities would be

$$\rho_{xx} = \frac{B}{eN} \cdot \frac{\omega_{c2}\tau}{1+\omega_{c2}^2\tau^2} \quad \text{and} \quad \rho_{xy} = \frac{B}{eN} \cdot \frac{1}{1+\omega_{c2}^2\tau^2}, \quad (II.30)$$

which is consistent with the usual flux flow resistivity $\rho_{xx} = \rho_n B/B_{c2}$ only provided $\omega_{c2}\tau \gg 1$ (superclean limit), but reproduces the NV Hall resistivity $\rho_{xy}$ only for $\omega_{c2}\tau \ll 1$ (dirty limit).

Ao then considers that vortex-vortex interactions are dominant, leading to a regular Abrikosov vortex lattice. He then supposes that this lattice is rigidly pinned, but has just sufficient thermal energy to contain some defects. After a comparison of theoretical energy scales for these defects and with some experimental support, he then restricts the considered lattice defects to interstitials and vacancies, arguing that they are the perturbations of the periodic lattice with the lowest energy. Moreover, vacancies are said to have even lower energy than interstitials, therefore dominating the dynamical response in this regime.

We can introduce the thermally activated vacancy density as

$$n_v = n_o \exp\left(-\frac{b_v E_v}{k_B T}\right), \quad (II.31)$$

where $E_v$ is the vacancy formation energy scale and $b_v = 1$ (note that we can also use $b_v = 0$ to reflect pinning-induced vacancies). The vacancy effective relaxation time can be defined through the effective vacancy viscosity $\eta_v = N e \phi_o \omega_{c2} \tau_v$ as

$$\eta_v = \eta_o \exp\left(\frac{a_v E_v}{k_B T}\right) \quad \text{or} \quad \tau_v = \tau_o \exp\left(\frac{a_v E_v}{k_B T}\right). \quad (II.32)$$

Here, $a_v$ is a sample-dependent numerical factor, again of the order of unity and insensitive to temperature. Note that the corresponding expressions could also be written for interstitials, but with factors $a_i$ and $b_i$ larger than $a_v$ and $b_v$, respectively, since interstitial have larger formation energy.

The key idea is to consider the vacancies as moving independently in the pinned vortex lattice, following the equation of motion (II.29). Since in Eq. (II.30) we have $B = n \phi_o$ where $n$ is the vortex density, we can by analogy directly deduce the vacancy contribution to the resistivities

$$\rho_{xx} = \frac{n_v \phi_o}{eN} \cdot \frac{\omega_{c2}\tau_v}{1+\omega_{c2}^2\tau_v^2} \quad \text{and} \quad \rho_{xy} = -\frac{n_v \phi_o}{eN} \cdot \frac{1}{1+\omega_{c2}^2\tau_v^2} \quad (II.33)$$

with the expressions for $n_v$ and $\tau_v$ given above. Obviously, with the "anti-vortex" nature of vacancies, the corresponding Hall resistivity has a sign opposite to the flux flow Hall effect by construction. This model is interesting in the sense that it introduces intervortex interactions and the notion of the vortex lattice. However, it still remains based on a "single vortex" equation of motion, since finally the vacancies are just moving independently in the rigid lattice.

The main difficulty with this theory is to investigate the behavior close to the free flux flow region, in the vortex liquid, where the Hall sign change actually occurs. The notion of vacancy in this disordered and fluctuating regime is indeed not clear at all. On the other hand, this model might be worth considering in the pinned vortex solid regime.

The discussion, within this model, on the scaling law between the longitudinal and Hall resistivities will be given in the next section (paragraph 4.3, page 39).

### 3.3 Time-dependent Ginzburg-Landau equations

Aside from the two preceding hydrodynamic approaches, the phenomenological Ginzburg-Landau theory has also been used to calculate the conductivity. Here we very briefly give the method of derivation of the conductivity in this framework, and then show how this formalism should be modified to be able to describe the Hall anomaly.

For dynamic systems the time-dependent Ginzburg-Landau equation has to be used. With $\Psi(\mathbf{r}, t)$ the complex order parameter and

$$F_{GL} = \iiint \left[ a|\Psi|^2 + \frac{b}{2}|\Psi|^4 + \frac{\hbar^2}{2m}\left|\left(\nabla - \frac{2ei}{\hbar}\mathbf{A}\right)\Psi\right|^2 + \frac{(\nabla \times \mathbf{A})^2}{2\mu_o} \right] d^3r , \quad \text{(II.34)}$$

the Ginzburg-Landau free energy (including the magnetic contribution), this equation is [66]

$$-\gamma(\hbar \partial_t + 2eiV)\Psi = \frac{\delta F_{GL}}{\delta \Psi^*} , \quad \text{(II.35)}$$

where $\gamma$ is the dimensionless relaxation time and $V$ the electric potential. The idea is to calculate from (II.34) the variation of the free energy $\delta F_{GL}$ (per unit cell of the vortex lattice) corresponding to a displacement of the vortex by an arbitrary vector $\mathbf{d}$, and relate it to the work done by the drag force which is balanced by the driving Lorentz force :

$$\delta F_{GL} = -L \mathbf{d} \cdot \hat{\mathbf{f}}_{drag} , \quad \text{(II.36)}$$

where $L$ is the length of the vortex segment with

$$\hat{\mathbf{f}}_{drag} + \mathbf{j}_T \times \boldsymbol{\phi}_o = 0 . \quad \text{(II.37)}$$

For a slow vortex motion of velocity $\mathbf{v}_L$, the time derivative $\partial_t$ can be replaced with the operator $-(\mathbf{v}_L \cdot \nabla)$ acting on the variables describing the vortex at rest. We can then solve the resulting system (II.34)-(II.37) to determine $\mathbf{j}_T$ as a function of $\mathbf{v}_L$, and then extract the longitudinal and Hall conductivities from the relation $\mathbf{j}_T = B(\sigma_{xy}\mathbf{v}_L - \sigma_{xx}\mathbf{v}_L \times \mathbf{e}_z)$ introduced in the preceding chapter (see page 10). Before going into the detailed result, we can directly note that if $\gamma$ is real, the Hall conductivity is identically zero. In this case, Eq. (II.35) indeed has a "particle-hole symmetry" : it is invariant under the simultaneous transformations $\Psi \to \Psi^*$, $V \to -V$ and $\mathbf{A} \to -\mathbf{A}$, from which we can deduce that $\sigma_{xy}(-\mathbf{H}) = \sigma_{xy}(\mathbf{H})$. Combined with the Onsager relation $\sigma_{xy}(-\mathbf{H}) = \sigma_{yx}(\mathbf{H})$ and the assumed rotational symmetry of the system $\sigma_{xy}(\mathbf{H}) = -\sigma_{yx}(\mathbf{H})$, we easily show that $\sigma_{xy}(\mathbf{H}) \equiv 0$.

Therefore, it was realized quite recently that a non-zero imaginary part of the relaxation time $\gamma = \gamma' + i\gamma''$ was necessary to get the flux-flow component of the Hall effect [67,68]. The mean field vortex contributions to the conductivity then become, with $c_1$ and $c_2$ constants of the order of unity,

$$\sigma_{xx}^v = c_1 \sigma_{xx}^n \frac{B_{c2}}{B} ,$$

the standard result of the phenomenological BS and NV models of section II.2, and

$$\sigma_{xy}^v = -\text{sign}(e) c_2 \frac{\gamma''}{\gamma'} \sigma_{xx}^n \frac{B_{c2}}{B} . \quad \text{(II.38)}$$

Expressions taking the additional fluctuation contributions into account can be found in Ref. [69].

The central question is then to know what sign should be expected for $\sigma_{xy}^v$, or in other words how to determine the $\gamma''$ factor from microscopic considerations. Fukuyama *et al.* have calculated this imaginary part of the relaxation time within the BCS model [70], reaching the result [68]

$$\frac{\gamma''}{\gamma'} = -\frac{4T}{\pi} \left( \frac{1 + V_{BCS} N(\varepsilon_F)}{V_{BCS} N(\varepsilon_F)} \right) \frac{1}{N(\varepsilon_F)} \left. \frac{\partial N(\varepsilon)}{\partial \varepsilon} \right|_{\varepsilon = \varepsilon_F}, \qquad (II.39)$$

where $V_{BCS}$ is the BCS pairing interaction. We see that the vortex contribution to the Hall conductivity (II.38) can have a sign opposite to the normal state Hall effect depending on the Fermi surface structure. However, as we will see later on, this prediction of sign is usually not in agreement with the experimental data [71].

In general terms, the advantage of this formulation of the Hall effect in the frame of the Ginzburg-Landau theory is that it allows for the introduction of fluctuations, absent from a classical mean field "hydrodynamic" picture like the two preceding models. A potential of randomly distributed pinning centers can also be included in the Ginzburg-Landau free energy. Ikeda [72-74] has analyzed such a very complete system by a perturbative approach of lowest Landau levels (LLL). The formalism of this model clearly goes far beyond the scope of this theoretical overview. However, we shall come back to the results of Ikeda's work to discuss the pinning dependence of the Hall effect in the light of the results of this work, in chapter VI.

### 3.4 Vortex charge

In the preceding hydrodynamic theories, the vortex core was modeled by a cylinder of radius $a \approx \xi$, with a density of states equal to that of a normal metal of the same total electron density as the superconductor, following the picture emerging from the Caroli-de Gennes-Matricon [33] calculations (see discussion on page 11). In other words, the electronic density at the vortex core, $n_o$, was the same as the density far from the core, $n_\infty$.

However, this calculation of the minigap $\hbar \omega_{c2} = \Delta^2 / \varepsilon_F$ and the consequent vortex bound states is valid only for the traditional superconductors which are well described by the BCS theory, and in which the chemical potential can be considered as temperature independent ($\mu(T) = \varepsilon_F$). Provided the standard weak coupling BCS theory can not satisfyingly describe the high-$T_c$ superconductors properties, this temperature dependence should be taken into account [75]. The result can be expressed as a function of the particle-hole asymmetry in the electronic band structure, since the superconducting chemical potential $\mu(T)$ is related to the normal state potential $\mu_n$ by :

$$\mu(T) = \mu_n - c \frac{\Delta^2(T)}{N(\varepsilon_F)} \left. \frac{\partial N(\varepsilon)}{\partial \varepsilon} \right|_{\varepsilon = \varepsilon_F} = \mu_n - \delta\mu,$$

where $N(\varepsilon)$ is the density of states and $c \sim 0.3$ is a constant. As we can see, a spatial variation of the gap $\Delta$ (as is the case in a vortex) then leads to a gradient in the chemical potential. Therefore, there is a difference $\delta\mu$ in the chemical potential between the bulk superconducting phase and the normal vortex core, which corresponds, for an uncharged system, to an additional core density [76]

$$\delta n = 2\pi \xi^2 N(\varepsilon_F) \, \delta\mu = -2\pi \xi^2 c \, \Delta^2(T) \left. \frac{\partial N(\varepsilon)}{\partial \varepsilon} \right|_{\varepsilon = \varepsilon_F}. \qquad (II.40)$$

Although a detailed discussion on the origin of this vortex charge can be found in Ref. [76], the subsequent derivation of its consequence on the Hall effect is erroneous, featuring incorrect signs.

A valid microscopic derivation of the vortex charge-related Hall properties has been done in Ref. [77], whereas a deeper quantitative discussion of the vortex charge itself and its Coulomb screening can be found in Ref. [78].

Without going into the detailed microscopic calculation of the vortex charge, we can still derive its consequence on the Hall effect from a phenomenologic point of view, according to the original papers of Feigel'man *et al.* [79,80]. For this we consider as given the additional core density $\delta n = n_o - n_\infty \neq 0$, and look at the influence it has on the calculations of the NV model presented above. Since the latter was directly derived for $n_o = n_\infty = N$, it has indeed to be slightly adapted to this new vortex charge.

For example, the equation of motion (II.19) now becomes

$$\frac{n_\infty e}{2}(\mathbf{v}_T - \mathbf{v}_L) \times \boldsymbol{\phi}_o - n_o \pi a^2 \frac{m}{\tau} \mathbf{v}_{nc} = 0 , \qquad (\text{II}.41)$$

and the relation $\mathbf{v}_{nc} = \mathbf{v}_T$ does not satisfy the current continuity equation anymore : the current far from the core is $\mathbf{j}_T = n_\infty e \mathbf{v}_T$ whereas it is $\mathbf{j}_{nc} = n_o e \mathbf{v}_{nc}$ inside the core. Therefore, the correct continuity equation is, in the vortex frame,

$$n_\infty e (\mathbf{v}_T - \mathbf{v}_L) = n_o e (\mathbf{v}_{nc} - \mathbf{v}_L)$$

from which we extract $\mathbf{v}_{nc}$ :

$$n_o \mathbf{v}_{nc} = \delta n \, \mathbf{v}_L + n_\infty \mathbf{v}_T . \qquad (\text{II}.42)$$

We can then introduce (II.42) into (II.41) and solve for $\mathbf{v}_T$. One easily gets

$$\mathbf{j}_T = n_\infty e \mathbf{v}_T = n_o e \frac{\omega_{c2} \tau}{1 + \omega_{c2}^2 \tau^2} \mathbf{e}_z \times \mathbf{v}_L + \left( n_o e \frac{\omega_{c2}^2 \tau^2}{1 + \omega_{c2}^2 \tau^2} - e \delta n \right) \mathbf{v}_L .$$

Again, we use the relation

$$\mathbf{j}_T = B \sigma_{xx} \mathbf{e}_z \times \mathbf{v}_L + B \sigma_{xy} \mathbf{v}_L \qquad (\text{II}.43)$$

to determine the flux flow Hall conductivity

$$\sigma_{xy}^v = \frac{n_o e}{B} \frac{\omega_{c2}^2 \tau^2}{1 + \omega_{c2}^2 \tau^2} - \frac{e}{B} \delta n , \qquad (\text{II}.44)$$

which is the standard result (II.21) from the NV theory (with $N = n_o$) plus the additional term $-e \delta n / B$ related to the vortex charge. The longitudinal conductivity is the same as in the standard models. It is noteworthy to add that the vortex-charge induced term $-e \delta n / B$ can be shown to be the same as the Hall term (II.38) of the time dependent Ginzburg-Landau theory presented in paragraph 3.3, which is proportional to $-\gamma''$ : the vortex charge $\delta n$ (II.40) has indeed the same dependence on the Fermi surface curvature $\partial N / \partial \varepsilon$ as the $\gamma''$ obtained from the weak coupling BCS theory (II.39), such that $\sigma_{xy}$ is opposite to the normal state Hall effect $\sigma_{xy}^n$ for a negative $\partial N / \partial \varepsilon$ in both cases. Note that the BCS relation between the critical temperature $T_c$ and the coupling parameter $V_{BCS} N(\varepsilon_F)$ can also be used to express the vortex charge [78]. The result is then proportional to $\partial \ln T_c / \partial \mu$ instead of $\partial N / \partial \varepsilon$.

As we have briefly noted previously, the density modulation from Eq. (II.40) is valid for uncharged systems only. In a charged fluid, the initial excess of density $\delta n$ will be screened by Coulomb in-teractions. Within the Thomas-Fermi approximation of the dielectric function $\varepsilon(\mathbf{q}) = 1 + (r_D q)^{-2}$, where $r_D \ll \xi$ is the screening length, we can approximate the real charge density difference by

$$\delta \tilde{n}(r) = (r_D/\xi)^2 \delta n(r) \ll \delta n.$$

In other words, this means that the true vortex charge $\delta \tilde{n}(r)$ should not be calculated by imposing a constant chemical potential as we have done for Eq. (II.40), but a constant *electrochemical* potential $\mu_{el} = \mu(r) + eV(r)$. From the expression above, we see that the vortex charge is strongly reduced by electrostatic screening. However, by an adiabatic action approach, Feigel'man et al. [80] show that only the value of the *unscreened* charge $\delta n = \delta n(r=0)$ enters the Hall conductivity. The reason for this is that the singular topological contribution to the action leading to the additional term $\sigma_{xy}^t = -e\delta n/B$ in the Hall conductivity is not affected by Coulomb screening, since the latter only depends on the longitudinal component of the electric field $\mathbf{E}_{//} = -\nabla V$, whereas the topological action term is related to the angular expression of the vortex-induced phase.

Therefore, depending on the sign and magnitude of the vortex charge $\delta n$, a sign change in the total vortex Hall conductivity $\sigma_{xy}^v = \sigma_{xy}^{NV} + \sigma_{xy}^t$ can happen. In their paper, Feigel'man et al. [80] show by considering different cases that this theory can explain the observed temperature and field dependences of the Hall conductivity. For instance, since $\omega_{c2}\tau$ is large at low temperature and vanishes close to $T_c$, we can see that if $\delta n$ is positive and large enough, the Hall conductivity can have two successive sign changes as a function of $T$.

Note finally that, just as the topological term $\sigma_{xy}^t$ is derived on microscopic basis in Ref. [77], the standard scattering term from the NV theory $\sigma_{xy}^{NV}$ can also be obtained from a quasiclassic approach on the microscopic level [81,82,83]. More precisely, whereas the first of these papers directly verifies the NV result [81], the following ones also consider the effects of delocalized quasiparticles [82,83] (*i.e.* the normal component of the fluid). The result is that the NV term is reduced by a factor $g$ which goes from 0 (close to $T_c$) to 1 (for $T=0$), as already introduced in Eq. (I.3). Moreover, the delocalized states add a fraction $(1-g)$ of the normal state Hall conductivity $\sigma_{xy}^n$. However, the vortex charge term $\sigma_{xy}^t$ is not affected by the (homogeneous) normal component. We finally get for the total Hall conductivity [80] :

$$\sigma_{xy} = g \frac{n_o e}{B} \frac{\omega_{c2}^2 \tau^2}{1 + \omega_{c2}^2 \tau^2} - \frac{e}{B} \delta n + (1-g) \sigma_{xy}^n. \tag{II.45}$$

In conclusion, we can summarize the two last models in the following manner. First, we have seen that there is a vortex charge due to the temperature dependence of the chemical potential. This charge is proportional to $\partial \ln T_c / \partial \mu$, and directly influences the Hall conductivity. Unfortunately, the consequent Hall sign is opposite to the experimental data in most cases. Secondly, and independently, an imaginary relaxation time $\gamma''$ had to be included in the phenomenological time dependent Ginzburg-Landau theory to allow for a finite Hall conductivity. Again, this parameter can be shown to depend on $\partial \ln T_c / \partial \mu$ under the assumptions of a weak coupling BCS theory, and moreover lead to the same $\partial \ln T_c / \partial \mu$ dependence of the Hall conductivity in this case. As a consequence, this conductivity obviously still has a sign contradicting the experimental findings. Therefore, the microscopic determination of $\gamma''$ should not be done within a usual BCS coupling model.

A promising alternative model has been presented in the preceding chapter (page 12). In this pairing scenario, non-conducting pairs are "preformed" far above the superconducting transition, and condense in a superconducting state only at $T_c$. This condensation is however not of pure Bose character, since the preformed bosons are always interacting with underlying remaining unpaired fermions. This new pairing model provides another way to calculate microscopically the imaginary time $\gamma''$ [43]. It can be shown that $\gamma''$ now has a sign *opposite* to that coming from the vortex charge picture presented above. The sign is then consistent with the experimental data.

## 4. The Hall resistivity scaling law

In this last section of the chapter, we briefly discuss the different theories featuring the experimentally observed scaling $\rho_{xy} = A\,\rho_{xx}^{\beta}$ between the Hall and longitudinal resistivities.

### 4.1 Average pinning force

One of the most referred to models for the Hall scaling law is a work from Vinokur *et al.* [85], in which this behavior is assigned to the presence of vortex pinning. This phenomenological theory is based on two key points. First, the form of the equation of motion

$$\eta\,\mathbf{v}_L + k_2\,\mathbf{v}_L \times \mathbf{e}_z = N\,e\,\phi_o\,\mathbf{v}_T \times \mathbf{e}_z + \langle \mathbf{F}_p \rangle\,,$$

which is very similar to Eq. (I.1), with $k_1 = N\,e\,\phi_o$ for the driving Lorentz force and an additional average pinning force. Note that the drag force only depends on the vortex velocity $\mathbf{v}_L$, without other drag terms related to the superfluid velocity nor the pinning force, in contrast to the WDT theory presented in paragraph 3.1 (page 29). Second, the average pinning force itself is considered, by symmetry arguments, as only depending on the vortex velocity, and is therefore oriented along the vortex motion, this time similarly to the WDT model of paragraph 3.1 :

$$\langle \mathbf{F}_p \rangle = -\Gamma_p(v_L)\,\mathbf{v}_L\,.$$

As a consequence, its effect on the vortex dynamics is only a renormalization of the viscous coefficient, $\tilde{\eta}(v_L) = \eta + \Gamma_p(v_L)$. The resulting equation of motion can then easily be solved. Taking into account the experimental fact that $\tan\theta_H = k_2/\tilde{\eta} \ll 1$, it is straightforward to verify that

$$\rho_{xx} = \phi_o\,B/\tilde{\eta} \qquad \text{and} \qquad \rho_{xy} = \phi_o\,B\,k_2/\tilde{\eta}^2\,,$$

from which we deduce the wanted relation

$$\rho_{xy} = \frac{\phi_o\,B}{k_2}\,\rho_{xx}^2\,. \tag{II.46}$$

Consequently, the scaling law is verified with the exponent $\beta \equiv 2$ provided $k_2$ does not depend on the temperature $T$ and is approximately proportional to $B$. Note that in this model the sign change of the Hall resistivity can only occur through a sign change of the transverse force parameter $k_2$.

It is also obvious (in general terms, not only from this model) that a power law scaling cannot hold around such a sign change. Therefore, two cases have to be considered here : first, $k_2$ can be strongly temperature (and field) dependent, and undergo a sign change. In this regime, no scaling relation is possible. Secondly, far enough from these sign changes (there might indeed be more than one of them), and mostly at low temperature (or low field) where the resistivity vanishes, there might be another regime in which $k_2$ does almost not depend on the temperature $T$ and is approximately proportional to $B$, such that the scaling is verified with $\beta \equiv 2$. This phenomenological discussion also provides an explanation for the experimental observation of various values $\beta < 2$, by considering an intermediate case, where $k_2$ slightly increases when the temperature $T$ is raised and grows faster than $B$, thus reducing the effective exponent $\beta$.

However, we believe that the experimental scaling (often showing $\beta < 2$) is still not explained in this way. A temperature dependence of $k_2$ would of course break the $\beta = 2$ scaling, but recovering *again* the scaling relation with a new, lower exponent $\beta$ then requires that $k_2$ itself should scale as a power of the longitudinal resistivity, which is no less surprising than the former Hall resistivity scaling, and hence would also require an explanation.

### 4.2 Pinning and backflow

In paragraph 3.1, Eq. (II.28) indicated that a Hall scaling behavior might come out of the theory of Wang, Dong and Ting, which takes into account pinning-induced backflow as well as thermal fluctuations in the single-vortex equation of motion. First, we recall that Eq. (II.28) was

$$\rho_{xy} = \frac{\omega_{c2} \tau}{\phi_o B} \left[ \eta \left(1 - \tilde{f}_\ell\right) - 2 \tilde{f}_\ell \, \Gamma_p \right] \rho_{xx}^2$$

with $\tilde{f}_\ell = f_\ell \left(1 - B/B_{c2}\right)$, where $0 \leq f_\ell \leq 1$ depends on the charge carrier mean free path $\ell(T)$, and is basically expected to increase rapidly with increasing $T$ around the temperature $T_o$ corresponding to $\ell(T_o) = \xi(T_o)$. $\Gamma_p$ represents the pinning intensity, and thus presumably depends on both the temperature and the magnetic field. The difference with the model from Vinokur *et al.* (see preceding paragraph) is that the more complex form of the drag force in WDT's initial vortex equation of motion does not allow to get a scaling relation independent of the pinning factor $\Gamma_p$, as was the case in Eq. (II.46). The main reason is that this factor is explicitly contained in the drag force, since the pinning induces the backflow, which in turn determines dissipation in the vortex core, a mechanism which is absent in Vinokur's model.

In order to know whether a scaling relation $\rho_{xy} \propto \rho_{xx}^\beta$ can be deduced from Eq. (II.28), the overall temperature and field dependences of the factor multiplying $\rho_{xx}^2$ have to be carefully analyzed. If this factor is constant (as in a system with weak pinning, at high temperature where $f_\ell = 1$, and in a moderate magnetic field), obviously $\beta = 2$.

The discussion of other cases is far from trivial, but some studies show that a scaling still holds, however, with a smaller value of the exponent $\beta$. From numerical simulations (already discussed on page 30), Dong and Wang [58] get an exponent $\beta \approx 1.7$. Phenomenological analytical relations can also be used to estimate this exponent. Vinokur *et al.* indeed used the experimental values of scaling exponents in voltage-current characteristics around a vortex glass transition to get an approximate expression for the pinning factor $\Gamma_p$ [85], yielding $\Gamma_p(v_L) \sim v_L^{-1/2}$, which leads to an exponent $\beta = 1.5$ [57].

Therefore, many different values of the exponent can be expected depending on the pinning properties. It is indeed interesting to note that different types of pinning would apparently lead to different scaling exponents. Again, similarly to the predictions for the Hall anomaly, it should be said that the results are quite sensitive to the parameters of the model, which are not very well known experimentally. As a consequence, numerical studies appear to be a promising tool for investigating the possible scaling behaviors within this model. Since only Ref. [58] provides such a study, much more has to be done in this regard.

## 4.3 Vortex lattice defects

In Ao's model of moving vacancies in a pinned vortex lattice (see paragraph 3.2), a scaling law is also present in the low-temperature limit $k_B T < E_v$, where $E_v$ is the energy scale of the vacancy formation. From Eq. (II.33), using the expressions (II.31) and (II.32), we see that

$$\rho_{xx} = \frac{n_o \phi_o}{e N} \cdot \frac{1}{\omega_{c2} \tau_o} \exp\left(\frac{-(a_v + b_v) E_v}{k_B T}\right) \tag{II.47}$$

and

$$\rho_{xy} = -\frac{n_o \phi_o}{e N} \cdot \frac{1}{\omega_{c2}^2 \tau_o^2} \exp\left(\frac{-(2 a_v + b_v) E_v}{k_B T}\right). \tag{II.48}$$

This obviously leads to the relation $\rho_{xy} = A \rho_{xx}^\beta$, with

$$\beta = \frac{2 a_v + b_v}{a_v + b_v}$$

varying between 1 and 2 (for the thermally activated vacancies of temperature-independent mobility and for the pinning-induced vacancies, respectively). $A$ is a constant independent of the magnetic field $B$, provided $n_o$ is itself field-independent.

## 4.4 Vortex glass scaling

Another approach to the scaling law is the vortex-glass model [22], obtained within the Ginzburg-Landau theory, and briefly presented in the preceding chapter (paragraph I.2.2). According to the usual scaling relations for the asymptotic behavior in critical phenomena, the longitudinal conductivity is expected to diverge around the vortex-glass transition as

$$\sigma_{xx} \propto \xi_g^{z+2-d}, \tag{II.49}$$

where the coherence length $\xi_g$ itself diverges as

$$\xi_g \propto |T - T_g|^{-\nu}$$

with $T_g$ the vortex-glass transition temperature and $d$ is the spatial dimension of the system. The exponents $\nu$ and $z$ are universal critical exponents, only depending on the nature of the considered critical phenomenon, and can be determined from experiments. For the vortex glass transition, for example, we have $z \approx 5$ (see for example Ref. [84]).

For the Hall conductivity, the asymptotic dependence (II.49) has to be modified, to take into account the imaginary part $\gamma''$ of the relaxation time required for the Hall effect (see paragraph 3.3), not included in the original vortex-glass theory from Fisher *et al* [22]. For this correction, Dorsey *et al*. [86] propose the form

$$\sigma_{xy} \propto \xi_g^{z+2-d} F_{xy}\left(\gamma'' \xi_g^{-\lambda_{\gamma''}}\right)$$

where $\lambda_{\gamma''}$ is a new critical exponent related to the particle-hole asymmetry at the vortex-glass transition, represented by $\gamma''$. We have noted in paragraph 3.3 that the Hall conductivity vanishes when the imaginary relaxation time $\gamma''$ is zero, and that the sign of $\sigma_{xy}$ is determined by the sign of $\gamma''$.

It is then natural to expect that the correction function $F_{xy}$ is linear in its argument $\gamma'' \xi_g^{-\lambda_{\gamma''}}$ for small values (close to the transition). Therefore, we have

$$\sigma_{xy} \propto \gamma'' \xi_g^{-\lambda_{\gamma''}} \xi_g^{z+2-d}$$

or with Eq. (II.49)

$$\sigma_{xy} \propto \gamma'' \xi_g^{-\lambda_{\gamma''}} \sigma_{xx}.$$

This scaling relation for the conductivity can be straightforwardly translated for the resistivity, provided $\rho_{xy} \ll \rho_{xx}$:

$$\rho_{xy} \propto \gamma'' \xi_g^{-\lambda_{\gamma''}} \rho_{xx},$$

which can be written as

$$\rho_{xy} \propto \rho_{xx}^\beta$$

with $\beta = 1 + \lambda_{\gamma''}/(z+2-d)$. In their original work, Dorsey *et al.* determine that $\lambda_{\gamma''} \approx 3$ in order to get $\beta \approx 1.7$, the only experimental value known at the time. However, whatever the value of this critical exponent is, it is supposed to be universal, only $\gamma''$ being sample-dependent. Therefore, $\beta$ should be sample independent, which is not confirmed by experiments.

Moreover, in this model the scaling law is directly related to critical behavior around the vortex-glass transition. As a consequence, no scaling is expected far above the transition temperature $T_g$. Again, experiments in which scaling is observable in any regime (of course far enough from sign changes in the Hall resistivity), including close to $T_c$, does not corroborate this prediction. This is mostly evident in BiSrCaCuO, in which $T_g$ is very low because of the quasi-2d layered structure, although the scaling is verified up to much higher temperatures [87].

| CHAPTER III | *HALL EFFECT REVIEW : EXPERIMENTS* |

In this chapter, we review the main observations of the Hall effect in type II superconductors. We will restrict ourselves to a discussion on the mixed state Hall effect, focusing on the Hall anomaly and the Hall scaling law. However, since the discussions of experimental data often refer to the existing theories, we will also make a series of comments on the concerned models.

## 1. Hall anomaly

### 1.1 First observations

Even before high temperature superconductors were known, a sign change in the mixed state Hall effect had already been observed, for example in niobium [88] and in vanadium [89]. However, in these materials this feature was found to be strongly sample dependent. It was realized for example that macroscopic linear defects introduced by the rolling process in the sample manufacture were leading to a strong guided motion of vortices, then dominating the transverse electric response of these materials. Moreover, this sign change was never seen in many other samples, *e.g.*, metallic alloys [90]. It is therefore not very clear whether this sign change is strictly related to these defects or if it is actually an intrinsic consequence of flux flow.

Later, the same kind of anomaly has been seen in high temperature superconductors, first in $YBa_2Cu_3O_{7-\delta}$ [45] and in $Bi_2Sr_2CaCu_2O_{8+\delta}$ [46] polycrystalline samples (ceramics and films, respectively). Again, there were first some doubts about the origin of this sign change, because of a possible influence of inhomogeneities. But shortly thereafter, a similar sign reversal was also observed in single crystals of the same two compounds by Forró *et al.* [91], and on epitaxial $YBa_2Cu_3O_{7-\delta}$ thin films by Hagen *et al.* [92], strengthening the idea of an intrinsic cause for this effect. Other early observations can be found in Ref. [93,94].

Note that after the discovery of the Hall anomaly in cuprates described above, new measurements have been performed in low temperature superconductors, confirming that this effect is also intrinsically present in conventional materials. One of the most convincing examples is the case of

# III

2$H$-NbSe$_2$ [95], since this anisotropic layered superconductor can be easily grown as high purity single crystals free of macroscopic inhomogeneities.

**1.2 Hall anomaly in different cuprates**

In the preceding paragraph, we have only mentioned the two most studied cuprate high temperature superconductors, namely YBa$_2$Cu$_3$O$_{7-\delta}$ (hereafter Y:123) and Bi$_2$Sr$_2$CaCu$_2$O$_{8+\delta}$ (Bi:2212). We briefly list here the other compounds in which the same type of anomalous Hall effect was observed.

First, Y:123 has been doped either with praseodymium (Y$_{1-x}$Pr$_x$Ba$_2$Cu$_3$O$_{7-\delta}$) [96,97] or with calcium (Y$_{1-x}$Ca$_x$Ba$_2$Cu$_3$O$_{7-\delta}$) [98,71]. Complete substitution of yttrium is also used to obtain other compounds of the 123 family, such as with erbium (Er:123) [46], holmium (Ho:123) [99,100] or europium (Eu:123) [100].

Bismuth-based compounds were also doped in various ways, for example by introducing lead on bismuth sites, leading to (Bi,Pb):1112 and (Bi,Pb):2234 [101], (Bi,Pb):2223 [102], or even in the calcium free variety, (Bi,Pb):2201 [71]. The same calcium free bismuth-based cuprate Bi:2201 can also be doped with lanthanum on the strontium sites, leading to Bi$_{1.95}$Sr$_{1.65}$La$_{0.4}$CuO$_{6+\delta}$ [63]. All of these materials have shown a Hall anomaly, at least in some doping range (see paragraph 1.4 for a discussion on the doping dependence).

Similar structures, either thallium or mercury-based, have a behavior very similar to bismuth-based compounds. Therefore, it is not surprising to find again a Hall anomaly in Tl:2212 [93,103-105], Tl:2223 [106], and Hg:1212 [107].

In the first discovered superconducting cuprate[1] family, La$_{2-x}$Sr$_x$CuO$_{4-\delta}$ [71,108] has also been shown to exhibit a Hall anomaly. Similarly, Nd$_{2-x}$Ce$_x$CuO$_{4-\delta}$ has an anomaly in its Hall behavior [109]. However, the latter compound has a striking peculiarity : it is the only known cuprate to have a *negative* (electron-like) Hall effect in the normal state, the anomaly therefore consisting of an incursion into positive territory in the mixed state.

However, measurements in Y:123, Ho:123 and Eu:123 have shown that, although the "usual" Hall effect (that is, with the magnetic field along the *c*-axis, and the current and the electric field in the *ab* plane) is always hole-like (namely *p*-type) in the normal state with a negative Hall anomaly, the "out of plane" Hall effect (magnetic field and current in the *ab* plane, perpendicular to each other, and Hall voltage measured along the *c*-axis) is negative (*n*-type, electron-like) in the normal state [100,110-112]. In one of these cases, the Hall anomaly (still observable in this orientation) is seen to consist of a *positive* mixed state Hall signal, therefore inducing a sign reversal [100], just as in Nd$_{2-x}$Ce$_x$CuO$_{4-\delta}$, but in other reports the anomaly is a *negative* mixed state Hall contribution adding to the already negative normal state Hall effect [111,112].

Even though all the materials cited here show a Hall anomaly, the sign change occurs only in some regime. For example, it always disappears at high enough magnetic fields (see next paragraph).

From those measurements known at that time, and including all the known the measurements in low temperature superconductors, Hagen *et al.* [109] noticed a correlation between the existence of a Hall sign reversal and the ratio $\ell/\xi$, where $\ell$ is the mean free path at $T_c$, as estimated from the normal state resistivity, and $\xi$ is the BCS coherence length. They reached the conclusion that the anomaly occurs when this ration is of the order of unity, that is in the crossover between the clean and dirty limits. This correlation was later confirmed by Colino *et al*. [99].

---

1. which was a La-Ba-Cu-O compound, see Ref. [3].

## 1.3 Multiple sign reversals

In the most anisotropic materials, the Hall effect often displays two successive sign changes in the mixed state. This has been observed in bismuth, thallium and mercury based compounds, more precisely in Bi:2212 [113,114,87], Tl:2212 [103-105], Tl:2223 [106], and Hg:1212 [107].

Since the second sign change was observed in these materials, it has of course also been looked for in yttrium-based compounds, but no clear evidence could be found for a long time. Finally, using pulsed high currents in a Y:123 thin film, Nakao *et al*. were able to show that at much lower temperature than the usual measurements of the Hall anomaly, the Hall effect sign is indeed again the same as in the normal state (for some appropriate values of the magnetic field and temperature), proving the existence of a second sign change [115].

Very recently, a third sign reversal has even been reported for Hg:1212 by Kang *et al*. [116]. It should be noted however that it could only be seen in samples in which columnar defects were added by sample irradiation.

As we have mentioned before, the Hall sign change only occurs in a given regime. For example, it cannot be observed at an arbitrarily high magnetic field. An illustration of this change of behavior is given in Fig. III-1 for a situation with two sign changes (but the same happens when only one inversion is observed, as in 123 compounds) : when the Hall resistivity is measured as a function of the temperature at an appropriate magnetic field, the sign reversals can be observed (curve *a*). At higher fields, the sign finally becomes constant (curve *b*), even though the anomaly is still observable, but is too small to induce a sign reversal. For example, the Hall resistivity is reported to become strictly positive at approximately 8 T for Y:123 [87], between 2 and 3 T in Tl:2212 [103]. Note that these values are slightly sample dependent, and some variations can be found in the literature : for Bi:2212, Ri *et al*. find a limit around 3 T [87], whereas Zavaritsky *et al*. place it between 4 and 5 T [113].

It is interesting to add that, although the first negative incursion (shown in Fig. III-1) therefore decreases with increasing magnetic field, the second one (after the third sign change) seems to be enhanced at high fields (at least up to 8 T) [116].

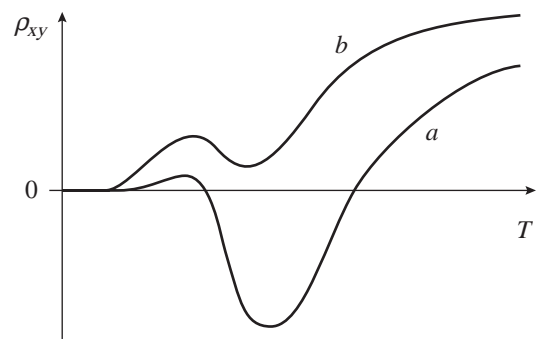

**Figure III-1 :** a) Double sign change in the Hall resistivity, as observed in some materials (see text for references). b) Usual behavior at higher magnetic fields : the sign is constant, but the Hall anomaly is still present.

### 1.4 Doping dependence

As we have already explained in chapter I (page 11), the oxygen content in the cuprates influences the hole concentration, allowing one to go from the underdoped to the overdoped regimes (Fig. I-6). This is also the case for the other kinds of doping, such as the lead inclusion in bismuth-based compounds, or the addition of praseodymium in Y:123 (see paragraph 1.2 above).

This doping level of the materials strongly affects the Hall effect. From a complete set of new measurements as a function of doping in different materials, and together with some other data, Nagaoka *et al.* reached the conclusion that only underdoped and slightly overdoped cuprates can have a Hall anomaly [71]. Significantly overdoped systems apparently have a strictly positive Hall effect.

However, a few contradicting results, not included in Nagaoka's analysis, can be found. The most significant one is a study from Jones *et al.* on different Y:123 epitaxial films with various oxygen contents [117]. Here, the opposite behavior is found : the anomaly disappears in underdoped films. Similarly, the sign change could not be seen in underdoped $Y_{1-x}Pr_xBa_2Cu_3O_{7-\delta}$, more precisely for $x < 0.24$ (or for $\delta > 0.27$ in praseodymium-free samples) [97,118]. These last results should nevertheless be considered carefully, since $Y_{1-x}Pr_xBa_2Cu_3O_{7-\delta}$ samples with $x < 0.2$ show a Hall anomaly in Ref. [96], and in Ref. [118], it seems that the measurements are performed only at a single magnetic field $B = 5.5$ T. Therefore, it is quite possible that for underdoped samples with low enough $T_c$ (70 K and lower), the anomaly would only be visible at lower fields, as we have explained above (paragraph 1.3).

Apart from these few exceptions, a clear correlation between the doping of the cuprates and their mixed state Hall effect still exists. At first sight, a theoretical explanation could be found in the vortex charge model (see preceding chapter, paragraphs 3.3 and 3.4), since in this theory the vortex contribution to the Hall effect is proportional to $\partial \ln T_c / \partial \mu$ (or $\partial N / \partial \varepsilon$), a parameter that indeed depends on the charge carriers concentration, and precisely changes sign around optimal doping. Unfortunately, the sign predicted from this model is exactly opposite to the observed one : $\partial \ln T_c / \partial \mu$ is positive (and so should be the Hall effect in the mixed state) for underdoped materials [71]. Therefore, another way to determine the imaginary part of the relaxation time for the time dependent Ginzburg-Landau (TDGL) formulation is required. As we have already discussed in the preceding chapter (page 36), such an alternative model can be found in the preformed pairs scenario from Geshkenbein *et al.* [43], which predicts the correct sign for underdoped systems.

### 1.5 Hall conductivity fit

In chapter I, we have mentioned that the Hall conductivity can be decomposed into two parts, one being the delocalized quasiparticles contribution, seen as the normal fluid Hall effect, the other being the vortex flow Hall effect (see Eq. (I.3) on page 10). Rigorous calculations of these terms can be done in the frame of the Ginzburg-Landau theory [67,68]. From the experimental point of view, many authors have tried to perform such a decomposition for the measurements of the Hall conductivity as a function of the magnetic field. Since in the normal state $\sigma_{xy} \propto H$ and in the mixed state $\sigma_{xy}$ generally diverges at low fields, it is rather natural to start from a decomposition

$$\sigma_{xy}(H) = \sigma_{xy}^v + \sigma_{xy}^{qp} = C_1/H + C_2 H,$$

where the factors $C_1$ and $C_2$ are in principle temperature dependent, and $\sigma_{xy}^{qp} = C_2 H$ denotes the normal (delocalized quasiparticles) component, written as $(1-g)\sigma_{xy}^n$ in Eq. (I.3). This expression has been satisfyingly used to fit the experimental data in some cases [119,120]. The vortex part of the conductivity can even be fitted with expressions for the fluctuation part of the conductivity resulting from the Ginzburg-Landau theory [121,122].

However, in some materials, the equation above is not adequate to fit the data. In many cases, a constant term $C_3$ had to be added to the conductivity [97,104,108,123] :

$$\sigma_{xy}(H) = C_1/H + C_2 H + C_3 ,$$

even though in Ref. [104] $C_2$ is then shown to be zero. A nice summary is given in the very recent reference [123].

Finally, note that in some cases a power law temperature dependence of the coefficients $C_1$ to $C_3$ have been reported [97,108,119,120,123]. However, since in our sample the "divergence" at low field is anyway much faster than $1/H$ and is current dependent (therefore related to pinning, see chapter VI), this decomposition is not really relevant for our measurements, so that we do not give more details on this topic.

## 2. Hall scaling

We will try now to summarize the available experimental data about the Hall resistivity scaling law $\rho_{xy} = A \rho_{xx}^\beta$. However, as we shall see, there is a large confusion in the analysis of the results. One of the objectives of this short section is therefore to make a rather exhaustive list of the references dealing with the subject, and attempt to give an objective digest of the whole.

The first observation of the scaling law was made from Hall measurements as a function of the temperature in epitaxial YBa$_2$Cu$_3$O$_{7-\delta}$ thin films by Luo *et al.* [47]. They found $\beta = 1.7 \pm 0.2$ at both $B = 1.4$ T and 3.7 T. Later, many other observations have been done in many different compounds and all sample types [50,87,104,107,112,114,120,124-130], for negative as well as positive Hall effects. Note that the "out of plane" Hall resistivity mentioned in paragraph 1.2 also follows the same scaling relation [112].

Vinokur's model (Ref. [85] and paragraph 4.1 in the preceding chapter) is often cited as the theoretical explanation of this scaling relation, and several authors indeed report an exponent of $\beta = 2$ [50,87,114,120,124], in agreement with this theory. However, some comments have to be made on several of these references.

First, in two of them, different values of $\beta$ are obtained, only one being $\beta = 2$ : in Ref. [87], which is a comparative study between Y:123 and Bi:2212, the authors get $\beta = 2$ for the first sample, but $\beta = 1.8$ for the second one. Similarly, in Ref. [124], in which Y:123 crystals are irradiated to create columnar defects, $\beta = 2$ is only found for the non-irradiated sample when measured at a field $B \geq 3$ T, whereas the other data (all the irradiated crystals, as well as the unirradiated one for $B = 1$ T) show $\beta = 1.5$.

In Ref. [120], the scaling law yields $\beta = 2$, but from the reported decomposition of the Hall conductivity $\sigma_{xy}(H) = C_1/H + C_2 H$ (paragraph 1.5 above), it is clear that $\sigma_{xy}$ diverges at low fields, since $C_1 \neq 0$. These two results are in some sense contradicting each other, since as we have already seen, the Hall conductivity can be expressed as $\sigma_{xy} = \rho_{xy}/(\rho_{xx}^2 + \rho_{xy}^2) \approx \rho_{xy}/\rho_{xx}^2 \propto \rho_{xx}^{\beta-2}$ knowing

# III

that $\rho_{xx} \gg \rho_{xy}$. Therefore, $\beta = 2$ should lead to a constant conductivity, or at most a conductivity that just follows the small residual temperature and/or field dependence of the prefactor $A$ of the scaling law, a factor which is in any case not expected to diverge at low fields. This contradiction could be explained by noting more generally that the numerical fit of the experimental data with a power law most often leads to statistical errors of the order of $\pm 0.1$ - $0.2$ for the exponent $\beta$, such that the actual value of $\beta$ in this work might indeed be slightly smaller than the reported value of 2.

Aside from these references where $\beta = 2$, other reports add to the confusion rising about the experimental values of $\beta$. In a work from Samoilov *et al*. [104], for example, lines corresponding to $\beta = 2$ are superimposed on the log-log plot of the data. However, these lines are just given for a visual reference, but do not in anyway correspond to actual fits of the data. A close examination of the figure (mostly at low fields) clearly reveals that a value of the order of $\beta \approx 1.5$ would much better represent the experimental results. In another Letter, the same author implicitly suggests that the scaling $\rho_{xy} \propto \rho_{xx}^2$ predicted by Vinokur is in agreement with the data [125]. However, on the only plot representing this scaling, lines corresponding to $\beta = 1.3$ to $1.4$ are used to fit the data (although this is not mentioned in the text). Since these papers are often incorrectly cited, we think that it is worth pointing out that the only work from the same group actually showing a set of data truly corresponding to $\beta = 2$ is their first experiment on Bi:2212 (Ref. [114]), in the Tesla range. Note that in this work, the behavior is also shown to be very different at much lower fields (a regime with $\beta \approx 1$ is found at $B = 0.1$ T).

After this discussion about the $\beta = 2$ value of the exponent, we now turn to the experimental data bringing different results. First, we note that these values are usually lying between 1.4 and 1.8 (quite often around 1.5 to 1.6 [112,127,128]), but values as low as 0.8 can be found [130]. Note that this scattering of data is apparently not related to the different compounds, but really seems to be the result of a sample dependence. Another important finding in some of these works is the field (or temperature) dependence of the exponent $\beta$, as already mentioned above in the case of Ref. [124]. The same kind of behavior is also observed by other authors. Most often, $\beta$ is seen to increase for increasing fields [107,126,128], whereas in Ref. [129] the inverse dependence is observed. In Ref. [128], $\beta$ decreases at high temperatures. Note that in Ref. [126] from Budhani *et al.*, lines for $\beta = 1.85$ are plotted on each different experimental data set, but obviously not all correspond to actual power law fits, just as already mentioned above concerning the work of Samoilov *et al*. [104]. Although this behavior indeed corresponds to the high field data, it clearly does not at all fit the curves for low fields, as has also already been commented by Kang *et al*. [107]. Again, a close inspection of the figure shows that the exponent is of the order of 1.5 at $B = 3$ T, and almost decreases down to 1 for $B = 1$ T.

To conclude the discussion about the temperature and field dependence of the scaling relation, we note that Wöltgens *et al*. report a analysis of the Hall resistivity measured as a function of the current density [50] in terms of Dorsey's scaling theory for a vortex glass transition [86] (which is discussed in paragraph 4.4 of the preceding chapter). However, they note that the scaling indeed holds *above* the glassy transition temperature $T_g$ (with $\beta = 2$), but *not* below, where it seems to become temperature dependent (even though the analysis in the text is somewhat confusing, giving contradicting conclusions).

In summary, despite the confusion prevailing about this subject, we believe that an exponent of $\beta = 2$ remains an exception in real samples (although it probably does occur in some cases), and that neither Vinokur's model nor Dorsey's vortex glass scaling can satisfactorily explain the observed Hall scaling relation.

Finally, we mention that the scaling is obtained for measurements as a function of the temperature (that is, by plotting log $\rho_{xy}(T)$ versus log $\rho_{xy}(T)$, as in most of the references mentioned here), as well as for measurements as a function of the magnetic field (namely reporting log $\rho_{xy}(B)$ versus log $\rho_{xy}(B)$ [120,127,130]), or even of the current density [50,128]. Note that data for which the scaling for $\rho_{xy}(T)$ leads to a field dependent $\beta$ (see above) would not scale if $\rho_{xy}(B)$ would be reported as a function of $\rho_{xx}(B)$.

## 3. Influence of disorder

### 3.1 Pinning dependence

After the Hall scaling exponent discussed in the preceding section, the influence of pinning on the Hall effect is another controversial subject. To address this question, two different methods are generally used : either increase the pinning by introducing columnar defects (by sample irradiation) [124-126,131-134], or use multilayered samples (or "superlattices") made of successive superconducting YBa$_2$Cu$_3$O$_{7-\delta}$ and semiconducting PrBa$_2$Cu$_3$O$_{7-\delta}$ layers, the respective thickness of which determines the pinning efficiency [120,128,129]. Increasing the current density (and thus the driving force on the vortex) to reduce the relative effectiveness of pinning forces, therefore approaching a "free flux flow" regime, has also been used [95,122,135]. An original method consisting in the rotation of the magnetic field with respect to the columnar defects to lower their pinning action has also been reported [132]. Finally, we note that the sign reversal of the Hall effect has been observed to happen even above (though very close to) $T_c$, first in YBa$_2$Cu$_3$O$_{7-\delta}$, then in bismuth-based compounds [63]. Of course, this regime is clearly even beyond free flux flow, and presumably no vortex (thus no pinning) can be considered here.

The only theory predicting pinning dependence of the Hall behavior discussed in these experimental works (but not the only theory existing, as we shall see in chapter VI) is WDT's model for pinning induced backflow (see paragraph 3.1 in the preceding chapter). This is actually the center of the controversy : there are approximately as many authors finding support for this theory in their experimental data [107,124,129,133] as the opposite [95,120,122,126]. However, in the latter references, the conclusion usually follows from the observation that the amplitude of the negative Hall resistivity *decreases* when the pinning is increased. Unfortunately, as we have already extensively discussed in the preceding chapter (page 30), the analysis in the frame of this theory is not so straightforward. The persistence of the controversy is probably partly related to the necessity to perform actually numerical simulations to investigate accurately the predictions of this model. As we have noted in the preceding chapter, the effect of disorder on the Hall resistivity depends on the competition between the decrease of $\rho_{xx}$ and the increase of the (negative) factor in brackets of Eq. (II.28). We notice here that in the only attempt to analytically use this model to quantitatively discuss experimental results, Smith *et al*. consider the limit $\tilde{f}_\ell \to \infty$ [136], whereas this parameter of the theory is strictly in the range $0 \leq \tilde{f}_\ell \leq 1$ (see paragraph 3.1 of chapter II). Since this choice precisely suppresses the key competition between the two terms indicated above, it is not surprising to see that this analysis cannot match the experimental data [136].

Actually, we believe that it would be better to discuss the pinning dependence of the *conductivity* $\sigma_{xy}$ rather than the resistivity $\rho_{xy}$ : the former, which can reasonably be written as $\sigma_{xy} \approx \rho_{xy}/\rho_{xx}^2$ (see section 2), is indeed directly proportional to the negative factor resulting from the effect of pinning (the term between brackets in Eq. (II.28)), which is not the case for the resistivity. According to WDT's model, the conductivity should then be more negative for stronger pinning (the very behavior which is often incorrectly expected for the resistivity), a fact usually experimentally verified in the low field/temperature regime (see *e.g.* Ref. [124,132]; note that Ref. [125] shows exactly the same Hall conductivity behavior as in Ref. [132], in contrast to the author's claim : the resolution is merely lower, and the use of too large symbols prevents the clear observation of this trend).

The experiments involving irradiation of samples deserve a few comments. First, we note that this irradiation should remain in reasonable limits : if the samples are overexposed, this process actually leads to an effect opposite to what is expected, that is a *reduction* of the critical current (hence of the pinning) because of excessive damage in the sample, as in Ref. [131,137]. Second, even when this limit is not reached, irradiation usually increases the normal state resistivity $\rho_n$ and reduces the critical temperature $T_c$. Therefore, a correct comparison of samples can only be done by representing $\rho/\rho_n$ versus $T/T_c$, a normalization which is often not respected, see *e.g.* Ref. [126].

Finally, we would like to raise a last question, yet almost never addressed, about irradiated samples : since the charge carrier's mean free path seems to be closely related to the mixed state Hall effect (see paragraph 1.2), one can wonder if the influence of the columnar defects is really only a question of pinning. More precisely, the mean free path is indeed reduced by the irradiation inside the defect : it is precisely this process which provides the pinning force. If one considers the column as a narrow and straight cylinder on which the vortex can be localized and pinned, and from which it can of course also unpin, then indeed the irradiation only provides additional pinning. However, the situation is much more complex : the columns actually have a diameter very large compared to the vortex core scale (they can be as much as three orders of magnitude larger than $\xi$ [126,133]).

Therefore, the vortex cores stand in large "damaged" regions of considerably reduced mean free path, which they have to cross (without feeling any pinning force, which is mainly localized at the boundary of the column where there is a strong mean free path gradient) before they can unpin from the defects. But since the mean free path presumably influences the flux flow Hall effect, this "intracolumnar motion" can significantly alter the Hall response of the system. In the case of a vortex unpinning from a column and moving to the next one, a part of the vortex path will be inside the column, while the other part is in the (almost) unaffected material, with the pinning force mainly occurring at the interface. The Hall effect is then influenced by two different values of the mean free path, in a ratio depending on the column size and density, both very much sample dependent. Unfortunately, although the defect density is always quite well controlled and included in the analysis, the column size is seldom reported, and almost never taken into account in the interpretation of data. The role of the mean free path and the diameter of the defects is in fact only discussed in the very recent work on the third Hall sign change in irradiated Hg:1212 [116].

Even if this description is very schematic and certainly oversimplified, we see that we have to consider a vortex trajectory with inhomogeneous mean free path plus increased pinning, and not only an irradiated superconductor with an enhanced pinning, which is otherwise similar to the unirradiated one. For this reason, the creation of columnar defects is probably not the best way to study the influence of pinning on the Hall effect as long as the respective importance of the two effects described above (increased pinning and inhomogeneous mean free path) is not better understood.

## 3.2 Effects of twins

Since the present work deals with a twinned $YBa_2Cu_3O_{7-\delta}$ crystal (see next chapter), it is worth briefly mentioning some previous works on the relationship between the Hall effect and the presence of twins. In fact, there are very few studies of this type.

First, we note that Rice *et al.* [127] performed measurements in an untwinned $YBa_2Cu_3O_{7-\delta}$ single crystal, and compared their results to those of twinned samples. They find no significant difference in the Hall anomaly of these two types of samples, indicating that the pinning is not the origin of the Hall sign change.

More recently, the work of Morgoon *et al.* [49] already discussed in chapter I (page 15) should be added the two very interesting contributions from Casaca *et al.* [138,139]. First, measurements of the Hall effect at high fields show that the pinning by twin boundaries indeed affects the Hall conductivity, even though this effect only happens at lower temperatures than the onset of the twins-related pinning effect obviously visible in both the longitudinal and the Hall resistivities [138]. Secondly, in unidirectionally twinned thin films, they also measure the Hall resistivity with the current applied successively along and perpendicularly to the twin planes [139]. They again observe that the Hall conductivity is influenced by twin boundary pinning, and also that the ratio between the Hall resistivity in the two directions is temperature dependent. They explain their results by introducing a modified WDT model for anisotropic pinning. However, the sign change occurs at the same magnetic field for the two directions, which is in apparent contradiction with this modified model.

Some more references and a related discussion on the general relations between the twins in $YBa_2Cu_3O_{7-\delta}$ and the vortex dynamics will be given in chapter V (paragraph 1.2, page 72).

## 3.3 Conclusion

In summary, we note that :

- unfortunately, too much confusion prevails concerning the values of the scaling exponent to draw a conclusion on its pinning dependence,
- the sign change of the Hall effect (the Hall anomaly) may happen in regimes where no pinning can be considered, such that vortex pinning is not the primary cause for this effect (but this does not necessarily prevent a pinning-based description to be adequate and successful in other regimes, see below),
- even though there is still no absolute consensus on that point, we affirm that the Hall conductivity $\sigma_{xy}$ is pinning dependent in some regimes (this is very clearly confirmed by our results presented in chapter VI), such that *the pinning definitely has to be taken into account for a complete description of the Hall effect*,
- even though the twin boundary pinning affects the Hall conductivity, vortex guided motion resulting from its anisotropic nature *does not* significantly influence the Hall angle (in the highest quality crystals at least), since twinned and untwinned samples show similar behaviors.

# III

| CHAPTER IV | *SAMPLE AND SETUP* |

## 1. Experimental setup and procedures

In this first section of the present chapter, we describe and comment all the experimental parts, devices and pieces of apparatus that are of importance for the present study, except for the sample itself, which is the object of the second section.

### 1.1 Cryostat, magnet and sample holder

The vacuum insulated cryostat is a commercial Oxford Spectromag 4000–7T system, operated with liquid helium (and a shielding liquid nitrogen vessel). Split pair superconducting magnets are lying in the helium bath, generating a magnetic field up to 7 Tesla. As we shall see later, it is noteworthy to mention that the whole cryostat is placed on an antivibration system, cutting off all external vibrations of frequency above 1 Hz. Since the magnet produces a horizontal field with vertical sample access (thanks to its split coils geometry, see Fig. IV-1), orientation of the sample with respect to the field can be straightforwardly achieved by a simple rotation of the sample rod, unlike systems with axial access in a vertical magnet coil.

On the upper part of the rod, a mirror is placed to determine the orientation of the sample, with the help of a laser beam. By measuring the position of the laser spot at a rather large distance from the mirror (about 4 meters), a very high accuracy and excellent reproducibility can be achieved : the angle is estimated to be known within a $\pm 0.01°$ error only. Of course, this accuracy is relevant for an angle relative to a given initial position only. The absolute value of the angle between the magnetic field and the sample orientation (direction of its crystallographic axes) is another issue, and will be discussed later (see page 66).

The sample holder is a home made, specifically designed part mounted at the end of a commercial Oxford sample rod. Before describing the sample holder itself, we just note that the rod is equipped with a bundle of wires, going down from the connector plate at the top to the sample position. These wires are simple insulated copper wires of 100 μm diameter[1], and are glued on the external wall of the thin inox tube forming the rod, in a side-by-side untwisted layout. In other words, no special measure is adopted to avoid ac coupling between the wires or noise generation by pick up of ex-ternal electromagnetic perturbations, as would be the case for example with twisted wires or specific cryogenic coaxial cables.

---

1. Except for the heater current leads which are 200 μm in diameter.

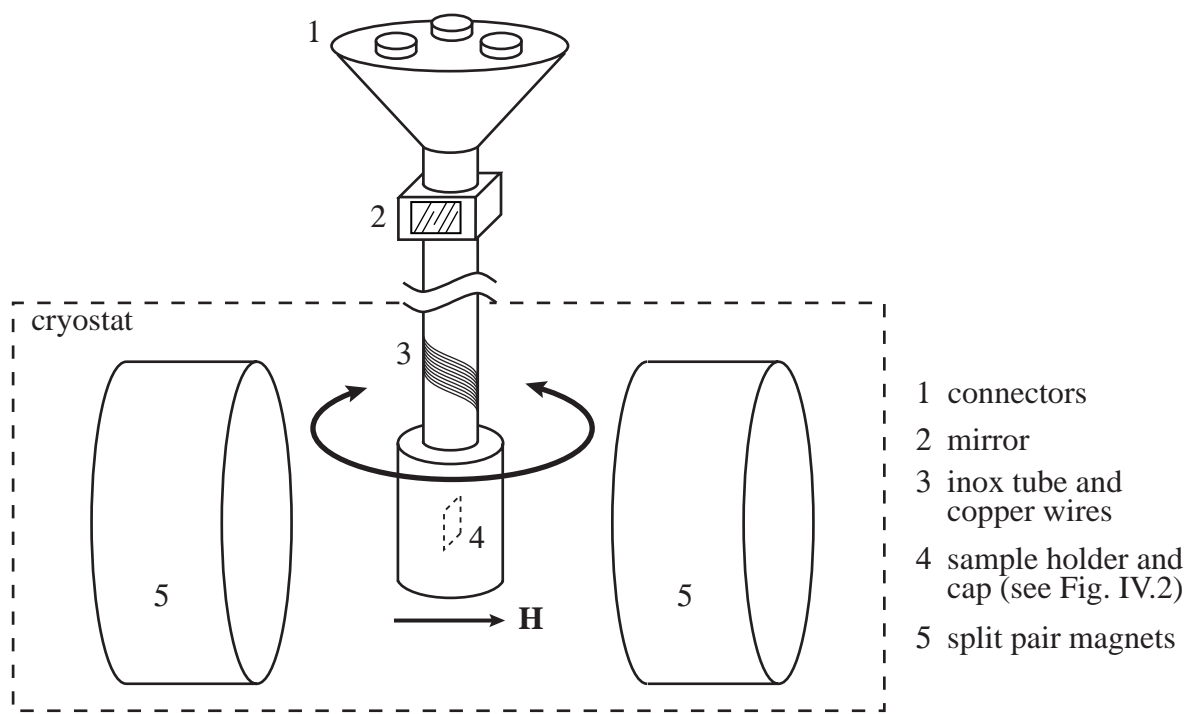

**Figure IV-1 :** Schematic representation of the split pair magnets and sample rod access geometry. The relative field-sample orientation can be changed by turning the sample rod, on which a mirror is fixed to measure accurately its orientation by deflection of a laser beam.

The sample holder is a copper part equipped with its own temperature probe and electric heater (see Fig. IV-2). The connection plate is an easily interchangeable sample support, on which is placed the sample, glued onto a sapphire for a better thermal contact with the copper parts. This glass-fiber plate for printed circuits is designed for the connection of up to 22 independent electric contacts on the sample. Using standard IC socket technology (not shown in Fig. IV-2), it provides the possibility to remove the sample together with its wiring without any soldering operation, therefore leaving the opportunity to replace it later to resume measurements exactly in the same conditions.

### 1.2 Temperature control

A "variable temperature insert" (VTI) is fitted vertically in the cryostat to provide a sample space with controlled temperature between the split coils of the magnet. The sample rod with its sample holder described above is simply inserted in this VTI. A capillary tube equipped with a needle valve brings helium from the liquid bath to the bottom of the VTI, generating a cooling gas flow around the sample holder, and is then pumped out of the VTI by an exhaust tube on top of the cryostat. The capillary ends in a heat exchanger, the temperature of which is adjusted with an Oxford ITC 503 PID temperature controller. Although the expanding helium flow is supposed to be very homogenous in time and space, the sample is protected from direct exposure to temperature or flow fluctuations by a copper cap fitted to the sample holder (see Fig. IV-2). Gas exchange into this al-most closed sample space is however always possible through the wire access passing through the heater coil, at the upper part of the sample holder. This passage is large enough to allow a preliminary thermal equilibrium by convection between inside and outside the cap. For most of the measurements, the pressure of the helium flowing gas was 20 ± 3 mbar.

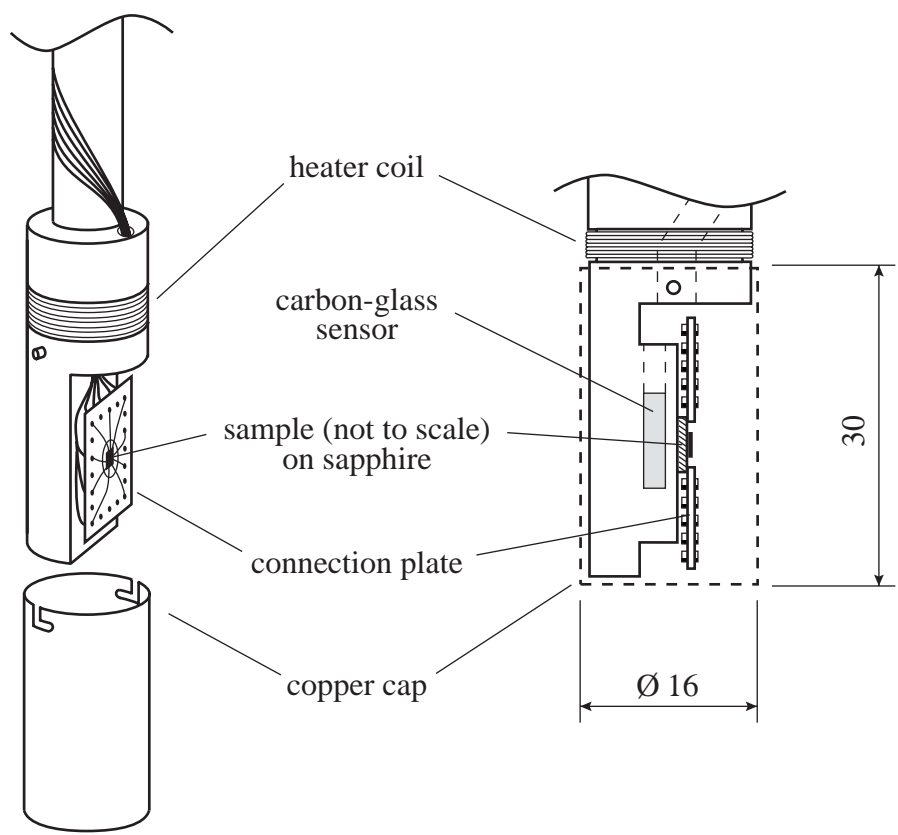

**Figure IV-2 :** Schematic view of the sample holder design. Left : 3D view of the copper sample holder (and open cap) at the end of the inox tube. Right : detail showing the relative positions of the sample, heater and temperature sensor. The main dimensions are given in millimeters.

The sample is glued onto a 0.5 mm thick sapphire plate which is in direct contact with the copper part of the sample holder. This sapphire provides both electric insulation and high thermal conduction. A carbon-glass temperature sensor is embedded in the copper, only 1 mm away from the sapphire position. Even though the gas flow temperature is already regulated, the sample temperature, as determined by this carbon-glass probe, is controlled by a second temperature controller (Lakeshore DRC 93-CA) with the heater shown in Fig. IV-2, achieving a much faster temperature stabilization. The temperature reproducibility (corresponding to the sensitivity of the probe measurement) is of the order of 10 mK, its stability is about 0.1 K (the regulation can be performed within ± 0.05 K of the setpoint), and its accuracy (mainly limited by the sensor calibration and the influence of the magnetic field on the sensor reading) is estimated to be lower than 0.1 K.

### 1.3 Currents and voltages

The current used for the Hall measurements is generated by an ac+dc voltage source (Stanford DS360 ultra low distortion function generator). Its output, a sine wave at 30 Hz added to an offset, is connected to the sample in series with a 1 kΩ resistor. This resistor plays two different roles.

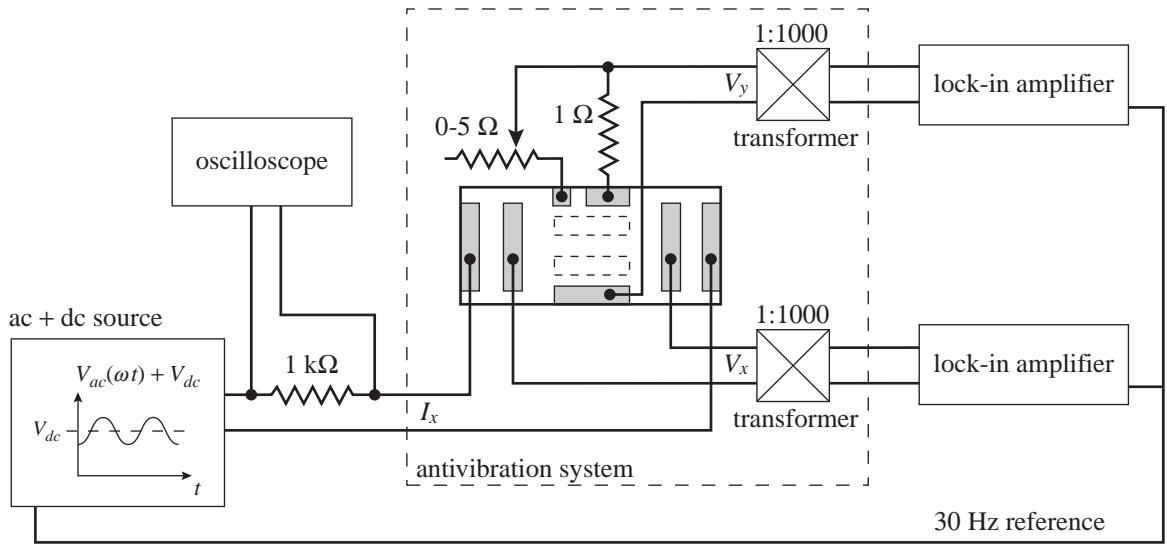

**Figure IV-3 :** Schematic connections for Hall effect measurements.

First, is provides a constant load for the source, since its resistance is much larger than the sample and contacts (the wires, contacts and sample resistances are less than 6 Ω altogether). Second, the voltage drop across this resistor can be accurately monitored by a digital oscilloscope to provide a measurement of the actual current ac and dc components (see Fig. IV-3). Note that a synchronous square signal is also provided by the function generator to serve as a reference for the lock-in amplifiers (see below).

From the connectors of the sample rod, using as short as possible twisted pairs of wires, the two voltage signals are directly fed to very low noise passive transformers (EG&G 1900), amplifying the signal by a factor of 1000. The transformers are placed on the cryostat, to take advantage of the antivibration system. With the sample and wires total resistance of about 4 to 5 Ω, the amplifiers have a rather narrow frequency response, with a maximum of −1 dB close to 30 Hz, but less than −4 dB at 100 Hz already. This has conditioned the choice of measurements at a single frequency of 30 Hz, for which the transformer gain is maximal and does not significantly depend on the input charge (sample and wires resistance). The transformers output are then connected to a lock-in amplifier (Stanford SR830) to measure the amplitude and phase of the voltage ac component.

Special care must be taken to avoid signal grounding (by the use of floating connections on devices and differential measurements on lock-in and oscilloscope, for example), which would lead to undesired ground loops generating large parasitic voltages.

### 1.4 Specific "orientable current" source

For some of the measurements done is this work (presented in the next chapter), we had to apply a current density oriented in any direction in the sample plane. This can be achieved by decomposing this total current into two perpendicular components (see Fig. IV-4).

More precisely, the application of an orientable ac+dc current requires two *in-phase* sources with ac and dc components in the same adjustable ratio. For this purpose, we have developed a simple double current source based on two analogical, four-quadrant multiplier, integrated circuits (Analog Devices AD 385). As before (paragraph 1.3), a single ac+dc source (DS360) provides the total current signal (corresponding then to the magnitude of the current). This signal is multiplied by two factors, corresponding to the cosine or sine of the angle $\beta_j$ (defined in Fig IV-4), to generate signals proportional to $I_x$ and $I_y$, respectively (Fig IV-5). Since the multiplier output is designed for a high impedance load and is limited to the ± 2 V range, these signals are then amplified (by two Stanford SR560 low noise preamplifiers) to match the required characteristics for a current source. The amplifiers outputs are then connected to the appropriate sample current contacts through resistors (Fig IV-6), just as described in paragraph 1.3.

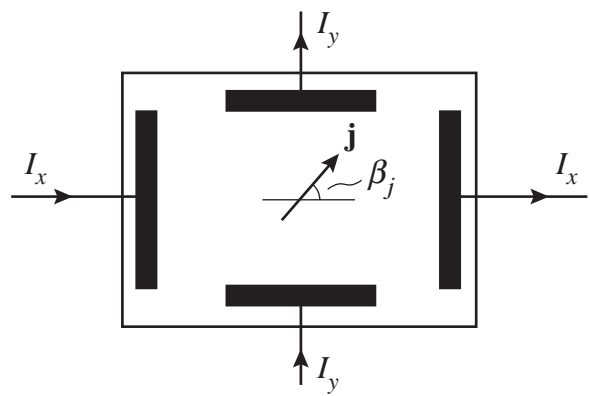

**Figure IV-4 :** Principle of an orientable current density **j** in the sample center by the application of two in-phase perpendicular current components *Ix* and *Iy*.

The dc signals for the multiplication factors are obtained from digital to analog converters (DACs), integrated into the two commercial lock-in amplifiers, and can thus be easily controlled through a computer interface. Note that the exact transfer function of the multiplier circuits has a gain $1/U$ (Fig. IV-5), unknown though very close to unity. However, since the *final* values of the currents $I_x$ and $I_y$ are accurately determined by the voltage drop across the resistors (see paragraph 1.3), it is not necessary to determine nor compensate accurately $U$. On the other hand, another imperfection of these multipliers had to be compensated : their output has a small offset, or in other words a small but significant dc output when either (or both) $x_1-x_2$ or $y_1-y_2$ is zero. An additional input $z$ is precisely provided for this kind of offset compensation, and was connected to two other DACs (see Fig. IV-5).

## 1.5 Measurements procedure

In the last paragraph of this section, we briefly describe the procedure used for the measurements reported in the next chapters. All the data collected correspond to an ac current component at 30 Hz for the reasons evoked in paragraph 1.3.

*a) Hall measurements*

For the Hall measurements, the ac+dc current is applied between contacts 1 and 2 (all the contact numbers refer to Fig. IV-13 on page 64). The longitudinal resistivity is obviously extracted from the voltage between contacts 3 and 4. According to the sample dimensions, a current of $I_x = 1$ mA corresponds roughly to a current density of $j = 10$ A/cm$^2$. Note that small systematic errors are probably present in the numeric estimations of the current densities, as well as the resistivities due to some uncertainty on the effective distances between contacts. The maximum used current den-

# IV

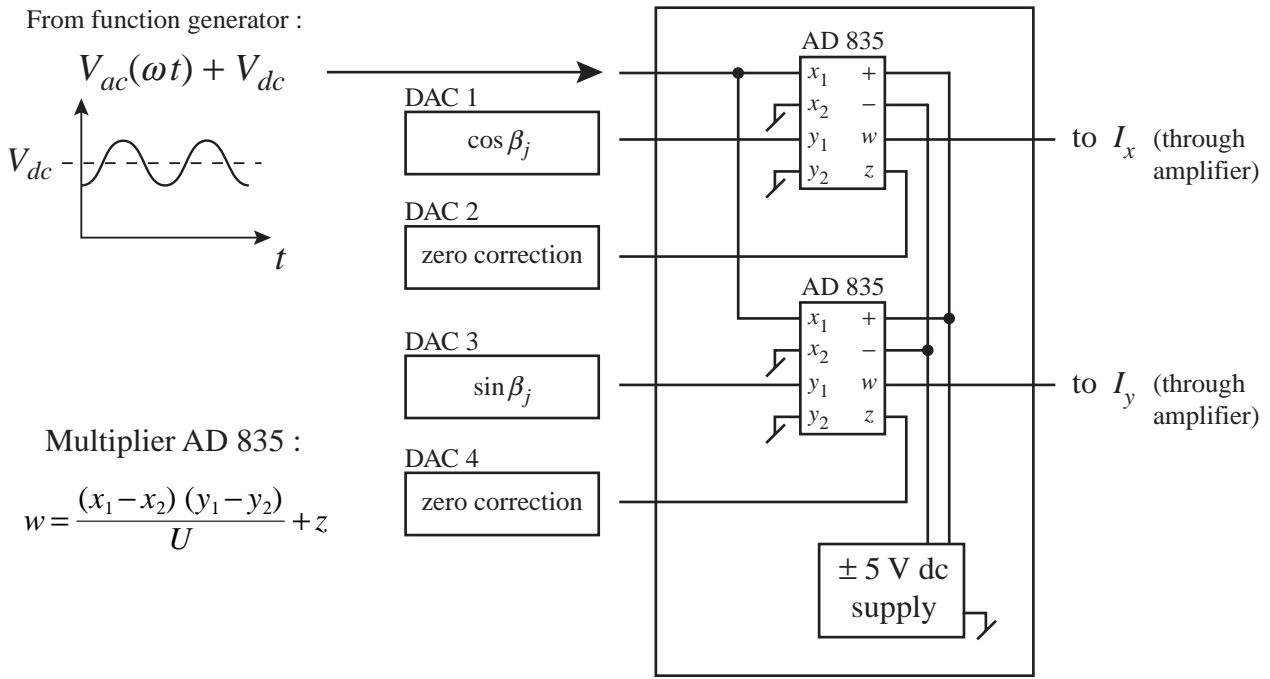

**Figure IV-5 :** Schema of the multiplier circuit used to provide the two in-phase current sources.

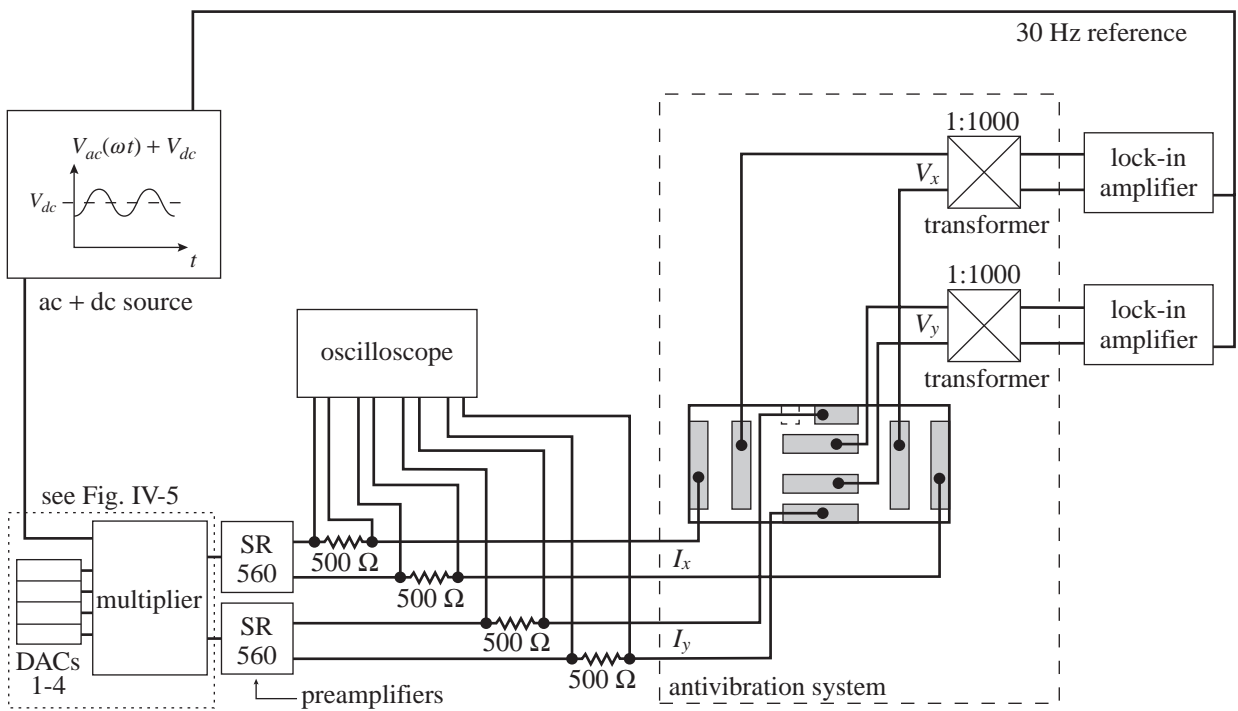

**Figure IV-6 :** Schematic connections for measurements as a function of the current orientation.

sity is of the order of 150 A/cm$^2$, since densities above 200 A/cm$^2$ have been observed to heat the sample slightly.

The transverse resistivity is measured between the contact 5 and a mixing of contacts 6 and 7. These two contacts are indeed linked by an adjustable resistor (Fig IV-6), in order to compensate as much as possible for the misalignment of the measuring direction, which should ideally be exactly perpendicular to the current direction. The resistor is adjusted to get a zero transverse voltage at zero magnetic field.

The two components of the resistivity are simultaneously measured during slow magnetic field or temperature sweeps, and the non-sweeping parameter is of course held constant. At very low field and temperature, deep in the vortex solid where the critical current is much larger than the applied current, the vortices are strongly pinned at rest. We therefore do not expect to measure any flux flow voltage. However, there is, surprisingly, a non-negligible signal at the output of the passive transformers amplifying the measured voltage. This background is mostly due to an inductive coupling between the wires on the sample rod and around the sample (remember that the wires are not twisted). Since we are interested here in the *in-phase, resistive* response only (without inductive part nor electronically induced phase shift influence), it is necessary to perform a background subtraction from the raw data.

If we assume that the total generated voltage corresponds to a resistor (our sample) *in series* with an inductance, we must then make a subtraction of a complex impedance from the complex response given by the voltage modulus and phase (both measured by the lock-in amplifiers). This operation is illustrated in Fig. IV-7, and is nicely justified *a posteriori* by noting that after this simple background subtraction processing, the corrected signal has a constant zero phase (or π if the in-phase voltage is negative), whereas the raw phase has some "arbitrary" – though approximately monotonic – variations of the order of $\pm \pi/2$.

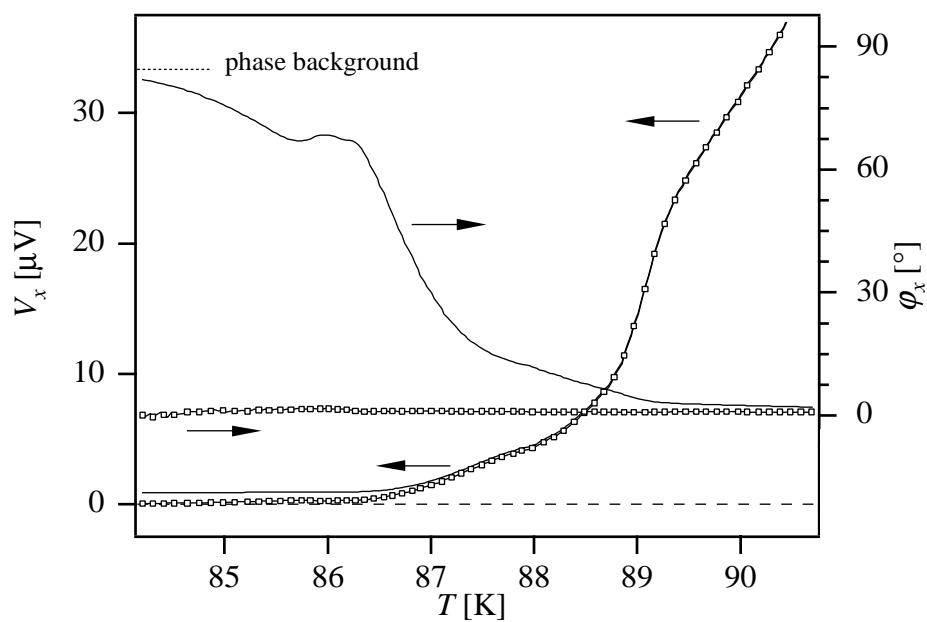

**Figure IV-7 :** Modulus $V_x$ and phase $\varphi_x$ of the longitudinal voltage at a magnetic field $B = 2$ T exactly aligned with the *c*-axis, and for a current density of $j = 150$ A/cm$^2$ dc + 50 A/cm$^2$ pp (peak to peak). Solid lines : uncorrected, raw data. Lines with symbols : result after the inductive background complex subtraction.

We see that this processing step mainly affects the results in the low level part of the signal, which is precisely the region of interest for this study. Moreover, as we shall see further on, this subtraction processing has even more dramatic effects in the case of measurements as a function of the current direction (Fig. IV-10).

Measurements are performed under the same conditions with both magnetic field polarities (by swapping the magnet current leads), and the Hall resistivity is identified with the antisymmetric part of the transverse resistivity. An example of such a pair of measurements is given in Fig. IV-8. We see that the difference between the transverse voltages $V_y$ for both field polarities (curves $a$ and $b$) is large around $T = 90$ K, roughly corresponding to the negative maximum of the Hall resistivity $\rho_{xy}$. At lower temperatures, this difference becomes very small. It can be realized from this figure to which extent it is fundamental to keep the noise as low as possible. We recall that the objective of the present work is to investigate the Hall effect in the vortex solid phase. Since the vortex phase transition is around $T_m = 88.2$ K for this magnetic field, we see that the Hall signal (the half difference between curves $a$ and $b$) becomes only a small fraction of the total measured voltage. The symmetric part of the transverse voltage $V_y$ (the average between curves $a$ and $b$) reflects a guided motion of vortices along twin boundaries, and is the subject of the next chapter. Note that, even though this operation has almost no effect on the results, the longitudinal resistivity $\rho_{xx}$ is rigorously computed from the average between the longitudinal voltages $V_x$ for both field polarities.

Since at high currents electromagnetic forces on the wires may move them and induce parasitic voltages, all the measurements (for each field polarity) are repeated after a current direction reversal, and the average of the data for both current polarities is taken. The importance of this operation is best seen on the Hall conductivity $\sigma_{xy}$ (Fig. IV-9), which is the most sensitive to noise and other undesired contributions to the signal. The strongest influence on the data occurs at low magnetic fields (or at low temperatures for a measurement as a function of the temperature), *i.e.* is signifi-cantly below the vortex phase transition, where the voltages become very low, eventually vanishing.

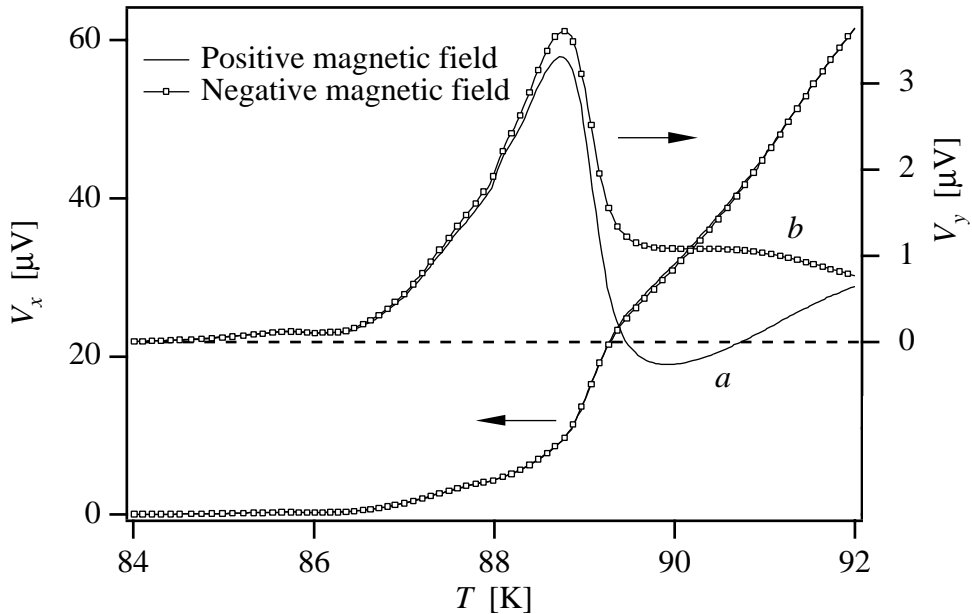

**Figure IV-8:** Longitudinal and transverse voltages (after inductive background subtraction, see Fig. IV-7) for the two polarities of the magnetic field, $B = \pm 2$ T. The field is exactly aligned with the $c$-axis, and the current density is $j = 150$ A/cm$^2$ dc + 50 A/cm$^2$ pp. The Hall effect is then obtained from the difference between the two transverse voltages $V_y$, curves $a$ and $b$.

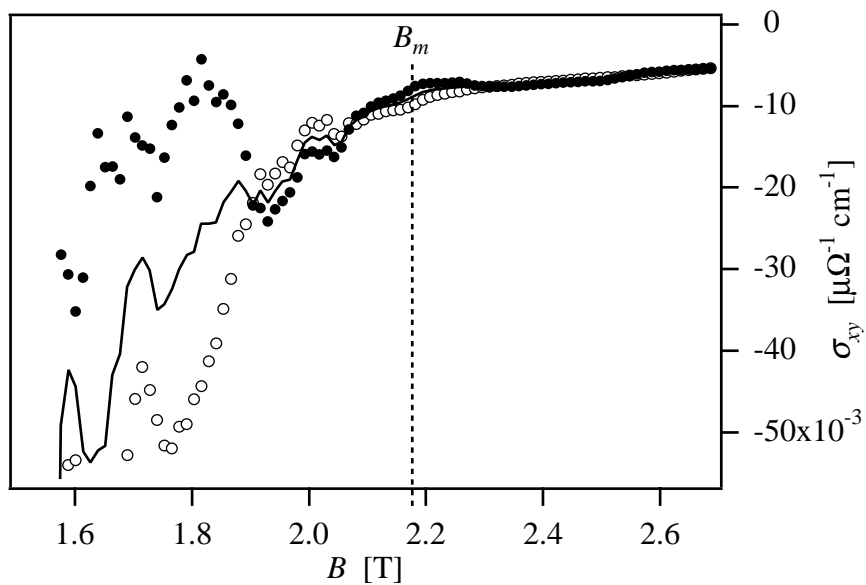

**Figure IV-9 :** Open and solid symbols : hall conductivity $\sigma_{xy}$ as a function of the magnetic field for the two polarities of the current, $j = \pm 150$ A/cm² dc + 50 A/cm² pp. The temperature is $T = 88$ K and the magnetic field is inclined at 4 ° from the *c*-axis. Solid line : hall conductivity $\sigma_{xy}$ averaged between the two current polarities.

In summary, the procedure for Hall measurements is the following. First, temperature or field sweeps are performed for both current *and* field polarities (hence four measurements in total), making sure that the sweeping parameter was reaching low enough values to go deep into the vortex solid, where there is only the inductive contribution in the signal. The inductive modulus and phase backgrounds of each of the longitudinal and transverse signals are then subtracted (following complex algebra) to the recorded raw modulus and phase of the corresponding voltages. The longitudinal resistivity is finally obtained from the average of the four longitudinal voltages, whereas the Hall resistivity is extracted from the antisymmetric part (half difference) with respect to the magnetic field, averaged between the two current polarities.

*b) Orientable current measurements*

Thanks to the numerous available contacts on the sample surface, we can also measure the direction of the electric field as a function of the direction of the current density in the sample. Using the specific source described in paragraph 1.4, $I_x$ is injected between contacts 1 and 2, $I_y$ between 5 and 6, and the corresponding electric field components were deduced from $V_x = V_{3-4}$ and $V_y = V_{8-9}$.

All the measurements are done at a fixed temperature and magnetic field, and both components of the electric field are recorded as the current direction is changed through a complete 360 ° turn. In this case, the inductive background is measured around such a whole 360 ° current turn at low magnetic field and temperature, deep in the vortex solid where vortices are strongly pinned. Its angle dependent complex value is then subtracted from the raw measurements, similarly to the Hall data processing described above. We can see in Fig. IV-10 that this operation is especially important to obtain the true angular dependence of the resistivity in the vortex solid, where the resistive part becomes smaller than the inductive background : the raw data for the desired temperature (87.5 K) and the low temperature inductive background (measured at $T = 80$ K) are extremely similar, the difference essentially lying in the phase of the total signal (solid line). After the complex subtraction, however, a significant resistive component is revealed (solid symbols).

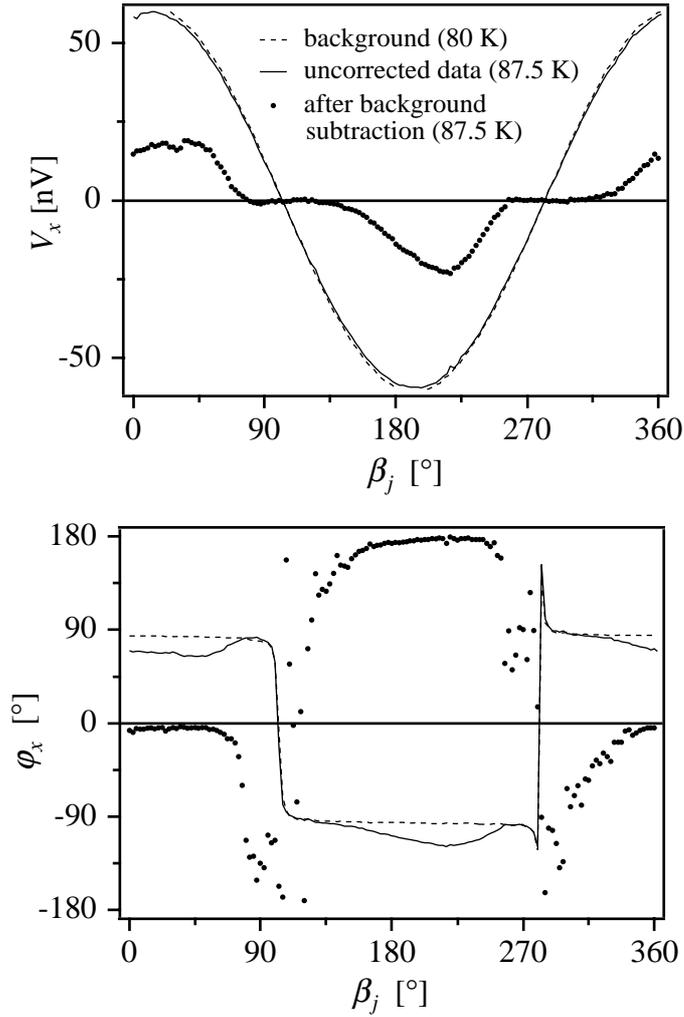

**Figure IV-10 :** Modulus $V_x$ and phase $\varphi_x$ of the longitudinal voltage at a magnetic field $B = 2$ T exactly aligned with the *c*-axis, and for a current density of $j = 50$ A/cm² dc + 12.5 A/cm² pp. Solid lines : uncorrected, raw data. Solid symbols : result after the complex subtraction of the inductive background measured at low temperature (dashed line).

## 2. Sample

The sample used for this study is a $YBa_2Cu_3O_{7-\delta}$ single crystal which has been grown in a $BaZrO_3$ crucible (by A. Erb *et al.* in Geneva), a method known to produce crystals of very high purity [140]. Its size is about $870 \times 400 \times 24$ µm. Because in $YBa_2Cu_3O_{7-\delta}$ some of the oxygen atoms are anisotropically distributed along chains, there is a small in-plane anisotropy, the *b* axis being very slightly greater than *a*. The lattice structure is therefore orthorhombic, and internal stresses, appearing when samples (grown at high temperatures) are cooled, usually lead to microtwinning of the crystal. The twins are adjacent domains with *a* and *b* successively exchanged, and the twin boundaries are oriented along the (110) directions. Among the consequences of the *a-b* anisotropy is the different optical reflectivity of twins of opposite orientation, such that the twin structure can be observed with polarized light. As shown in Fig. IV-11, our crystal is indeed twinned, with one dominant twin plane family. Rather large monodomains are also present.

As we shall see, these twin planes can have under some conditions a rather strong influence on the magnetic field distribution inside the sample. It is then important to check to what extent these magnetic field inhomogeneities might in turn influence the resistivity measurements. For this we have used magnetooptic observations, a very useful tool [141] based on the Faraday effect, namely the rotation of the light polarization in some adequate medium under the influence of a magnetic field. More precisely, the total angle of the polarization rotation is proportional to the magnitude of the field component *parallel* to the light propagation direction. The idea is then to chose a material showing a very strong Faraday effect, in this case a ferrimagnetic doped iron garnet (with a Faraday rotation of almost $10°$ for about 1 kG, the saturation field of the material). This "magnetooptic indicator" (actually a film deposited on a reflecting substrate) is then placed on top of the sample, and observed with a polarized light microscope. The areas of high magnetic field will then be brighter than the zones with low magnetic field.

This has been done for our sample at a temperature close to the conditions of our subsequent resistivity measurements, in a perpendicular applied field (Fig. IV-12). At low fields ($B = 90$ G), the effect of the twins is easily observable, resulting in a very inhomogeneous field distribution clearly correlated to the twin structure observable in Fig. IV-11 (other magnetooptic observations of twinned samples can also be found in [142]). Note that the inclined "rhombs" at the lower edge of the sample visible in Fig. IV-12a are artifacts related to magnetic domains in the magnetooptic indicator itself, and are not directly related to the field distribution in the sample. When the field is increased to $B = 260$ G (Fig. IV-12b), the flux inhomogeneities are obviously smoothed out. Since the magnetic fields used for the resistivity measurements are in the Tesla range, and usually the temperature is even a few degrees higher (which strongly contributes to a homogeneous flux distribution, due to rapidly vanishing critical currents in this temperature range), it can be assumed that the twin effects on the field distribution are negligible in the frame of the resistivity measurements.

More generally, it is noteworthy to specify that the same sample has already been used for other types of extensive magnetooptic observations [143] which have proven its very high quality. However, even though we have just seen that the flux distribution is almost unperturbed by twins at high enough fields, the question of whether a small[1] additional applied current would lead to a vortex motion unperturbed by twin boundaries is another issue, and will be experimentally addressed in the next chapter.

## 2.1 Contacts

In order to perform resistivity and Hall measurements, we obviously need electric contacts on the sample surface. This step has been a major difficulty of this work. Many attempts have been made on different similar single crystals. The different techniques tried were for example a direct "soldering" of gold wires on the sample surface using a conducting silver epoxy paste, or a silver paint.

---

1. Small with respect to the magnetization screening currents.

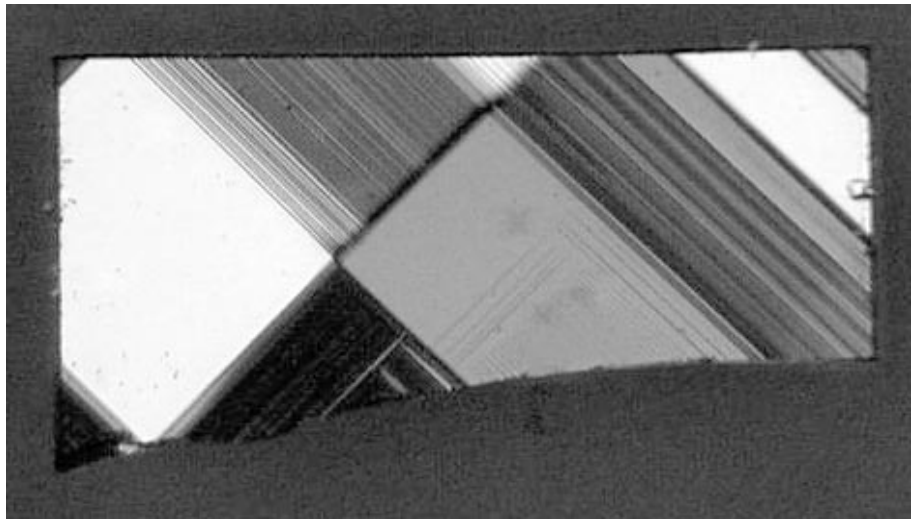

**Figure IV-11 :** Image of one sample surface in polarized light, revealing the twins structure (the sample length is 870 µm, see Fig. IV-13 for more detailed dimensions). One orientation of twin planes is clearly dominant (from top left to bottom right).

Gold contact pads have also been deposited through a mask by sputtering, and gold wires were then again attached to these contact pads with silver epoxy. In all these cases, the resulting contacts were of too high resistivity (above 1 Ω, sometimes up to about 100 Ω), inducing a very high noise in voltage measurements (in the worst case even masking the superconducting resistive transition).

There are probably many reasons for this high contact resistivity. First of all, the quite small dimensions of the used samples imply the need to make rather small contacts, provided we want to have many of them (the chosen pattern includes nine contacts, see Fig. IV-13). Some of them have then a very small contact surface with the sample material. This small size makes it also difficult to manipulate masks and wires to actually build these contacts.

Another source of problems was probably the quality of the surface for some of these samples. First of all, a surface layer with a possibly different oxygen content might influence its conductivity. But also any type of other substance polluting the surface might play an insulating role. This has probably been the case, for example, when the solvent used to glue the sample onto the sapphire was not properly removed before evaporation, leaving traces of the non-conducting glue on the sample top surface.

A heat treatment is also definitely necessary after the contacts have been made, to let the gold diffuse into the first layers of the crystal, in order to reduce the contact resistivity as much as possible. The adequate parameters for this annealing have to be carefully selected.

Finally, the successful attempt has been done according to the following procedure. After the sample had been glued onto a glass plate, its top surface was thoroughly cleaned under a microscope with the help of a razor blade. Then gold contacts have been evaporated through a mask in a high vacuum evaporation chamber. More precisely, two successive evaporations with different masks were necessary to obtain the desired contact pattern. The masks were hand made by cutting long and very narrow bands of aluminum foil (down to about 0.1 mm width), or simply using 50 µm diameter gold wires to make separations between the closest contacts. All these elements were placed in the correct position, then anchored with glue beside the sample, on the glass surface. The evaporated gold layer is about 50 nm thick.

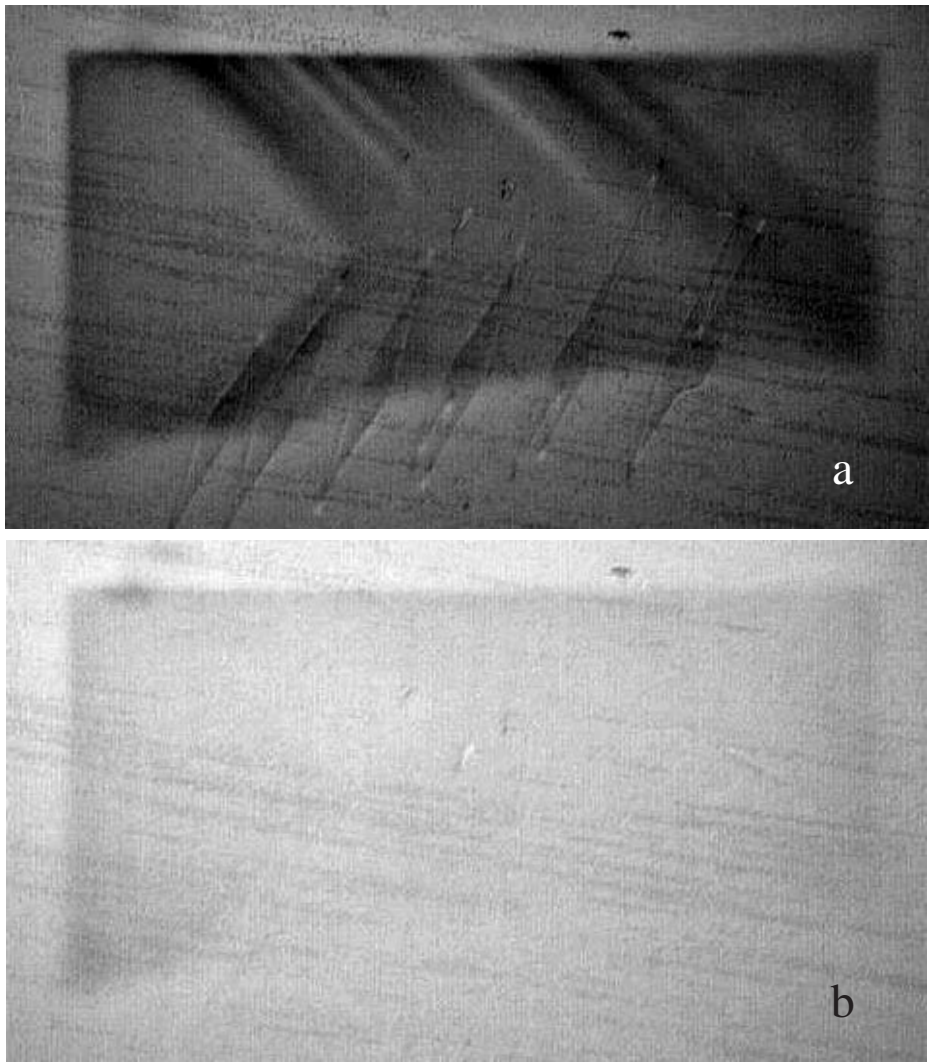

**Figure IV-12 :** Magnetooptical imaging of the perpendicular magnetic flux penetration into the sample at 81 K. a) Applied field 90 G. The influence of twins is clearly visible. b) Applied field 260 G. The contrast is strongly reduced, indicating a much more homogeneous flux distribution.

After the two evaporations, the sample (removed from its glass support) has been annealed for 46 hours at a temperature of $405 \pm 3$ °C, in a continuous oxygen flow at ambient pressure. It was then glued on the sapphire of the sample holder, making sure the glue layer is as thin as possible, in order to have the sample as parallel as possible to the sapphire surface, and also to have the best thermal contact between them.

Finally, gold wires of 12 μm diameter have been attached to the contact pads with conducting silver epoxy (Du Pont 4929). The other end of these wires were soldered on the connection plate (Fig. IV-2) with indium.

The resulting contacts have a very low resistivity ($< 0.5$ Ω), and the achieved noise level in the measured voltage is as low as 0.1 nV to 1 nV (depending on the considered contacts). Unfortunately, the adhesion force of the silver epoxy is rather weak, and after some thermal cycling of the sample, some wires became lose from the sample surface a couple of times. However, each time new wires were placed on the same contact pads following the same procedure, and the resulting contact quality was always fully recovered.

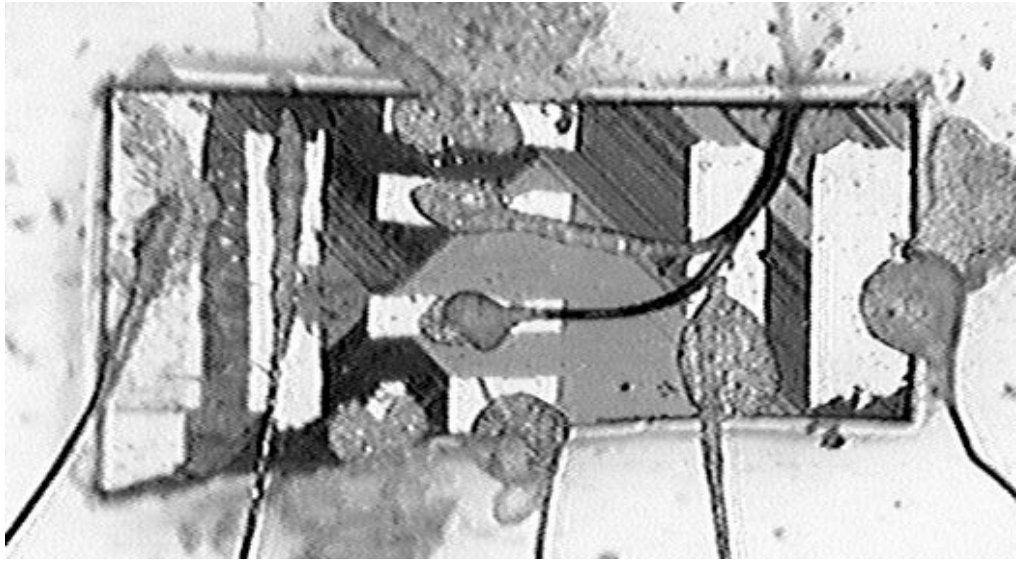

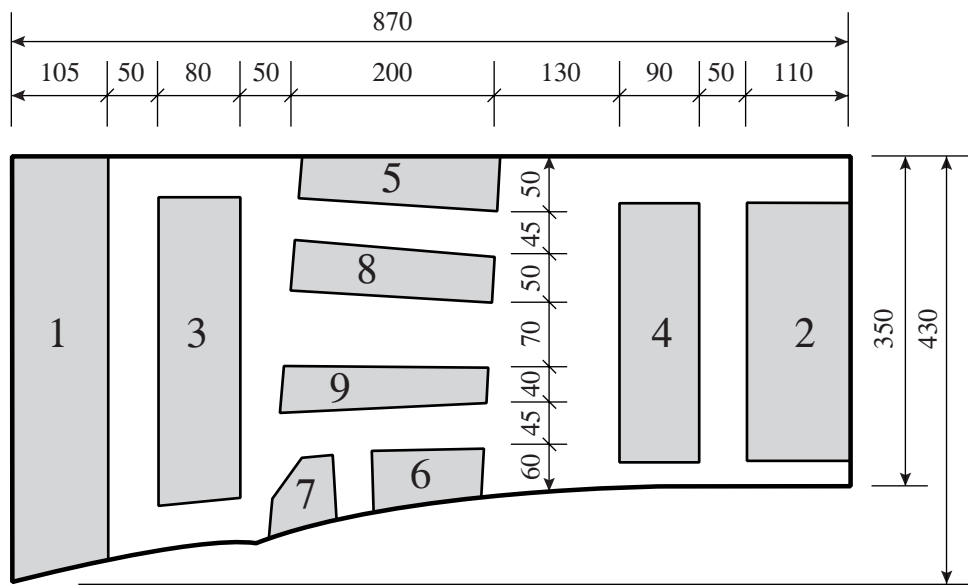

**Figure IV-13 :** Top : Picture of the sample surface after deposition of gold contact pads and bonding of the gold wires. Bottom : schematic representation of the contacts pattern, with their characteristic dimensions in microns. The sample thickness is 24 µm.

Note that, aside from the contacts quality, we believe that another factor is determinant in the very low noise level : the antivibration system. A risk of noise generation by electromagnetic pick up indeed exists, as a consequence of vibrations of the loose gold microwires around the sample, and to some extent of the untwisted wires along the sample rod as well. Thanks to the suppression of vibrations in all of the cryostat parts (including the sample and its gold wires, the sample rod, and, most importantly, the passive signal transformers), the only parts of the wiring subject to pick up noise are the coaxial cables in which the signal is already amplified by a factor of 1000, reducing the relative importance of such noise.

## 2.2 Preliminary measurements

*a) Critical temperature*

Since the sample had to be annealed in oxygen atmosphere in order to get contacts of sufficient quality (see paragraph 2.1), it is legitimate to wonder whether this process has significantly altered the oxygen concentration. To answer this question, we can experimentally estimate the critical temperature $T_c$ of the sample by measuring the resistive transition at zero magnetic field (Fig. IV-14).

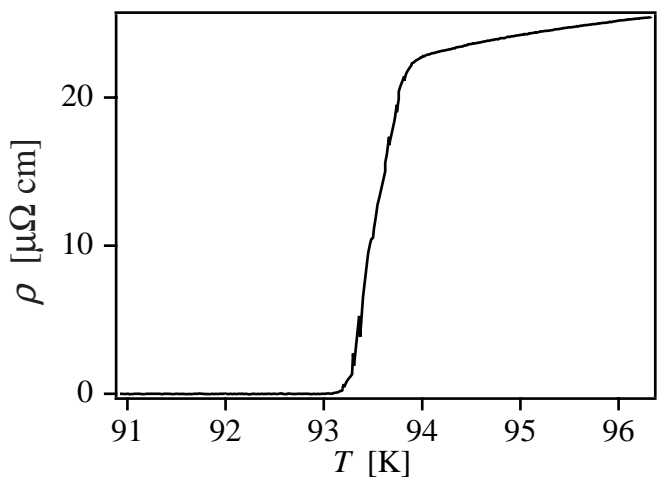

**Figure IV-14 :** Measurement of the resistive transition at $H \approx 0$, showing that 93 K $< T_c <$ 94 K. The current density is 0.2 A/cm$^2$ rms and the temperature is decreased at a rate of 0.1 K/min.

At first sight, we see that $T_c \approx 93.5$ K with a transition width of $\Delta T_c \approx 0.5$ K. The temperature sweep rate is slow enough so that there is no measurable difference between ramping up or down. Note that the transition width is rather large for a high quality crystal. This is however an artifact unrelated to the sample itself. Unfortunately, since the magnet present in the cryostat is superconducting, it is indeed itself subject to vortex pinning from the first time it is energized until it is raised above its own critical temperature. In other words, a remanent field is intrinsically present after a first application of the magnetic field, and can only be suppressed by letting the liquid helium bath evaporate, warming up the whole system. Knowing that the bonding of the wires onto the sample surface is extremely fragile, and could break by simple thermal cycling to and from room temperature, it was chosen not to let the magnet warm up. Therefore, the only solution is to reduce the remanent field as much as possible by successive applications of opposite field directions of decreasing magnitudes. It is estimated that the usual remanence of about 100 G can be reduced by about 90 % with this process. However, in no way can it be affirmed that the field is strictly zero during this measurement, and hence the result cannot be used to reliably determine the transition width $\Delta T_c$ for example. The value of $T_c$ itself has to be considered as an indication of limited accuracy ($\pm$ 0.2 K). In any case, we conclude from this estimation of $T_c$ that the sample is practically at optimal doping, so that the oxygen content has not been affected by the thermal treatment of the contacts.

*b) Field alignment*

As we have mentioned at the beginning of this chapter, the sample can be oriented with respect to the magnetic field produced by the superconducting magnet. The question is then to know the absolute value of the angle between the field and the sample *c* axis. For this, we use the fact that in the vortex liquid, the resistivity is minimum when field is precisely aligned to the *c* axis (because of the influence of twin boundaries on the vortex dynamics [144]). We therefore measure the resistivity for different angles at constant appropriate magnetic field and temperature. Typical results are shown in Fig. IV-15.

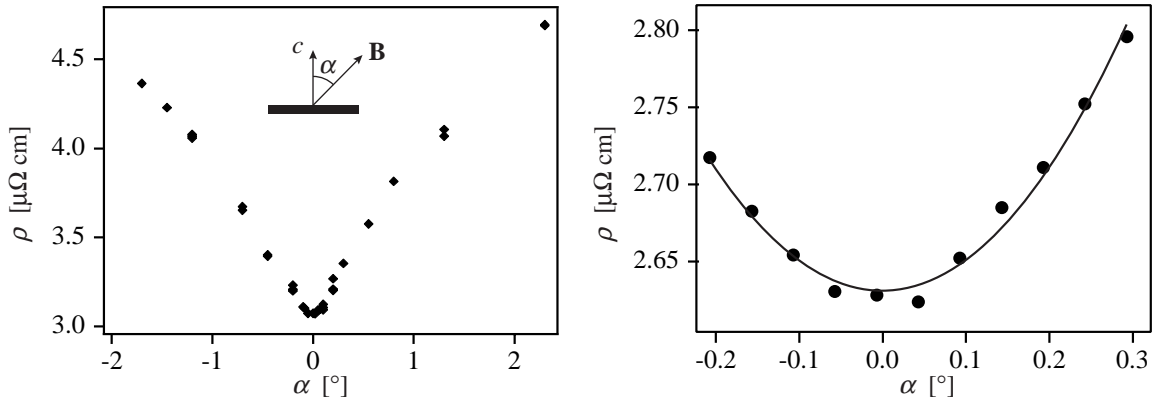

**Figure IV-15 :** Angular dependence of the resistivity used to align the sample with the magnetic field. Left : 89 K, 2 T and 7 A/cm² rms. Right : more precise set of data taken at 85.5 K, 4 T and 10 A/cm² rms; the solid line is a parabolic fit. $\alpha = 0$ is identified to the minimum of the parabola.

The main difficulty for this operation is the stability of temperature, since even the small oscillations of about 0.02 K induce resistivity variations larger than the increase corresponding to a rotation from 0° to ±0.1°. Averages over times much longer than the temperature oscillations were necessary to localize the minimum with an accuracy better than 0.05° (with the help of a parabolic fit on the small angle data, see Fig. IV-15 left). This operation had to be performed every time the sample was unmounted from the sample rod (to fix loose wires, for example) and then reinserted into the cryostat.

*c) Vortex lattice melting*

A good test of the sample quality is to see if the vortex liquid to solid phase transition is observable by resistivity measurements. If it is not the case, this means that the irreversibility line (the onset of non-zero critical currents) is above the melting transition line in the (*H,T*) phase diagram, such that the vortex phase-dependent dynamics is masked by pinning effects. The measurements at low current densities shown in Fig. IV-16 clearly reveal the presence of a rather sharp vortex phase transition, the signature of which is the steep drop to zero of the resistivity when temperature or magnetic field are reduced. In the solid phase (at low temperature and low field), the vortex dynamics can still be probed with the help of larger currents, overcoming the critical current (Fig. IV-16 right).

A still open question is to determine whether this transition is indeed a true first order lattice to liquid phase transition, or if the solid phase is rather a disordered glass, the transition being of second order (see chapter I, page 5). Specific heat and magnetization measurements have shown that in clean enough samples, a first order transition occurs. A critical point is often seen to limit this regime, yielding a second order transition at high fields [17]. However, according to the latest developments on this topic, the position of this critical point is sample dependent, and lies apparently above 50 T in the cleanest crystals [145].

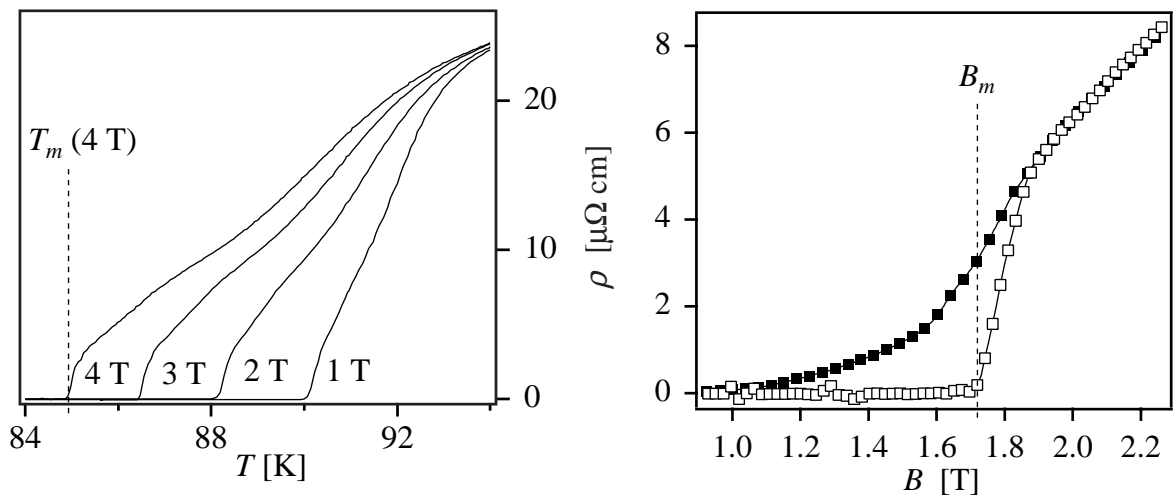

**Figure IV-16 :** Left : Resistive transition associated to the vortex lattice melting, as a function of temperature and for different applied magnetic fields. The current density is 1 A/cm² rms, and the magnetic field angle is $\alpha = 3°$. Right : Equivalent measurement as a function of magnetic field at T = 89 K, same current (open symbols). With a larger current (150 A/cm² dc + 5 A/cm² ac), the vortex solid can be set into motion (solid symbols).

On the other hand, the low field part of the phase diagram sometimes also seems to be influenced by some defects that suppress the first order nature of the transition [17]. In this case, the defects are possibly the twin boundaries, since this lower critical point on the phase transition line is usually not seen in naturally untwinned crystals [146]. Aside from this role of the twins in the low field data scattering, we believe that there is also a relation with the use of very large samples. The peaks in calorimetric measurements are indeed extremely close to the experimental resolution, and their observation is improved by the selection of rather massive samples, very thick (usually more than 200 μm). Unfortunately, it is well known that these thick samples feature non-superconducting intergrowths or simply interplanar cracks which, even if they are not apparent on any sample lateral surface, still dramatically influence magnetic flux penetration, as revealed by magnetooptic observations [147].

For our sample, we have investigated the role of twin boundaries by measuring the resistive step associated to the vortex phase transition as a function of the angle $\alpha$ between the magnetic field and the twin planes. The curves obtained at different angles are given in Fig. IV-17. We see that the resistivity is strongly reduced at low angles in the vortex liquid. This is precisely what we have used to align the sample with the magnetic field (see Fig. IV-15 above). Note that we will come back to this reduction of the longitudinal resistivity in the next chapter. Above the temperature $T_{TB}$, the curves at moderate angles are superimposed, indicating that the twins have no significant influence anymore. The curve for larger angles (for example for $\alpha = 20°$) are clearly lying lower, because the magnetic field component perpendicular to the current starts to decrease as the field is further rotated away from the c-axis.

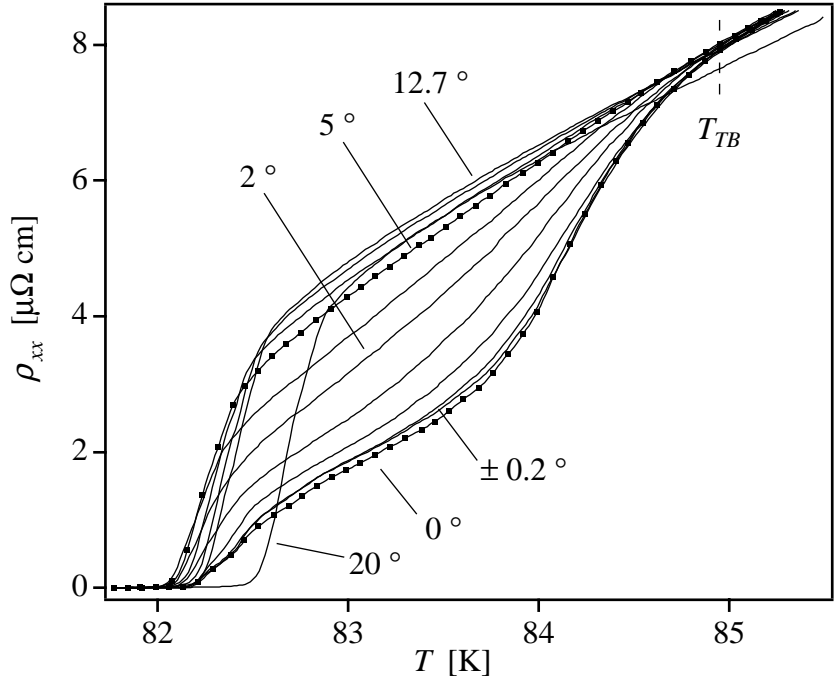

**Figure IV-17 :** Resistive transition as a function of the temperature, at $B = 6$ T and $j = 1$ A/cm$^2$ rms, for different angles $\alpha$ between the magnetic field and the twin planes ($\alpha = 0°, \pm 0.2°, 0.5°, 1°, 2°, 3°, 5°, 7°, 10°, 12.7°, 20°$). $T_{TB}$ marks approximately the temperature above which the twins do not affect the resistivity anymore.

We have then estimated the temperature of the resistivity onset for each of the angles $a$. For this, an intersect of a linear fit of the resistivity step in the neighborhood of $T = 82.3$ K with the axis of zero resistivity was used. A plot showing this onset temperature as a function of the angle $\alpha$ for two different magnetic field intensities is given in Fig. IV-18. This figure should be compared to Fig. I-3 on page 7. It is indeed very similar to the recent results from Grigera *et al.* on an analogous YBa$_2$Cu$_3$O$_{7-\delta}$ single crystal [25], who have interpreted the peak of $T_{onset}$ at low angles as the signature of a Bose-glass transition due to the twin boundaries. Following their suggestion, we have fitted this part of the data with the expression

$$T_{onset}(\alpha) = T_{onset,0} \left(1 - \frac{\sin \alpha}{x_c}\right)^{1/3\nu},$$

where $T_{onset,0}$, $x_c$ and $\nu$ are free parameters. The critical exponent $\nu$ is then $\nu = 1.23$ for $B = 2$ T and $\nu = 1.09$ for $B = 6$ T, in excellent agreement with the value of $\nu = 1 \pm 0.2$ found by Grigera *et al.* at $B = 6$ T. At larger angles, the behavior is the one expected from simple anisotropic scaling [148,6], as revealed by the parabolic fit in Fig. IV-18. The transition between the two different regimes happens at an angle $\alpha^*$ of the order of 2° to 3°, again consistent with the crossover of $\alpha^* \approx 2.1°$ of Ref. [25].

In conclusion, since we only have access to resistivity measurements to characterize our sample, and have no "thermodynamic" data like specific heat or magnetization, we cannot affirm rigorously what is the order of the vortex phase transition we observe. However, considering that the angular dependence is just the same as in Ref. [25], that our sample is much thinner than the crystals usually used for calorimetric measurements (see discussion above), has a very low normal state resistivity (less than 25 µΩ cm, meaning that it is extremely clean), and is not heavily twinned (with large untwinned domains, see Fig. IV-11), we have strong reasons to believe that the solid phase is a Bose-glass for $\alpha < \alpha^*$, and an ordered vortex lattice (Bragg-glass) for $\alpha > \alpha^*$, separated from the vortex liquid by a first order phase transition. In any case, the issue of the exact nature of the solid phase for $\alpha > \alpha^*$ is not determinant for our study. The most important point is to be able to precisely localize this transition in the magnetic field and temperature, which is clearly the case as can be seen from Fig. IV-16 and IV-17 for example.

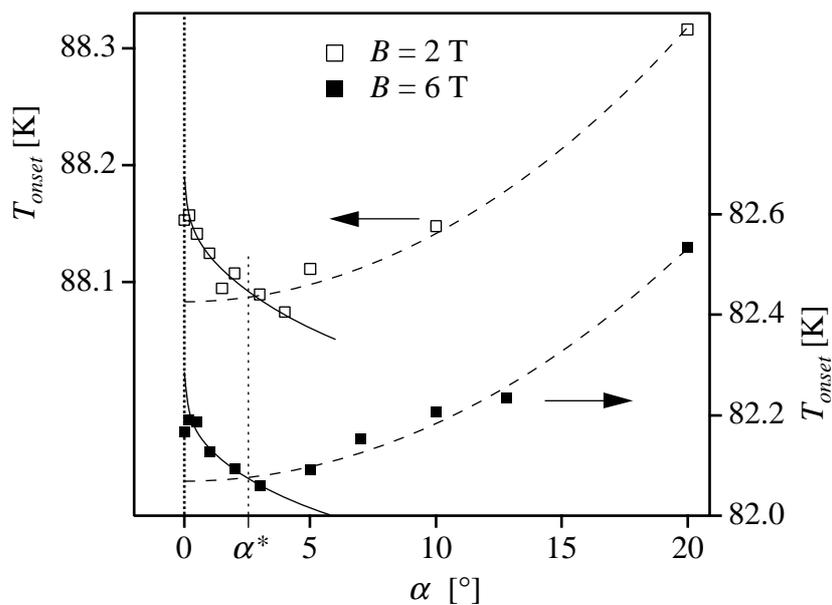

**Figure IV-18 :** Onset temperature of the resistive transition as a function of the magnetic field orientation $\alpha$ with respect to the twin boundaries. Data for $B = 6$ T (solid symbols) correspond to Fig. IV-17. Similar data for $B = 2$ T are also given (open symbols). Dashed lines are parabolic fits representing the anisotropic scaling, solid lines are fits corresponding to a Bose-glass transition (see text).

# **IV**

# CHAPTER V  *RESULTS : GUIDED MOTION*

## 1. General idea

### 1.1 Even Hall effect

Since the early work on the Hall effect in the mixed state of type II superconductors, it has been observed that the transverse resistivity has not only an antisymmetric part (with respect to the magnetic field), namely the Hall component, but also usually a symmetric part (sometimes also called "even Hall effect"), indicating that the vortices actually have a direction of motion influenced by material inhomogeneities. In low $T_c$ materials for example, this guided motion was seen to depend on the details of sample processing, which induces extended defects in preferential directions.

When performing the first preliminary measurements mentioned in the preceding chapter, we have indeed found out that a large transverse voltage is also present in our sample (an example is given in Fig. IV-8). Of course a part of this signal is antisymmetric and represents the Hall effect, but after averaging the data for both field polarities, an important symmetric part remains. Actually, this symmetric part can even be of the same order of magnitude as the longitudinal resistivity, whereas the Hall resistivity is usually one to two orders of magnitude smaller.

This shows that vortices undergo a guided motion, such that the electric field is at a rather large angle (much larger than the Hall angle) from the current direction. The question now is to know more precisely in which direction do the vortices exactly flow depending on the current orientation, what happens as a function of the vortex phase (solid or liquid), and what is the dependence on the angle $\alpha$ (the direction of the magnetic field with respect to the twin planes, see page 66). Thanks to the contact pattern on our sample, we have the opportunity to actually determine the direction of the electric field in response to a current density of various orientations. This is the objective of the present chapter.

### 1.2 Twins and vortex dynamics

It is evident that in twinned samples, twin planes should play an important role in this guided motion. The understanding of the effect of twin boundaries on the vortex dynamics have been the subject of extensive work. On the theoretical side, we can find, for example, a Fokker-Planck equation approach [149], or various numerical simulations, like molecular dynamics simulations [150] or algorithms based on time dependent Ginzburg-Landau equations [151].

From the experimental point of view, the role of twins in $YBa_2Cu_3O_{7-\delta}$ has been discussed in details on the basis of magnetooptic observations [142], as already cited in section 2 of the preceding chapter. But resistivity measurements have also brought much information to the understanding of this topic. The first measurements in unidirectionally twinned samples [152,24] usually focused on the magnetic field orientation (angle $\alpha$) dependence, and were performed with a single current orientation (usually at 0°, 45° or 90° from the twin planes), while recording only one voltage component along the current direction. Later, an important work has been the direct comparison of different current orientations with respect to the twins (0°, 45° *and* 90° under the same conditions), however in different similar samples (with one fixed current direction in each sample) and still only measuring the longitudinal voltage [153]. Much more recently, two different perpendicular current directions (along and across the twin planes) were used in the same film sample, and both longitudinal and transverse voltage were measured [139], but only the usual antisymmetric Hall data were reported, not the even Hall effect reflecting guided motion.

A direct study of the guided vortex motion in unidirectionally twinned samples has been accomplished for the first time only very recently, by Morgoon *et al.* [48]. In this work, three different samples where prepared, each with a different angle between the applied current and the twin planes (+90°, +45° and −30°), and the transverse voltage has been recorded to extract the even Hall effect. The authors can then track the transverse behavior in terms of guided motion. Unfortunately, very few data are given in this short report. We propose here a new, more versatile method and present new results in the continuation of this work. This method is based on the reconstruction of the total electric field, both in amplitude *and* direction, as a function of the direction of the current, which is *continuously* adjustable with respect to the twin planes.

## 2. Results

### 2.1 Detailed procedure

As briefly described in the preceding chapter, a specific double current source is used to feed two orthogonal currents $I_x$ and $I_y$ into the sample, in a ratio determining the orientation $\beta_j$ of the total current density (Fig. IV-4).

For each current source ($I_x$ and $I_y$), a 510 Ω resistor was placed on both current leads to check by the voltage drop across these resistors that all the current leaving one of the sources was indeed only returning to that same source. In other words, it was checked that no cross-feed happens between the preamplifiers providing the currents $I_x$ and $I_y$ (page 56).

The principle of the measurements reported here is to determine the electric field modulus $E$ and its orientation $\beta_E$ (Fig. V-1) from two orthogonal voltage components $V_x = V_{3-4}$ and $V_y = V_{8-9}$ (see

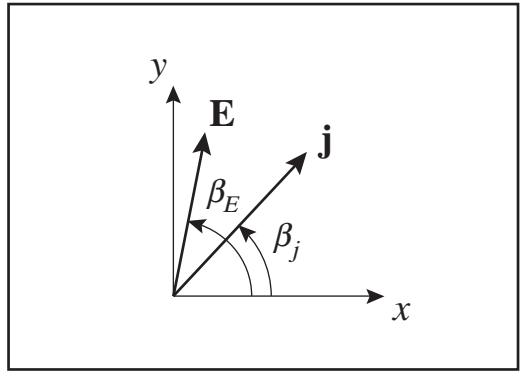

**Figure V-1 :** Definition of the two angles used in this chapter : $\beta_j$ (related to the orientation of the current density in the sample plane $x$-$y$) and $\beta_E$ (related to the electric field orientation in the same $x$-$y$ reference frame).

Fig. IV-13 for contact numbering) as $\beta_j$ goes over a complete turn[1]. After the subtraction of the inductive background (as already described in paragraph IV.1.5), $V_x$ and $V_y$ are converted into electric field units, taking into account the distances between contacts.

Two different series of such measurements have been done. The first one is performed at a constant temperature of 89 K, and angular dependences are recorded for many different magnetic field values, from the vortex liquid state down to the solid phase. For this set of data, the total current density is 60 A/cm$^2$ dc + 10 A/cm$^2$ ac peak-to-peak (pp), and the inductive background is obtained by a measurement at 0.8 T. The other series is measured at 2 T, for several different temperatures (also crossing the vortex phase transition), with a current density of 100 A/cm$^2$ dc + 25 A/cm$^2$ pp. The inductive background is then measured at 80 K. In each series, all the measurements are repeated under the same conditions for two different magnetic field orientations : once with $\alpha = 0°$ (field exactly aligned with the twin planes) and once for $\alpha = 4°$.

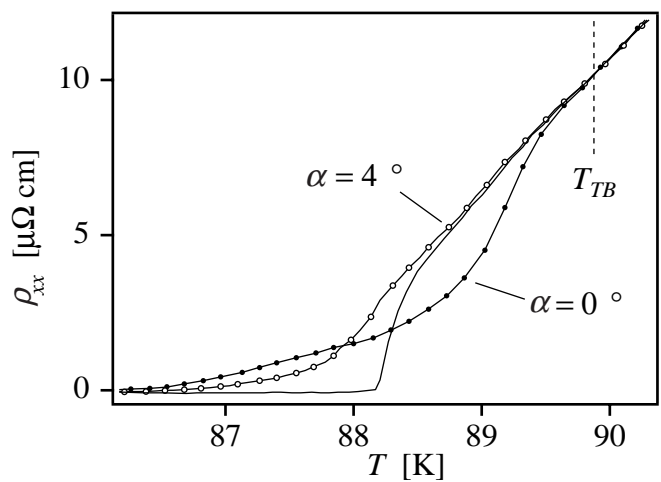

**Figure V-2 :** Longitudinal resistivity at 2 T for two different angles $\alpha$ between the magnetic field and the twin planes. The data represented with a solid line without markers is measured at low current density (about 3 A/cm$^2$ rms) to localize the vortex phase transition ($T_m \approx 88.2$ K). The other data are measured at a higher current, namely 150 A/cm$^2$ dc + 50 A/cm$^2$ pp. The twin boundaries influence the resistivity up to approximately $T_{TB} = 89.9$ K.

---

1. Actually, the measurements are performed over 380 ° for $\beta_j$, the overlapping of data reflecting the reproducibility after a whole turn.

Before presenting the actual results for angular dependences, we first show what the effects of these two field orientations on the longitudinal resistivity $\rho_{xx}$ are, both in the vortex liquid and solid phases. Fig. V-2 shows comparative measurements for $\alpha = 0°$ and $\alpha = 4°$. It appears that when the field is right along the twin planes, the resistivity is reduced in the liquid phase, but slightly increased in the solid phase [144]. Note that some comments will be given on this point at the end of this chapter, in the light of the results presented below.

The difference between the two curves, obviously induced by the twin boundaries, is observable up to the temperature $T_{TB} \approx 89.9$ K lying significantly above the vortex phase transition temperature $T_m \approx 88.2$ K. Of course, the same qualitative picture can be obtained from measurements as a function of magnetic field at constant temperature. For a temperature of 89 K, the two corresponding magnetic field values are $B_m \approx 1.65$ T and $B_{TB} \approx 2.5$ T.

### 2.2 Measurements at constant temperature

*a) Data for $\alpha = 0°$*

We first present the results for the magnetic field parallel to the twin planes, namely for $\alpha = 0°$. In Fig. V-3 are shown the data for magnetic fields ranging from 3 T to 1 T. Due to the very fast decrease of the resistivity below the melting field, the curves have to be traced on two different plots, with linear scale at high fields, but a logarithmic scale for the low field data. These polar plots are representations of the modulus of the total electric field $E$ both as a function of the direction of the current and the direction of the electric field itself, namely the $E(\beta_j)$ and $E(\beta_E)$ curves.

The first general observation that can be made from these two first plots is that there are clearly two axial symmetries, with the axes roughly in the (110) (or $45°-225°$) and the $(1\bar{1}0)$ (or $135°-315°$) directions, obviously corresponding to the twin plane orientations. However, we observe only a twofold symmetry in the data, meaning that only one of the two possible twin plane families is dominating, consistent with the magnetooptic observations of the preceding chapter (section 2). More precisely, we see that at low enough magnetic fields, the electric field is larger (meaning a faster vortex motion) when the current is oriented in the (110) direction. Since it was observed in Fig. IV-11 and IV-12 (page 62 and page 63, respectively) that the dominating twin family is along $(1\bar{1}0)$, *the data show that the twin planes act as a barrier for vortex motion across them, but facilitate the motion along the planes*, as already reported in the literature [142,153] (remember that the driving Lorentz force is perpendicular to the current density, and that from Josephson's relation the electric field is also perpendicular to the vortex motion [32]).

When looking at the low field data, it becomes apparent that the symmetry axes rotates away from the exact (110) and $(1\bar{1}0)$ directions. This is most probably the consequence of the combined effect of contact misalignments (the pairs of contacts are not exactly along the *x* and *y* axes) and variations in the current distributions, due to the increasing non-linearity in the voltage-current characteristics in the vortex solid. Therefore, we believe that the physical preferential direction is still completely determined by the twin planes in this regime, and only the measured angles are biased by these extrinsic processes.

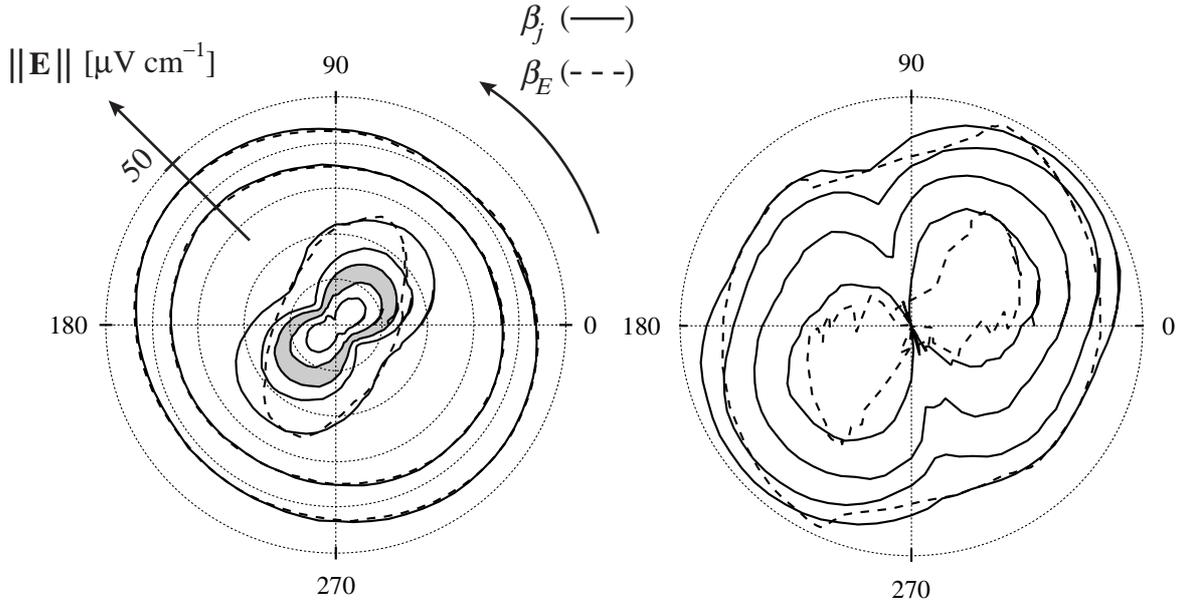

**Figure V-3 :** Polar plots of the electric field as the current is rotated in the sample plane. The magnetic field is parallel to the twin planes ($\alpha = 0°$), the total current density is $j = 60$ A/cm$^2$ dc + 10 A/cm$^2$ ac pp and the temperature is $T = 89$ K. Solid curves show the modulus of the electric field as a function of the current orientation ($\beta_j$), dashed lines are the actual electric field values (in modulus *and* argument $\beta_E$) visited during a complete turn of the current. Left : The radial scale is linear from 0 (center) to 50 μV cm$^{-1}$. From outside towards the center, the solid curves correspond to magnetic fields of $B = 3$ T, 2.5 T, 2 T, 1.8 T, 1.7 T, 1.6 T and 1.5 T. The vortex phase transition happens between the two curves separated by a shaded area. Corresponding dashed curves are shown for $B = 3$ T, 2.5 T and 2 T. Right : Similar representation of data at lower magnetic field in logarithmic radial scale (from 0.01 μV cm$^{-1}$ at the origin to 10 μV cm$^{-1}$). Shown are $B = 1.5$ T, 1.4 T, 1.2 T and 1 T (dashed lines only for $B = 1.5$ T and 1 T).

Another interesting feature is that, at the highest magnetic field values, *e.g.* 3 T and 2.5 T, the curves are not perfect circles, even though the influence of twins on the vortex dynamics is observed to be negligible deep in the vortex liquid (Fig. V-2). Moreover, the small observable ovalization is opposite to the strong asymmetry observable at lower fields. It is very interesting to note that this is perfectly consistent with measurements done in the normal state by Villard *et al.* [154], where it was shown that the normal state resistivity for the current applied perpendicularly to the twin planes is much larger than when the current is along the planes, in direct opposition to the mixed state resistivity. These two opposite behaviors can be summarized by saying that the mobility of both normal electrons *and* vortices (even though they are quite different objects) is larger for a motion *along* the twin planes. Therefore, a large electron mobility obviously leads to a small normal resistivity, whereas a large vortex mobility corresponds to a large vortex velocity, inducing a large electric field, and thus a large flux flow resistivity. It is then possible that, when measuring the current angular dependence of the resistivity above $B_{TB}$, where the twins contribution to flux flow is negligible, only the normal electrons contributes to the (smaller) observable anisotropy.

Finally, it can be said from Fig. V-3 that the $E(\beta_E)$ and $E(\beta_j)$ curves almost perfectly coincide at high fields, but no longer when below $B_{TB}$. This difference means that $\beta_E \neq \beta_j$, or in other words that *the electric field is not parallel to the current, which is nothing but the signature of guided motion*. Unfortunately, the visualization of this relative angle from such a figure is far from easy. To illustrate more explicitly the guided motion, we have therefore tried to present the same data in a different way, focusing on constant directions of the current (constant $\beta_j$), and looking at the direction (and modulus) of the electric field.

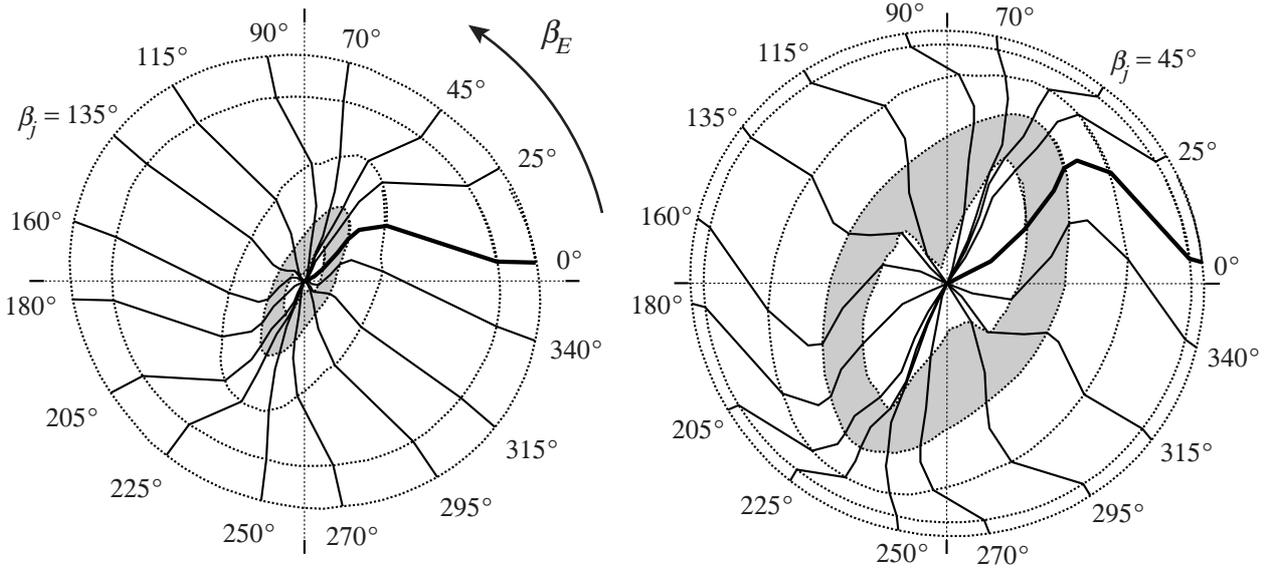

**Figure V-4 :** Left : From the same data as Fig. V-3 (with $\alpha = 0°$, $j = 60$ A/cm² dc + 10 A/cm² ac pp and $T = 89$ K), representation of the evolution of the modulus and argument $\beta_E$ of the electric field as the magnetic field is reduced from $B = 3$ T to 1 T, for a series of given current orientations $\beta_j$ (indicated all around the plot). The dotted lines show the values of the electric field for some specific magnetic fields (and thus correspond to the dashed lines of Fig. V-3) which are $B = 3$ T, 2.5 T, 2 T, 1.7 T and 1.5 T (from outside towards the center). The vortex phase transition happens between the two curves separated by a shaded area. On the right is exactly the same plot in logarithmic radial scale (the origin corresponds to 1 µV cm⁻¹).

This has been done in Fig. V-4, where a series of lines have been drawn between the values of the electric field (in modulus and argument, $E(\beta_E)$) taken as the magnetic field is reduced, each line corresponding to a given orientation of current $\beta_j$ (indicated close to each of these lines, all around the plot). Some curves of $E(\beta_E)$ for constant values of the magnetic field have also been traced, which are exactly equivalent to the dashed lines in Fig. V-3. The same plot is represented twice, the right version having a logarithmic radial scale to reveal the details at low magnetic fields.

From this representation, we first note that, even at high magnetic field in the vortex solid, the angle $\beta_E$ is slightly shifted by about $+ 5°$ (when the current is along $x$, namely at $\beta_j = 0°$ or $180°$) to almost $+ 10°$ (when $\beta_j = 90°$ or $270°$) with respect to $\beta_j$. This is an indication of the respective contact pairs misalignment with the $x$ and $y$ axes.

But the most impressive feature on this graphic is the very clear signature of guided motion along the dominant twin boundaries family. When the Lorentz force is along these twin planes (for $\beta_j = 45°$ or $225°$), the electric field slowly vanishes as the magnetic field is reduced, but keeps always the same orientation, with $\beta_E \approx \beta_j$ (although a small reorientation is visible, again because of the above mentioned probable small changes in the current density distribution). Similarly, when the current is along the twins (hence the driving force on vortices is just orthogonal to the twin planes, for $\beta_j = 135°$ and $315°$), the direction of the electric field is also unchanged as the magnetic field decreases from the vortex liquid to the vortex solid. The only difference is that the electric field vanishes much more rapidly in the solid (as can be seen from the anisotropic shape of the dotted lines), indicating a stronger vortex pinning for this current orientation.

On the other hand, when the current is at roughly ± 45 ° from the twin planes, there is a very obvious onset of guided motion as the magnetic field decreases, observable in the strong variation of the electric field orientation for a constant current direction $\beta_j$ (see for example the thicker line corresponding to $\beta_j = 0°$). This guided motion becomes effective quite precisely at the field $B_{TB}$ at which the twin boundaries have been observed to influence the longitudinal resistivity (Fig. V-2). However, the transition is progressive : although the electric field tends to rotate towards the (110) direction (meaning that the vortices tend to move along the twin boundaries), a complete vortex canalization along the twin planes occurs only at lower magnetic fields. Moreover, for the currents oriented far away from the (110) direction, it is not even clear whether this true canalization eventually happens.

*b) Data for $\alpha = 4°$*

Exactly similar measurements have been performed with the magnetic field slightly tilted away from the twin planes, and the results are presented in the same manner in the two next figures (Fig. V-5 and V-6).

We can note three major differences between these figures and the data presented above for $\alpha = 0°$ (Fig. V-3 and V-4). First, the reduction of the electric field, as the magnetic field is decreased, is more regular than for $\alpha = 0°$, where the electric field was mainly dropping sharply between 2.5 and 2 T (Fig. V-3). Of course, this is the consequence of the different behaviors of the resistivity for the two magnetic field orientations shown in Fig. V-2.

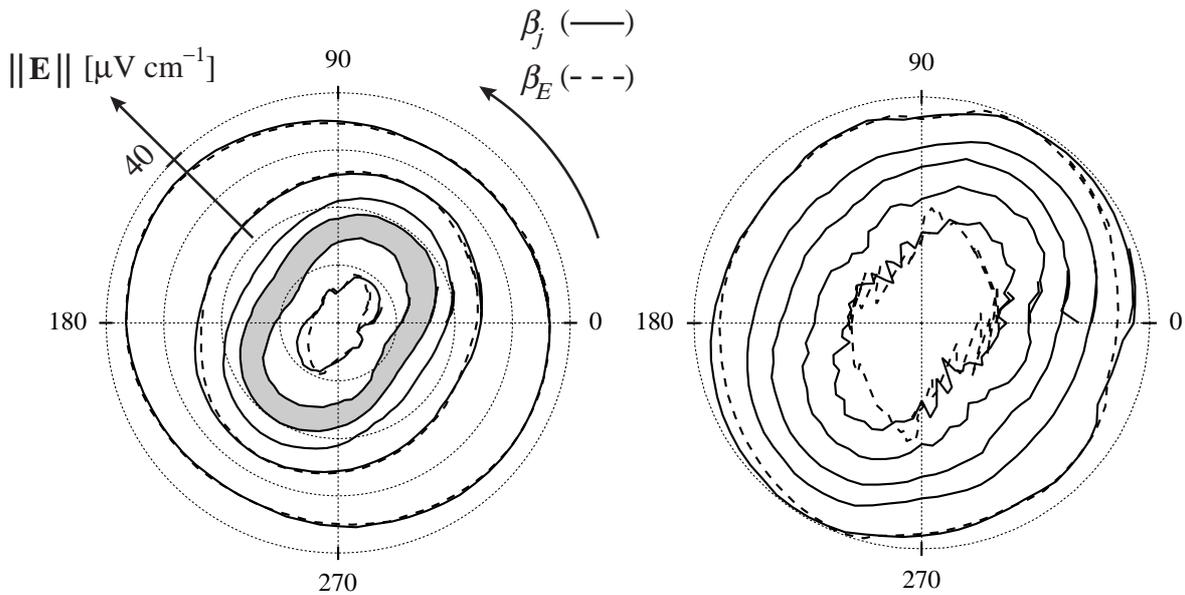

**Figure V-5 :** Polar plots of the electric field as the current is rotated in the sample plane. The magnetic field is tilted at $\alpha = 4°$ from the twin planes, the total current density is $j = 60$ A/cm$^2$ dc + 10 A/cm$^2$ ac pp and the temperature is $T = 89$ K. Solid curves show the modulus of the electric field as a function of the current orientation ($\beta_j$), dashed lines are the actual electric field values (in modulus *and* argument $\beta_E$) visited during a complete turn of the current. Left : Linear radial scale from 0 to 40 μV cm$^{-1}$. From outside towards the center, the magnetic fields are $B = 2.5$ T, 2 T, 1.8 T, 1.7 T, 1.6 T and 1.5 T. The vortex phase transition happens between the two curves separated by a shaded area. Corresponding dashed curves are shown for $B = 2.5$ T, 2 T and 1.5 T. Right : Logarithmic radial scale (from 0.01 μV cm$^{-1}$ to 10 μV cm$^{-1}$), with $B = 1.5$ T, 1.4 T, 1.3 T, 1.2 T and 1.1 T (dashed lines only for $B = 1.5$ T and 1.1 T).

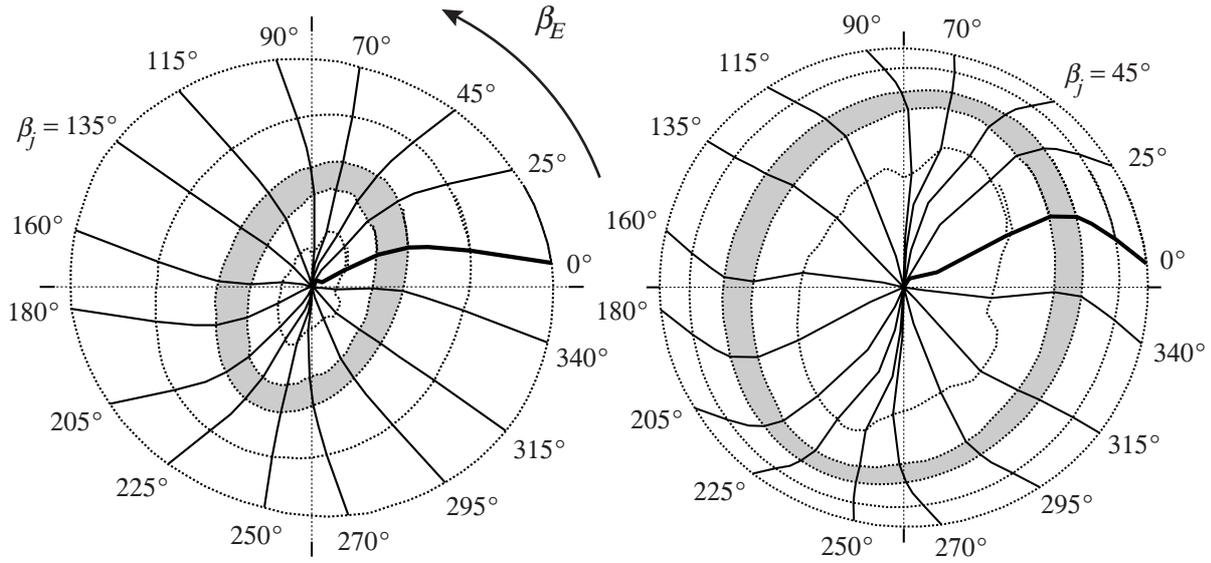

**Figure V-6 :** Left : From the same data as Fig. V-5 (with $\alpha = 4°$, $j = 60$ A/cm$^2$ dc + 10 A/cm$^2$ ac pp and $T = 89$ K), representation of the evolution of the modulus and argument $\beta_E$ of the electric field as the magnetic field is reduced from $B = 2.5$ T to $1.1$ T, for a series of given current orientations $\beta_j$ (indicated all around the plot). The dotted lines show the values of the electric field for some specific magnetic fields (and thus correspond to the dashed lines of Fig. V-5) which are $B = 2.5$ T, 2 T, 1.7 T, 1.6 T and 1.5 T (from outside towards the center). The vortex phase transition happens between the two curves separated by a shaded area. On the right is exactly the same plot in logarithmic radial scale (the origin corresponds to 1 µV cm$^{-1}$).

Secondly, even though the response is still clearly anisotropic, the strong reduction of the resistivity (or electric field) at low magnetic field when the current is along the twin planes ($\beta_j \approx 135°$ or $315°$) has almost completely vanished. We can reasonably assume that at such small angles $\alpha$ between the twin planes and the applied magnetic field, the vortices are composed of segments parallel to the $c$ axis "trapped" in the twin boundaries, linked by inclined parts between the twin planes [155]. The difference between Fig. V-3 and Fig. V-5 might then mean that the inclined kinks are sufficient to suppress the strong and sharp resistivity drop present at $\alpha = 0°$ when the current is along the twin planes, but have almost no influence on the smaller and smoother 360° anisotropy, which is present on both figures. However, this picture is somehow oversimplified, as we will see in paragraph 2.3.

Thirdly, Fig. V-6 indicates that the transition to guided motion still occurs when vortices are tilted away from the twin planes, even though the magnetic field dependence is much smoother. However, in this case no absolute canalization of vortex motion along twin boundaries is observable, in the sense that, at the lowest magnetic field values, the electric field reaches a different orientation for each different direction of the current density $\beta_j$. Finally, recalling that the vortex phase transition field is $B_m \approx 1.65$ T, it is evident that for a constant current orientation, the electric field rotates away from the current mainly in the vortex liquid phase, its direction then remaining almost constant in the vortex solid (note that this observation is also partly valid for $\alpha = 0°$, at least as long as the Lorentz force is not oriented almost perpendicularly to the direction of guided motion, see Fig. V-4).

## 2.3 Measurements at constant field

The difference between what we have presented in paragraph 2.2 and the set of data presented here is not only that the magnetic field instead of the temperature is now kept constant, but also the measurement and data processing procedures have been slightly modified. First, we have tried to compensate the small contact misalignment observed in the preceding paragraph, by introducing correction factors in the mixing of $I_x$ and $I_y$ used to generate a current density with a given orientation $\beta_j$, as well as a different calculation of $E_x$ and $E_y$ from the measured voltages, again introducing some mixed contributions of $V_x$ and $V_y$. We have also measured the angular dependence of the electric field for both magnetic field polarities, calculating the symmetric part (or in other terms the average of the data for both fields) to rigorously obtain the even Hall effect (even though the asymmetric Hall effect is usually negligible compared to the total signal, as we have noted at the beginning of this chapter). Finally, note that a larger current density was used, in order to enhance the signal-to-noise ratio.

In Fig. V-7 are shown the data for four different temperatures, $T$ = 90 K, 88.5 K, 88 K and 87.5 K. We recall that the vortex phase transition temperature is $T_m \approx 88.2$ K and the onset of the twin boundaries influence on the vortex dynamics is $T_{TB} \approx 89.9$ K (see page 73). The main difference with Fig. V-3 and V-5 is that even above $T_{TB}$, at $T$ = 90 K, the anisotropy is still very well apparent, and is not reversed with respect to the vortex solid behavior, as was the case for the "normal state-like" anisotropy observed at high magnetic fields in the preceding paragraph (page 75). It would be interesting to know whether the reason is that the temperature is too close to $T_{TB}$, such that the twins actually still act on vortices, or if this is the result of a contact misalignment inaccu-rate compensation. Unfortunately no measurement have been performed in the same conditions at higher temperatures up to the normal state, so this question cannot be answered.

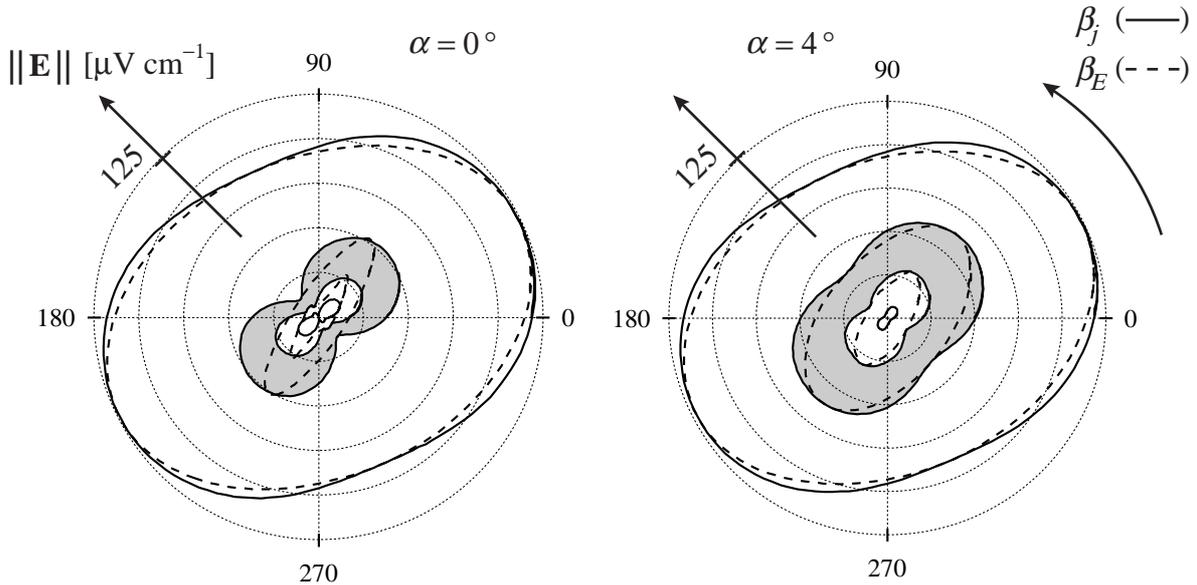

**Figure V-7 :** Polar plots of the electric field as the current is rotated in the sample plane, for two different orientations of the magnetic field. Left : $\alpha$ = 0 ° (field along the twin planes). Right : $\alpha$ = 4 °. In both cases, the magnetic field intensity is $B$ = 2 T and the total current density is $j$ = 100 A/cm$^2$ dc + 25 A/cm$^2$ ac pp. Solid curves show the modulus of the electric field as a function of the current orientation ($\beta_j$), dashed lines are the actual electric field values (in modulus *and* argument $\beta_E$) visited during a complete turn of the current. From outside towards the center, the temperatures are $T$ = 90 K, 88.5 K, 88 K and 87.5 K. The vortex phase transition happens between the two curves separated by a shaded area. The dashed curves E($\beta_E$) are not shown for the lowest temperature.

Aside from this striking difference, the data for $\alpha = 0°$ (Fig. V-7 left) form a picture very similar to the results of paragraph 2.2 : a well marked resistivity drop still occurs at lower temperatures when the current is parallel to the twin planes, and a strong difference between the current and electric field directions, observable through the separation of the $E(\beta_j)$ and $E(\beta_E)$ curves, still reflects the existence of guided motion. For $\alpha = 4°$ however, (Fig. V-7 right), the same resistivity drop in the $(1\bar{1}0)$ direction of current is now also clearly visible, whereas it was almost undetectable in Fig. V-5 (except maybe on the curve for 1.6 T). Therefore, the remark formulated in Fig. V-5 about the apparent absence of this dip in the $(1\bar{1}0)$ direction (see page 78) should be reconsidered : the dip, although less marked at inclined magnetic fields, is still present.

We can also note that the data of Fig. V-7 are better than those of paragraph 2.2 (the noise level is lower), in part due to the field symmetrization involving an averaging between two sets of measurements. It is then worth looking in more detail at the low field data, as is done in Fig. V-8. A small feature, which was scarcely visible on the measurements of paragraph 2.2, is now very nicely observable : a kind of partial fourfold symmetry shows up at the lowest magnetic fields. It is more marked when vortices are exactly aligned with the twin planes ($\alpha = 0°$, Fig. V-8 left), but is also detectable for inclined field ($\alpha = 4°$, Fig. V-8 right). It is very surprising to see that this feature is not at all symmetric with respect to the $(1\bar{1}0)$ axis (it is mostly evident from the 87.5 K curve for $\alpha = 0°$), suggesting a kind of "history dependence" related to the direction of current rotation during the measurement, even though the current is rotated very slowly, stopping for many seconds in each orientation to record stabilized averages of the voltage signals. Unfortunately, no data is available for the opposite direction of current rotation, so more measurements are required to answer this question.

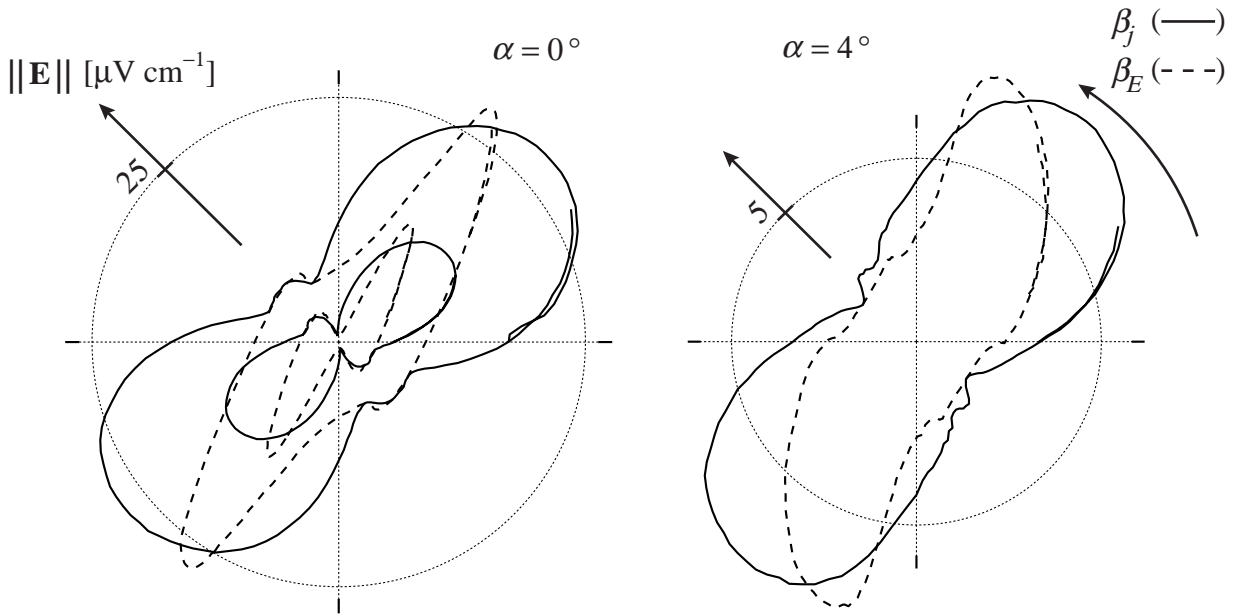

**Figure V-8 :** Enlarged view of the low field data of Fig. V-7, showing both the $E(\beta_j)$ and $E(\beta_E)$ curves in the vortex solid phase. Left : For $\alpha = 0°$, $T = 88$ K and $T = 87.5$ K. Right : at $\alpha = 4°$, only $T = 87.5$ K.

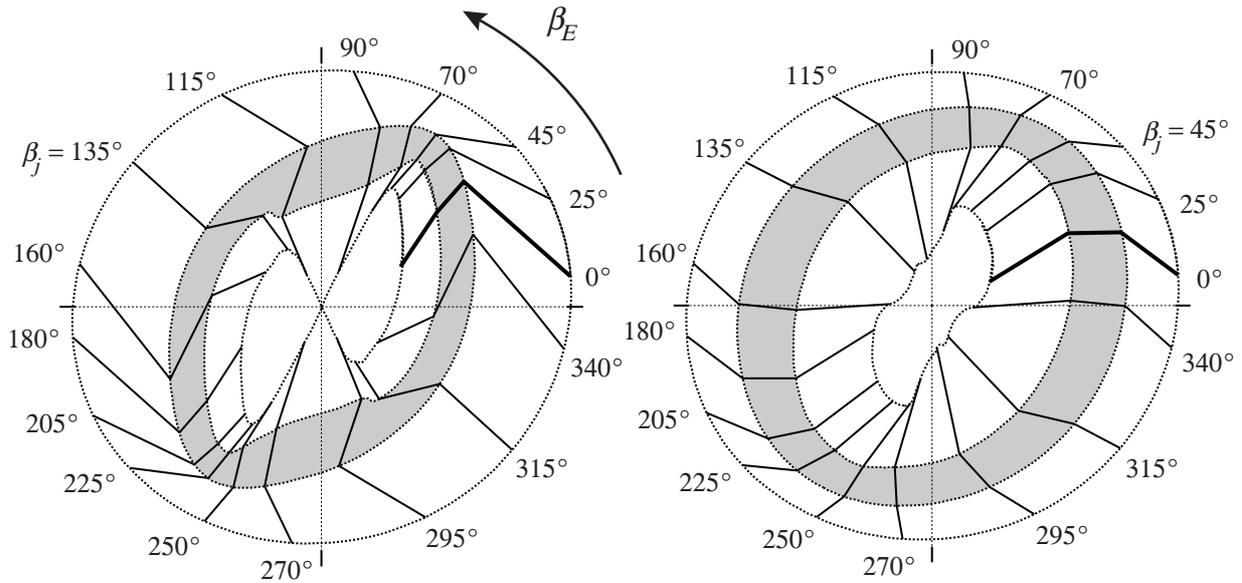

**Figure V-9 :** From the same data as Fig. V-7 (with $B = 2$ T, $j = 100$ A/cm$^2$ dc + 25 A/cm$^2$ ac pp, and for $\alpha = 0°$ on the left and $\alpha = 4°$ on the right), representation of the evolution of the modulus and argument $\beta_E$ of the electric field as the temperature is reduced, for a series of given current orientations $\beta_j$ (indicated around the plots). The dotted lines show the values of the electric field for the different temperatures (and thus correspond to the dashed lines of Fig. V-7 and V-8), which are $T = 90$ K, 88.5 K, 88 K and 87.5 K (from outside towards the center). The vortex phase transition happens between the two curves separated by a shaded area. The radial scale is logarithmic, the origin corresponding to 1 µV cm$^{-1}$.

In Fig. V-9, we also provide the plots of $E(\beta_E)$ lines at constant current directions $\beta_j$, just as for the representations in Fig. V-4 and V-6. Similar to the observations made in these two figures, we note that guided motion is quite visible, and again smoother at tilted magnetic fields, even though data for more temperatures are lacking to reach pictures as complete as Fig. V-4 and V-6.

Fig. V-9 also allows us to establish that the correction of contact misalignment is not conclusive : $\beta_E$ is now shifted the other way compared to $\beta_j$, by $-10°$ (in the liquid phase, at 90 K) almost isotropically instead of $+5°$ to $+10°$ as in Fig. V-4 and V-6, suggesting that the small errors have been overcompensated. It is difficult to know if this is also the reason for the important remanent anisotropy at high temperatures, above $T_{TB}$.

## 2.4 Implications for the interpretation of resistivity measurements

From the measurements shown here, it is also possible to reconstruct the equivalent of resistivity measurements at two different magnetic field orientations, like what is presented in Fig. V-2. Here, we have the advantage of having access to more information, like the total magnitude of the electric field, instead of only its component parallel to the current, and also the behavior for various current directions, *all in the same sample*. Note that in Ref. [153], Fleshler *et al.* already compared the resistivity for three different current directions, but as we mentioned already at the beginning of this chapter, only the longitudinal component (along the current) of the electric field was measured, and a different sample was used for each current direction. Since the pinning effects are often sample dependent, it is therefore difficult to know exactly what is the importance of the current direction with respect to the twin boundaries for the vortex dynamics.

# V

For this purpose, we use the measurements as a function of the magnetic field presented in paragraph 2.2, since we have more data in this case. First, we extract from these data the equivalent of a standard resistivity measurement, as was shown in Fig. V-2, *i.e.* the longitudinal electric field $E_x$ for a current along the *x*-axis (namely for $\beta_j = 0°$). The resulting curves are represented in Fig. V-10 and V-11 with dashed lines for $\alpha = 0°$ (solid symbols) and $\alpha = 4°$ (open symbols). As we expected, the picture is very much like the curves of Fig. V-2.

The idea is now to see what is the influence of guided motion on the difference between the two curves, and what is really a intrinsic difference in the vortex mobility between $\alpha = 0°$ and $\alpha = 4°$. First, we plot in Fig. V-10, together with the curves described above, the values of the total magnitude of the electric field in the same conditions (solid lines). The result is a sort of indication of the amplitude of guided motion : if the solid and dashed lines are perfectly superimposed, then the total electric field is indeed parallel to the current ($E_x = E_{tot}$), so that there is no guided motion, and the difference between $\alpha = 0°$ and $\alpha = 4°$ is truly only a consequence of a change in the vortex mobility. Of course, since we have already demonstrated the existence of guided motion above, it is not surprising to see that this is not the case here : the longitudinal field is only a part of the total electric field. However, the effect is much stronger for $\alpha = 0°$ than for $\alpha = 4°$, since the guided motion has been observed to be more efficient when vortices are exactly along the twin planes. As a consequence, this means that a part (about one half to one third) of the resistivity difference between the two orientations of the magnetic field reflects a difference in the vortex motion *direction*, the other part resulting from a different vortex mobility for this current orientation.

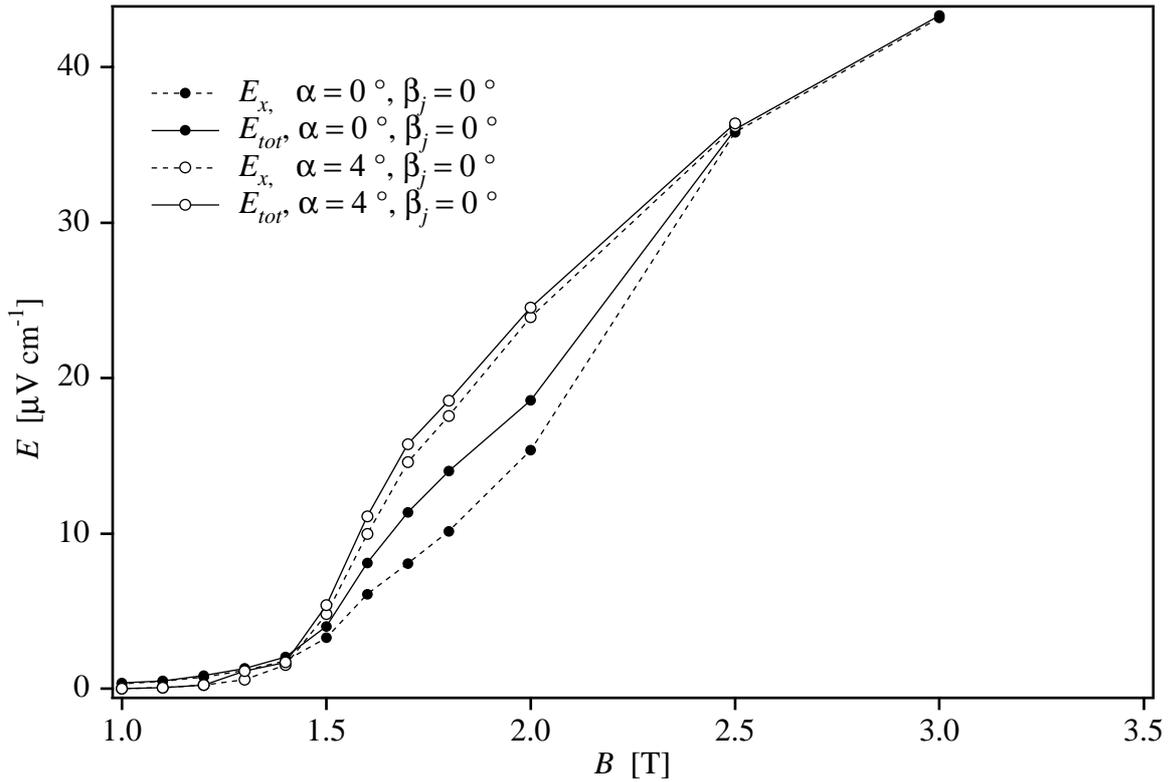

**Figure V-10 :** Selected data of paragraph 2.2 reported as a function of the magnetic field. The dashed lines represent what would be measured in a standard resistivity measurement (longitudinal component of the electric field $E_x$ for a current at 45° from the twins, corresponding to $\beta_j = 0°$). The solid lines show what the total magnitude of the electric field $E_{tot}$ actually is in this case. The temperature is $T = 89$ K and the current is $j = 60$ A/cm$^2$ dc + 10 A/cm$^2$ ac pp. Solid symbols are for the magnetic field aligned with the twin boundaries ($\alpha = 0°$), open symbols correspond to $\alpha = 4°$.

In the discussion above, we have just considered a case of a current oriented at 45 ° from the twin planes, so that the vortices also have to move at an angle from the twins (even if this angle is smaller because of guided motion). The question is now to know what is the influence of the magnetic field orientation on the vortex mobility if the current is perpendicular to the twins, so that the vortices will move along the twins (and so that, even for an inclined magnetic field, the vortex segments located *inside* the twin boundaries will always remain in them).

For this, we compare the usual "longitudinal resistivity" curves (in dashed lines, the same as in Fig. V-10) to the data obtained for $\beta_j = 45°$ in the present chapter. Note that we show the total magnitude of the electric field $E_{tot}$ for $\beta_j = 45°$ (solid lines in Fig. V-11). Interestingly enough, we see that there is only a very small residual difference between the two magnetic field orientations $\alpha = 0°$ and $\alpha = 4°$ in this case : the vortex mobility becomes very few dependent on the magnetic field orientation (that is on the presence of inclined segments of the vortex) when the motion happens along the twin planes. Note however from the inset of Fig. V-11 that the crossing of the two

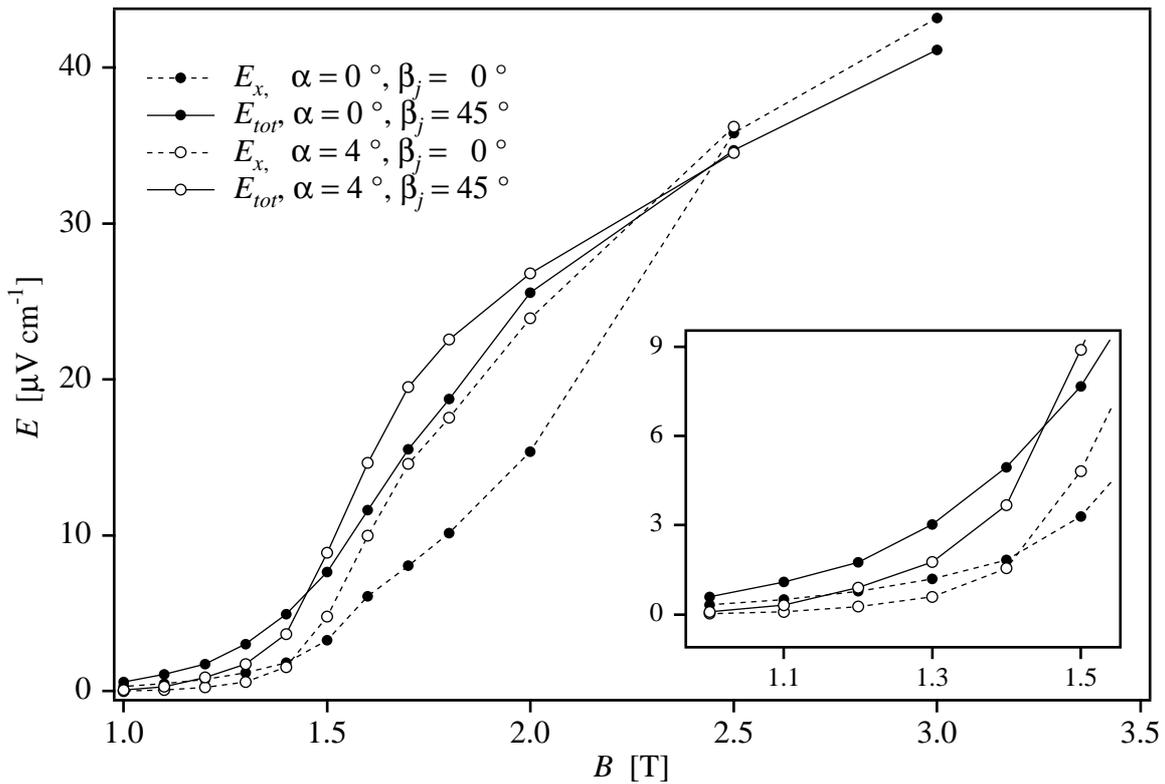

**Figure V-11 :**  Selected data of paragraph 2.2 reported as a function of the magnetic field. The dashed lines represent what would be measured in a standard resistivity measurement (longitudinal component of the electric field $E_x$ for a current at 45 ° from the twins, corresponding to $\beta_j = 0°$). The solid lines show the total magnitude of the electric field $E_{tot}$ for a current perpendicular to the twins, corresponding to $\beta_j = 45°$ (that is for maximum vortex mobility, forcing them along the twins). The temperature is $T = 89$ K and the current is $j = 60$ A/cm$^2$ dc + 10 A/cm$^2$ ac pp. Solid symbols are for the magnetic field aligned with the twin boundaries ($\alpha = 0°$), open symbols correspond to $\alpha = 4°$. Inset : Detailed view of the vortex solid behavior.

# V

curves still happens at the same position, the resistivity being *smaller* in the vortex solid for *larger* angles ($\alpha = 4°$). Regrettably, the small alignment errors mentioned in the previous paragraphs make it difficult to know whether the small residual difference between $\alpha = 0°$ and $\alpha = 4°$ is observed because the current is actually not exactly across the twin planes, or if a dependence in the vortex mobility on the magnetic field orientation would always remain, whatever the current direction is.

Obviously, the data shown here are only a preliminary test for this method, showing however that it is really feasible, with a simple eight contact geometry, to study thoroughly, quantitatively and reliably the influence of twins on vortex dynamics (see next section below), and for example check precisely the predictions of models like the one of Ref. [155] for inclined vortices.

## 3. Conclusion

We have presented here a study of vortex guided motion in a twinned YBa$_2$Cu$_3$O$_{7-\delta}$ crystal as a function of a *continuously orientable* current density. The experimental method consists of resistivity measurements, measuring two perpendicular electric field components, while applying two orthogonal and synchronous current components. We have also demonstrated how this effect has a strong influence on what can be measured in standard longitudinal resistivity measurements, as they are very often performed in similar samples.

The observed guided motion is clearly related to the twin structure of the sample, vortices moving preferentially along the twin planes, having a very poor mobility across them. In our particular sample, one of the two twin boundary families is dominating, a characteristic also confirmed by the magnetooptic observations of the preceding chapter. The resulting (110) anisotropy is seen to manifest itself already in the vortex liquid phase, becoming more and more pronounced as approaching the vortex phase transition. Once in the vortex solid, the behavior tends to remain constant, except for the emergence of a small ($1\bar{1}0$) anisotropy, probably due to a minor influence of the second twin plane family. We have seen no sharp feature corresponding to the crossing of the vortex phase transition. This result will have some importance for the next chapter.

As we have seen on various occasions from the results of the present chapter, the data may be difficult to discuss quantitatively with great precision, since the information is biased by imperfections like contact misalignments. Of course, a compensation can in principle be easily done by introducing experimentally determined trigonometric correction factors, corresponding to a change from an arbitrary to an orthogonal *x-y* frame of reference, as is tempted for the measurements of paragraph 2.3. Even though this correction was not performed precisely enough in the case of these results, it is most probable that such a method would not lead to a perfectly satisfying outcome in any case. Indeed, another imperfection certainly plays a role, namely the modification of the current distribution in the sample as the voltage-current characteristic becomes strongly non-linear. One of the consequences observable on our data is the apparent progressive rotation of the (110) and ($1\bar{1}0$) directions as the magnetic field or temperature is reduced, reflecting a change in the effective ratio of *x* and *y* current components in the region of the voltage measurement.

This last problem could be solved by a more symmetric contact configuration, as illustrated in Fig. V-12. We see that our contact pattern on the sample is not really adapted to this type of investigation, since for example the $V_x$ voltage is not measured in between the contacts providing the $I_y$ current. Therefore, depending on the local spatial distribution of the current flowing between the $I_y$ pads, and with only a slight asymmetry in the voltage contacts, the measured $V_x$ could easily be a mixed contribution of both $E_x$ and $E_y$. If the contact pattern is made as illustrated in Fig.V-12 b, these coupled effects will essentially be suppressed, since then both voltage components are measured in a zone where the current distributions are more homogeneous.

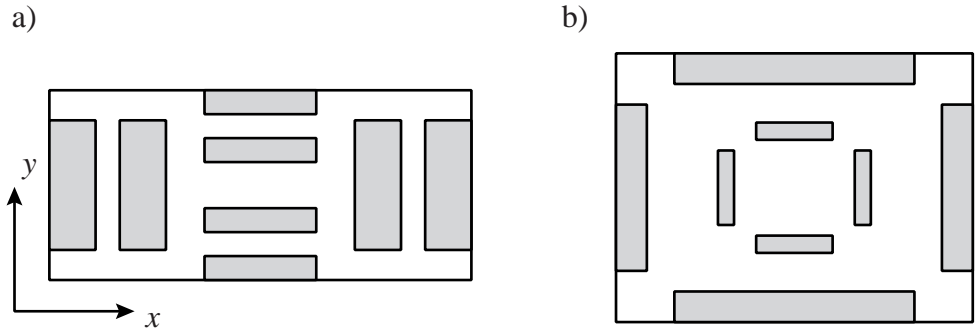

**Figure V-12 :** a) Schematic contact pattern used in this work : the *x* and *y* axis are not equivalent, since, for example, the voltage taps for the *x* direction are not placed properly in between the feeds for the *y* current component. b) Example of a more ideal contact configuration for the study of guided motion as a function of the current orientation, with a better reciprocity in both *x* and *y* directions.

Also, a second cause of perturbed current distributions is that our sample is not uniformly twinned (see Fig. IV-11). Hence vortex guided motion probably have a quite complex and inhomogeneous form. In this sense, the overall qualitative results are even surprisingly coherent.

However, because of these homogeneities, some observations (such as the existence of a true canalization exactly along the twin planes, or the appearance of a partial fourfold symmetry) might be dependent on the current intensity, since these various influences might be overcome by different critical currents. This can be a possible explanation for some of the differences observed between the data of paragraphs 2.2 and 2.3 in the vortex solid phase, since the measurements have been obtained at different current densities.

It would also be extremely interesting to extract from our data the ordinary (antisymmetric) Hall effect as a function of current orientation with respect to the twin boundaries. Unfortunately, we have tried to perform this data processing, but the results were completely inconsistent, not even showing any general trend, nor being reproducible after a complete 360° current rotation. The problem is that this requires the determination of the angles $\beta_j$ and $\beta_E$ with very high accuracy, since we then have to project the electric field on a direction perpendicular to the current. Therefore, non-constant errors in $\beta_j$ or $\beta_E$ would result in a variable mixing of longitudinal and transverse components. Since the Hall effect is very small, only a little parasitic contribution of other components is enough to completely dominate the result. A contact pattern like the one drawn in Fig.V-12 b, with a fairly good misalignment correction, would probably provide much more useful results, since it would provide a more accurate and reproducible determination of $\beta_j$ and $\beta_E$.

In any case, this new experimental method is obviously a very good way to investigate the influence of twin planes on vortex dynamics in $RBa_2Cu_3O_7$ – type samples. Many questions can be addressed this way. For example, more data should be collected on the transition from the normal state anisotropy (as measured by Villard *et al.* [154]) to the opposite guided motion anisotropy, since the present results are not conclusive.

# V

Deeper in the mixed state, we have seen that our sample mainly shows a twofold symmetric angular response, with some indication of a fourfold symmetry lower in the vortex solid phase. It would be interesting to look at other types of samples, either unidirectionally twinned or on the other hand more uniformly twinned in both directions, to see how the angular response correlates to the twin structure as a function of the temperature and magnetic field. We could then evaluate the influence of the guided motion on standard longitudinal resistivity measurements, in which the response of only one current direction is analyzed through only one, parallel, voltage component.

A very important issue is also to understand what happens exactly in originally twinned crystals that are submitted to a uniaxial stress at rather high temperature in order to be detwinned. Even though the contrast in polarized light pictures of the sample surface[1] is observed to disappear during this process, it is difficult to know if the samples are then really completely twin free. For example, the critical current is not significantly reduced in the detwinned samples, whereas an important source of vortex pinning should have been removed [156]. Moreover, when looking at the surface more carefully, dark lines in the (110) and ($1\bar{1}0$) directions are still slightly visible [157], indicating that the uniaxial stress might only have moved the twin boundaries to promote one of the two twin orientations, without actually resulting in the annihilation of the boundaries themselves [157]. In some cases, two successive twin planes would then be too close to each other to be detectable with the help of an optical microscope. Clearly, the method used here might be a good way to characterize these samples, by looking at a possible remanent anisotropic pinning.

As a complement to the results presented here, a confirmation that the even Hall effect in $YBa_2Cu_3O_{7-\delta}$ is indeed related to guided motion of vortices along the twin boundaries can be found in Ref. [127]. In these measurements of the Hall effect in an untwinned crystal, the transverse voltage has been reported to have a negligible part symmetric with respect to the magnetic field.

Finally, we also want to point out that this experimental method can find another very interesting application. We have mentioned in chapter I (page 5) that the Bragg glass, the vortex solid phase occurring in sufficiently clean systems, has a characteristic transverse dynamics which should allow us to clearly identify this particular vortex phase. The idea is that once this glass has been put in motion in a given direction, a change of this direction requires an intrinsic transverse force specific to this vortex phase, since its motion occurs along well defined channels. Therefore, even if the longitudinal motion is itself non-linear, the transverse component of the response should have a different non-linear characteristic due to this effect. A nice way to detect it could then be to apply a current rotating in the sample plane, and measure the resulting electric field, in magnitude *and direction*. The difference between the longitudinal and the transverse components of the response would then induce a lag $\Delta\beta = \beta_E - \beta_j$ between the current and the electric field directions. This lag should then change its sign when switching from a clockwise to a counterclockwise current rotation, and should depend on the current intensity. Of course, an untwinned crystal is required for this kind of measurement, since the lag $\Delta\beta$ is very probably much smaller than the contribution of guided motion by twin boundaries observed in our sample.

---

1. corresponding to the alternation of twins of opposite orientations (see Fig. IV-11).

# CHAPTER VI    *RESULTS : HALL EFFECT*

## 1. Hall effect in the vortex solid

In this chapter, we present and discuss the results for the Hall effect in the mixed state of our $YBa_2Cu_3O_{7-\delta}$ sample, already partially presented elsewhere [158]. As we have mentioned previously, the main achievement of this work with respect to the already existing experimental studies is that we have performed the measurements both in the vortex liquid and the vortex solid phase, across the well identified vortex phase transition.

All the measurements and data processing have been performed following the procedures commented in chapter IV, paragraph 1.5. Among these procedures, we recall that we extract the Hall effect from the part of the transverse resistivity which is *antisymmetric* upon reversal of the magnetic field polarity, in order to reduce as much as possible the influence of the guided motion (discussed in the preceding chapter) on the results. The only non-systematic step in the data processing procedures is the averaging between two equivalent sets of measurements performed with opposite current polarities : only some of the data shown here are the result of such a process. An indication of whether or not this current symmetrization has actually been done will be given for each of the reported results.

### 1.1 Measurements as a function of the temperature

We start with measurements at constant magnetic field, done as the temperature is slowly increased at a rate of 0.4 K per minute, with the field inclined by $\alpha = 4°$ away from the *c*-axis, in order to reduce the influence of the twin boundaries. In Fig. VI-1 the longitudinal and Hall resistivities are first shown for $B = 2$ T. The open symbols correspond to measurements at low current density, clearly revealing the vortex phase transition at $T_m \approx 88.2$ K. Solid symbols represent data for a much higher current (with only one current polarity). The vortex solid is then set into motion, allowing us to probe its dynamics.

In both cases, the Hall anomaly is clearly visible : the Hall resistivity $\rho_{xy}$ is indeed positive in the normal state and slightly below, but changes sign just above 92 K, only slightly below $T_c \approx 93.5$ K. It then reaches a negative maximum before returning to a value of zero because of the vanishing vortex mobility at lower temperatures. For low current, both resistivities drop to zero quite sharply at the vortex phase transition, since at that point the critical current steeply increases. On the other hand, at higher current, the Hall resistivity shows a shoulder extending into the vortex solid, much like the corresponding well known behavior of the longitudinal resistivity. A careful examination of the data in the vortex solid reveals no significant difference in the onset temperatures of both resistivities $\rho_{xx}$ and $\rho_{xy}$.

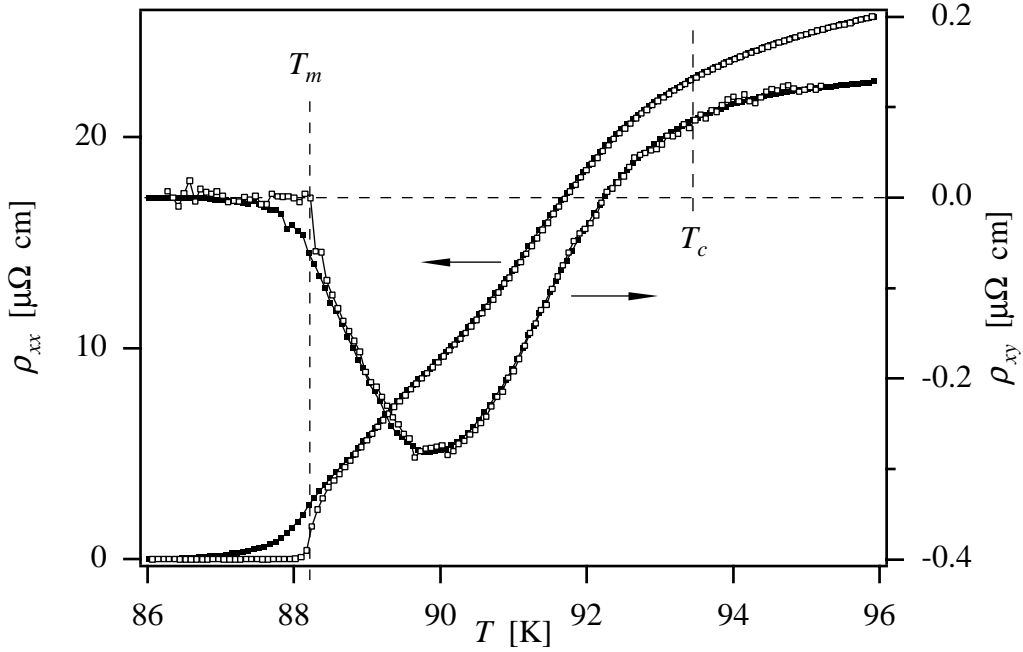

**Figure VI-1 :** Longitudinal resistivity $\rho_{xx}$ and Hall resistivity $\rho_{xy}$ as a function of the temperature, in a magnetic field $B = 2$ T inclined at $\alpha = 4°$ away from the *c*-axis. Open symbols correspond to a current density of $j = 1$ A/cm$^2$ rms, and reveal the vortex melting at $T_m \approx 88.2$ K. Solid symbols are measured with a higher current $j = 150$ A/cm$^2$ dc + 50 A/cm$^2$ pp of single polarity.

From these data, we can then extract the Hall angle and the Hall conductivity, both shown in Fig. VI-2. From the expressions relating these two quantities to the resistivities ($\tan \theta_H = \rho_{xy} / \rho_{xx}$ and $\sigma_{xy} = \rho_{xy} / (\rho_{xx}^2 + \rho_{xy}^2) \approx \rho_{xy} / \rho_{xx}^2$), it is evident that when the measured $\rho_{xx}$ and $\rho_{xy}$ decrease below the noise level, the calculated values of $\tan \theta_H$ and $\sigma_{xy}$ are scattered and are no more significant. Therefore, the representation of the data for low current is interrupted when the large scattering starts, namely at the vortex phase transition, where the resistivity vanishes. From the data at high current, it is also interesting to note that the noise becomes more important just as the vortex phase transition is crossed from the vortex liquid to the vortex solid. This noise is intrinsic to the vortex solid, which moves by irregular motion of bundles, or avalanches, of correlated vortices.

If we now focus on the conductivity $\sigma_{xy}$, we immediately see what is the most striking result of this work : the slope of the Hall conductivity is much steeper in the vortex solid than in the vortex liquid. Moreover, the transition from the usual vortex liquid behavior to the fast drop of the conductivity at low temperatures is very sharp, and happens right at the vortex phase transition. *This is the first experimental evidence that the vortex phase transition affects the Hall behavior.*

In Fig. VI-3 are represented equivalent data for magnetic fields ranging from 1 to 4 Tesla. In all the cases, the steep drop of the Hall conductivity is clearly visible, and its position can be checked to correspond precisely to the vortex phase transition temperature. The sudden appearance of noise at this point is also systematic, even though the use of only one current polarity might be responsible for some of the peaks in $\sigma_{xy}$ (see Fig. IV-9 on page 59).

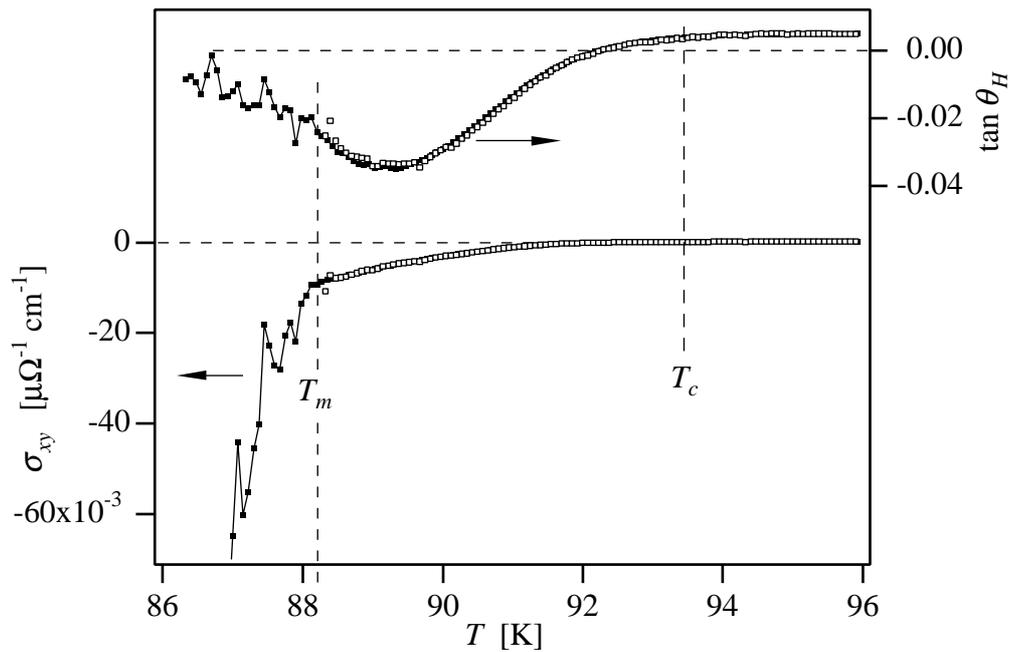

**Figure VI-2 :** Hall angle tan $\theta_H$ and Hall conductivity $\sigma_{xy}$ obtained from the data of Fig. VI-1. $B = 2$ T, $\alpha = 4°$, $j = 1$ A/cm$^2$ rms (open symbols) and $j = 150$ A/cm$^2$ dc + 50 A/cm$^2$ pp (solid symbols, single polarity). The dramatic change of behavior of the conductivity at the vortex phase transition is clearly visible.

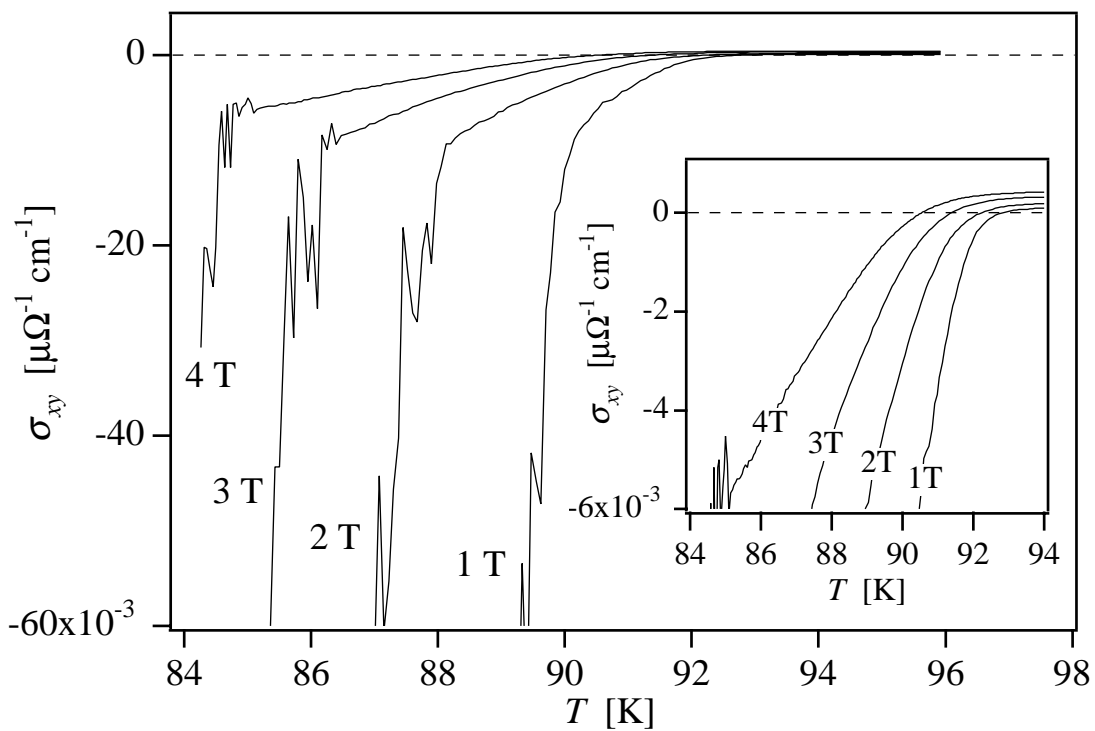

**Figure VI-3 :** Hall conductivity $\sigma_{xy}$ as a function of the temperature for various magnetic fields, always inclined by $\alpha = 4°$ from the $c$-axis. The current is $j = 150$ A/cm$^2$ dc + 50 A/cm$^2$ pp (single polarity). Inset : detail of the same data showing the sign change.

# VI

The inset of Fig. VI-3 shows the same data, but on a reduced scale, only concerning the vortex liquid phase. The sign change is then well observable, and the asymptotic normal state Hall conductivity (only weakly dependent on the temperature and proportional to the magnetic field) is apparent. Note that the scale of the inset provides a good idea of the improvement made in the present work compared to the data existing in the literature : to our knowledge, the lowest values precedently reported for $\sigma_{xy}$ in $YBa_2Cu_3O_{7-\delta}$ were between $-2.5 \times 10^{-3}$ and $-3.5 \times 10^{-3}$ $\mu\Omega^{-1}$ $cm^{-1}$ [122,124,125,132], about a factor two too small to reveal the change of behavior related to the vortex phase transition reported here. Note, however, that values with a magnitude similar to our data ($60 \times 10^{-3}$ $\mu\Omega^{-1}$ $cm^{-1}$) can more easily be obtained in $HgBa_2CaCu_2O_{6+\delta}$, which have a much larger longitudinal resistivity for a comparable (though positive) Hall angle [107].

Therefore, it is evident why the striking behavior of the Hall conductivity of Fig. VI-3 has never been reported before. Even though some experimental data for $\rho_{xy}$ or $\tan \theta_H$, other than those cited above, might have sufficient accuracy to reach large enough negative values of the conductivity $\sigma_{xy} \approx \rho_{xy} / \rho_{xx}^2 = \tan \theta_H / \rho_{xx}$, the fact that the conductivity is not explicitly calculated and shown prevents the observation of this effect. The signature of the vortex phase transition is indeed only present in $\sigma_{xy}$, whereas nothing can be detected in the Hall resistivity $\rho_{xy}$ nor in the Hall angle $\tan \theta_H$ (except for the larger noise in the vortex solid).

Finally, from the same set of data, we can also study the scaling relation $|\rho_{xy}| = A \rho_{xx}^\beta$ between the Hall conductivity $\rho_{xy}$ and the longitudinal resistivity $\rho_{xx}$. In Fig. VI-4 are shown the data measured as a function of the temperature for three different magnetic fields. Observe that the sharp dips at the right of the plot correspond to the sign changes of $\rho_{xy}$, corresponding to divergences when the absolute value is represented with a logarithmic scale.

First of all, it is remarkable to see how well the scaling law is verified over many orders of magnitude, as has already been reported in the literature (see chapter III for a review). Note also that the data for all the different magnetic fields not only form a straight line in the log-log plot, but also collapse all on the *same* line. This means that the prefactor *A* in the scaling relation is not only temperature independent in this regime, such that the power law is verified for $\rho_{xy}(T)$ versus $\rho_{xx}(T)$, but is also magnetic field independent, such that data measured as a function of the magnetic field at a fixed temperature would also verify the same scaling relation for $\rho_{xy}(B)$ versus $\rho_{xx}(B)$ with the same exponent (at least in an equivalent magnetic field range). This will indeed be directly verified in the next paragraph. Finally, the value $\beta = 1.4$ found here is consistent with the widely scattered data found in the literature.

But the most interesting result coming out of Fig. VI-4 is that *the scaling relation is absolutely continuous through the vortex lattice melting transition*, unperturbed by the change of vortex phase. Even though the scaling law has been observed to be verified in a vortex solid phase for data measured as a function of the current by Wöltgens *et al.* [50], it was shown that the parameters *A* and *β* were then temperature dependent (although they were constant in the liquid phase), meaning that data measured as a function of the temperature would not scale through the phase transition. Moreover, in this thin film sample, the solid phase was a disordered vortex glass, such that the transition was not sharp, first order-type, and was not reported to be associated to a dramatic change of the Hall behavior (like the sharp drop of the Hall conductivity that we see here). It is indeed very surprising to see in our case how a big change of slope of the conductivity can remain without any effect on the scaling relation. This certainly has to be explained.

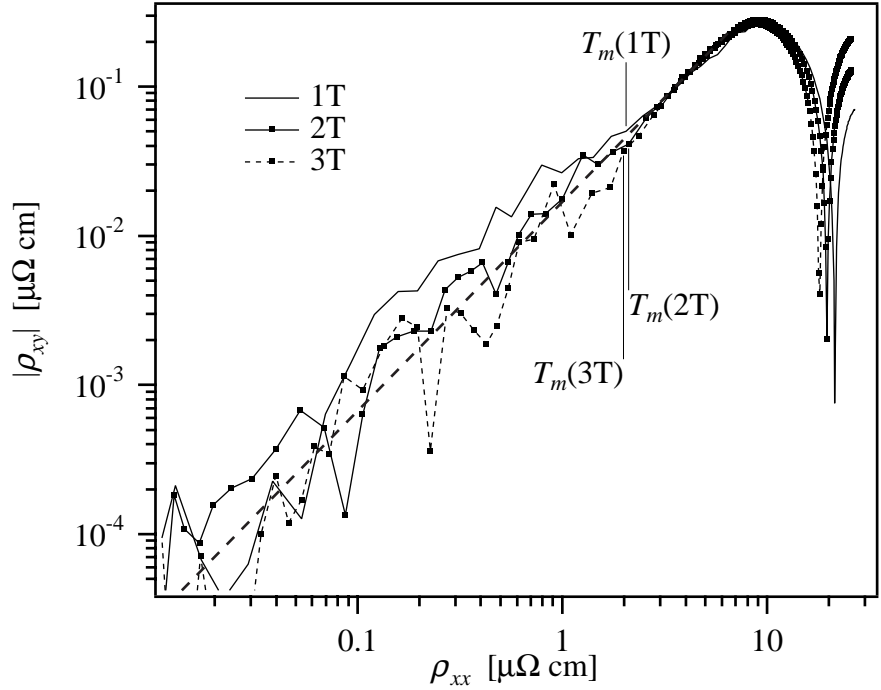

**Figure VI-4 :** Log-log plot of the absolute value of the Hall conductivity $\rho_{xy}$ as a function of the longitudinal resistivity $\rho_{xx}$ for $\alpha = 4°$, $j = 150$ A/cm$^2$ dc + 50 A/cm$^2$ pp (single polarity), and three different magnetic fields : $B = 1$, 2 and 3 T. The temperature is the implicit running parameter for each of the curves. The power-law scaling relation $|\rho_{xy}| \propto \rho_{xx}^{\beta}$ is clearly apparent, and is underlined by a dashed straight line, corresponding to an exponent of $\beta = 1.4$.

## 1.2 Measurements as a function of the magnetic field

If we try to interpret the results for the Hall conductivity shown above, we will certainly reach an important question : apart from the vortex melting corresponding to the sharp drop of $\sigma_{xy}$, which other varying parameter can play a role in this striking behavior ? As first noted by Hagen *et al.* [109] (see page 42), some microscopic parameters indeed also seem to be strongly correlated to the Hall anomaly, namely the ratio between the electronic mean free path $\ell$ and the coherence length $\xi$. Since these two quantities depend on the temperature in opposite ways, this ratio changes during the measurements of the preceding paragraph : $\xi / \ell$ is presumably larger than unity at high temperature (dirty limit), while it probably drops much lower when the temperature is sufficiently reduced (superclean limit).

Provided the expected behavior in the superclean limit precisely corresponds to a Hall angle of $\theta_H = \pi / 2$, or in other words to diverging $\tan \theta_H$ and $\sigma_{xy}$, which is at first sight consistent with our data. However, in superclean systems, the Hall angle can actually only be positive, which means that the vortices are at rest in the supercurrent frame of reference (vortices flow together with the current), thus not really compatible with the negative Hall anomaly, in which the vortices have to flow *against* the current density[1]. Moreover, even though our $\sigma_{xy}$ seems to indeed diverge, the Hall angle rather seems to vanish. But still the possible role of the temperature dependent microscopic parameters cited above is worth checking.

---

1. Note for example that in underdoped Y:123, Harris *et al.* incorrectly interpret a very large negative Hall angle as an evidence for a superclean regime [159]. In their case, a preformed pairs scenario seems more adapted to their data [43].

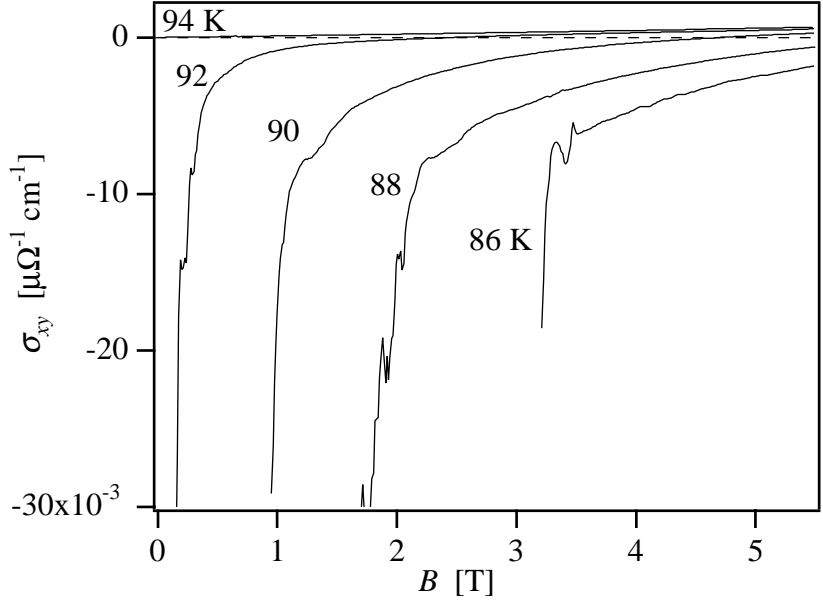

**Figure VI-5 :** Hall conductivity $\sigma_{xy}$ as a function of the magnetic field for various temperatures, from the normal state (at $T = 94$ K) down to $T = 86$ K. The magnetic field is inclined by $\alpha = 4°$ from the *c*-axis. The current is $j = 150$ A/cm$^2$ dc + 50 A/cm$^2$ pp, with a single polarity except for the data at 88 K, which are averaged between both current polarities.

This question can be very simply addressed by performing the same type of measurement for the conductivity as in the preceding paragraph, this time at constant temperature, as a function of the magnetic field. Therefore, these measurements can be assumed to be done at constant values of the ratio $\xi / \ell$. The data are shown in Fig. VI-5.

The answer is immediate : the fast change of slope of the Hall conductivity is again visible, and is still associated to the vortex phase transition. Aside from this fact, it is also noteworthy to stress in light of Fig. VI-5 that the normal state conductivity, measured here at $T = 94$ K, is linear in the magnetic field and strictly positive, even though this temperature is only half a degree above $T_c$, therefore still in the fluctuation region of the superconducting transition.

Finally, from the similarity between measurements done at constant temperature and at constant magnetic field, we can conclude that the new observed behavior for the Hall conductivity is not caused by the temperature dependence of the ratio $\xi / \ell$, but is really *directly related to the vortex phase*.

### 1.3 Current dependence

All the results presented up to now for the vortex solid phase have been performed with the same current density, large enough to set the solid into motion. However, as we have seen in Fig. VI-1, the system is already non-linear close to, but above the melting, in the vortex liquid phase. Whereas Fig. VI-2 shows that the Hall conductivity in the liquid phase is not significantly altered by this non-linearity, we also have to study the behavior of the vortex solid, where the non-linearity is much stronger.

For this, we have measured the longitudinal and Hall resistivities at various different current densities. The results are presented in Fig. VI-6. As we have already stated at the beginning of the chapter, the behavior of $\rho_{xx}$ and $\rho_{xy}$ in the solid phase are very similar : both resistivities drop rather sharply at low current densities, but extend more and more into the solid phase as the current is increased.

More interesting is the Hall conductivity that can be calculated from these data, even though they are more noisy than the curves presented before. The result is shown in Fig. VI-7 for a selection of three different dc current densities. Obviously, the Hall conductivity is not at all linear in the vortex solid phase : the change of slope at the vortex phase transition $B_m$ is dramatic at low currents, but is smoother for larger currents. At $j = 200$ A/cm$^2$, the change is strongly reduced – though still present – at the melting field, but it appears, despite the rapidly increasing noise, that the conductivity decreases gradually faster at lower fields to finally approach a slope similar to that observed at smaller currents.

This strong current dependence suggests that *the observed change of the conductivity slope at the vortex phase transition is related to pinning forces*. The effect is indeed the most apparent at low currents, where the material defects have the largest impact on vortex motion. When the current rises, the flux flow becomes less influenced by pinning forces, and the conductivity retains a behavior more similar to the liquid state, until the magnetic field is reduced enough to let again the disorder dominate the vortex dynamics. We shall come back to this discussion later on.

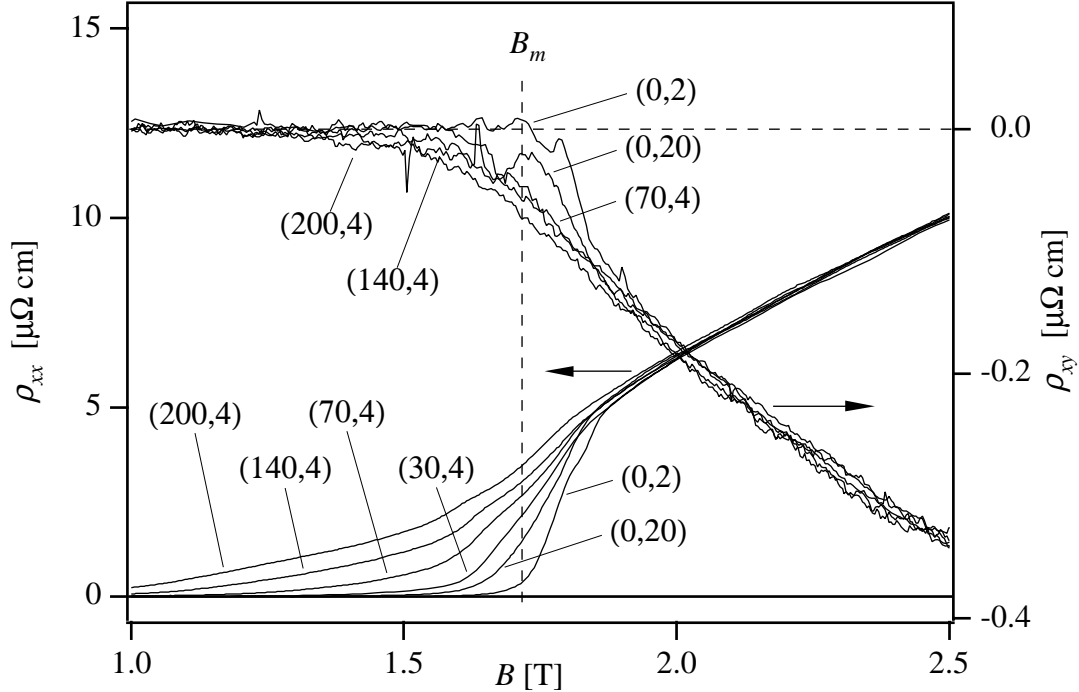

**Figure VI-6 :** Longitudinal resistivity $\rho_{xx}$ and Hall resistivity $\rho_{xy}$ at $T = 89$ K as a function of the magnetic field, for a set of different current densities. The currents are indicated in A/cm$^2$ following the notation $(j_{dc}, j_{ac})$, the measurements at non-zero dc currents being done with a single polarity. The field is inclined at $\alpha = 3°$ away from the *c*-axis.

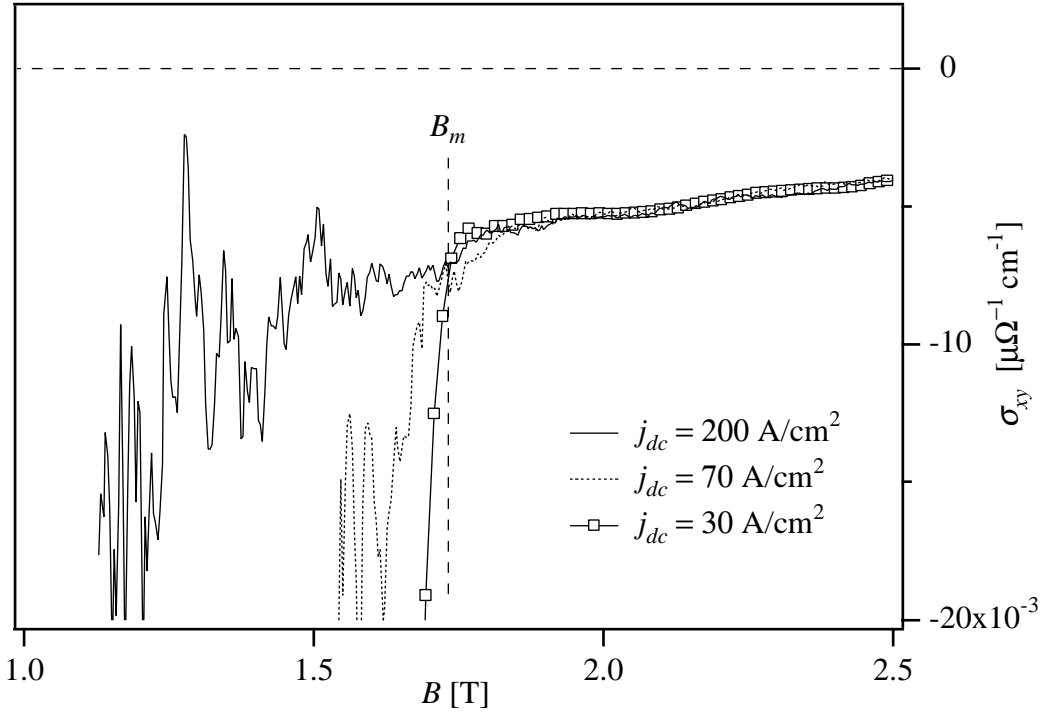

**Figure VI-7 :** Hall conductivity $\sigma_{xy}$ calculated from the data of Fig. VI-6, for three selected values of the dc current component (of single polarity). The ac component is always $j_{ac} = 4$ A/cm² pp, $T = 89$ K and $\alpha = 3°$.

We now turn to the scaling relation between the Hall and the longitudinal resistivities, more precisely to its current dependence. In Fig. VI-8 are plotted three curves from selected data of Fig. VI-6. Surprisingly, the large differences observed in Fig VI-7 are not apparent here : the scaling is systematically verified, still with the constant value of the exponent $\beta = 1.4$. Of course, this is not completely unexpected, since we had already seen in paragraph 1.1 that the scaling was continuous at the vortex phase transition for $j = 150$ A/cm². Therefore, if the scaling behavior was current dependent in the vortex solid (remaining current independent in the liquid, where both resistivities are linear), it would be very unlikely to have identical liquid and solid scaling laws precisely for this specific value of the current only.

Actually, this omnipresent scaling law and the peculiar value of the exponent $\beta$ brings us a new way to consider the data of Fig. VI-7. On the one hand, we know that the conductivity is very well approximated by $\sigma_{xy} \approx \rho_{xy} / \rho_{xx}^2$. On the other hand, we have checked that the scaling relation $|\rho_{xy}| \propto \rho_{xx}^\beta$ holds across the vortex phase transition with $\beta = 1.4$ for all of our measurements. Therefore, we get $\sigma_{xy} \propto \rho_{xx}^{-0.6}$ as a direct consequence of the general scaling relation. Provided that the drop to large negative values of the Hall conductivity at the vortex phase transition is very sharp at low currents, but almost disappears at high current density (see Fig. VI-7), it might be argued that it is only an extrinsic effect, a sort of pinning-induced artifact. However, we now see from the line of argument above that the sink of $\sigma_{xy}$ is no more an "artifact" than the sharp drop of $\rho_{xx}$ at the melting transition : both are indeed directly related to the vortex phase transition, and are simply similarly smoothed out at large current densities, as a direct consequence of the Hall scaling law binding them together. In some way, understanding the origin of the scaling law is a central challenge, the rest of the Hall behavior following from the longitudinal resistivity behavior. Of course, this does not mean that deriving the scaling law is the only theoretical difficulty, since the description of $\rho_{xx}$ is non-trivial in itself : to predict its behavior at the melting transition or in the vortex solid is a demanding, still open question in collective vortex dynamics.

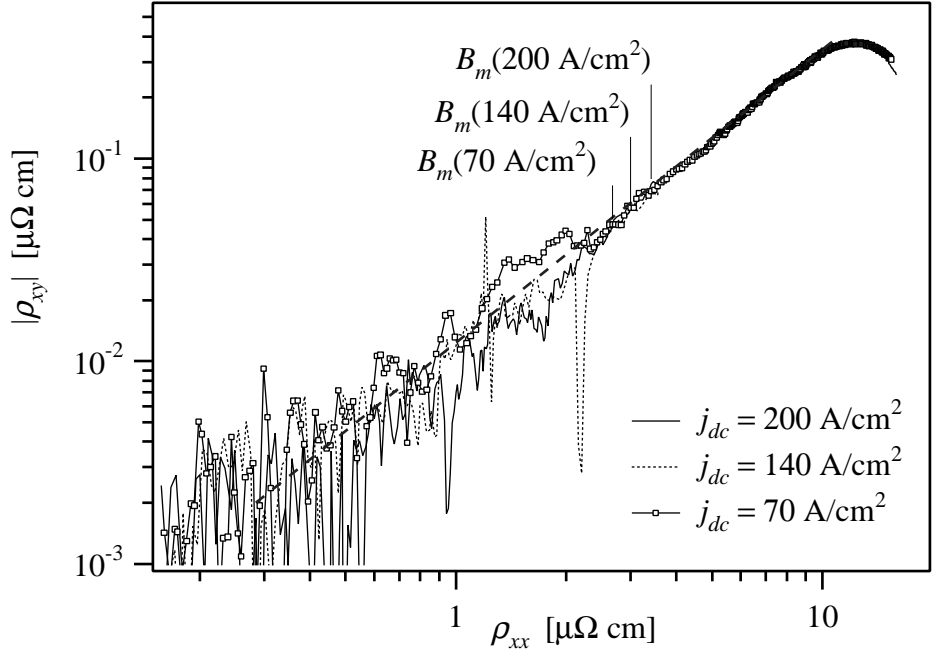

**Figure VI-8 :** Log-log plot of the absolute value of the Hall conductivity $\rho_{xy}$ as a function of the longitudinal resistivity $\rho_{xx}$ for $T = 89$ K, $j_{ac} = 4$ A/cm$^2$ pp and three different, indicated $j_{dc}$ (of single polarity). The magnetic field, inclined at $\alpha = 3°$, is the implicit running parameter for each of the curves. The power-law scaling relation $|\rho_{xy}| \propto \rho_{xx}^{\beta}$ is clearly apparent, and is underlined by a dashed straight line, corresponding to an exponent of $\beta = 1.4$.

The current dependence of $\sigma_{xy}$ brings us a second reason why this new behavior of the Hall conductivity at the vortex phase transition is rather hard to observe. Aside from the necessity to reach the large negative values of $\sigma_{xy}$ corresponding to the vortex lattice melting, as noted on page 90, it is also necessary to perform the Hall measurements at reasonably low current to observe the sharp slope change of the conductivity at the transition. Measuring at low current in the vortex solid means a very low signal, such that an extremely low noise level is absolutely essential for such work.

### 1.4 Field direction dependence

In paragraph 1.3 above, by changing the current density we saw that the Hall conductivity is pinning dependent. Another way to modify the pinning influence in our sample is to change the orientation of the magnetic field. As seen in the preceding chapter, the twin boundaries indeed have a strong impact on vortex dynamics, depending significantly on the angle between the magnetic field and the twin planes. Therefore, we give here a last series of measurements of the Hall effect for different field orientations $\alpha$.

The results presented in Fig. VI-9 allows us to compare the Hall behavior at $\alpha = 0°$, $4°$ and $7°$. Note that the data for $\alpha = 4°$ are just the same as those at high current of Fig. VI-1 and VI-2, as well as those labelled $B = 2$ T in Fig. VI-3 and VI-4.

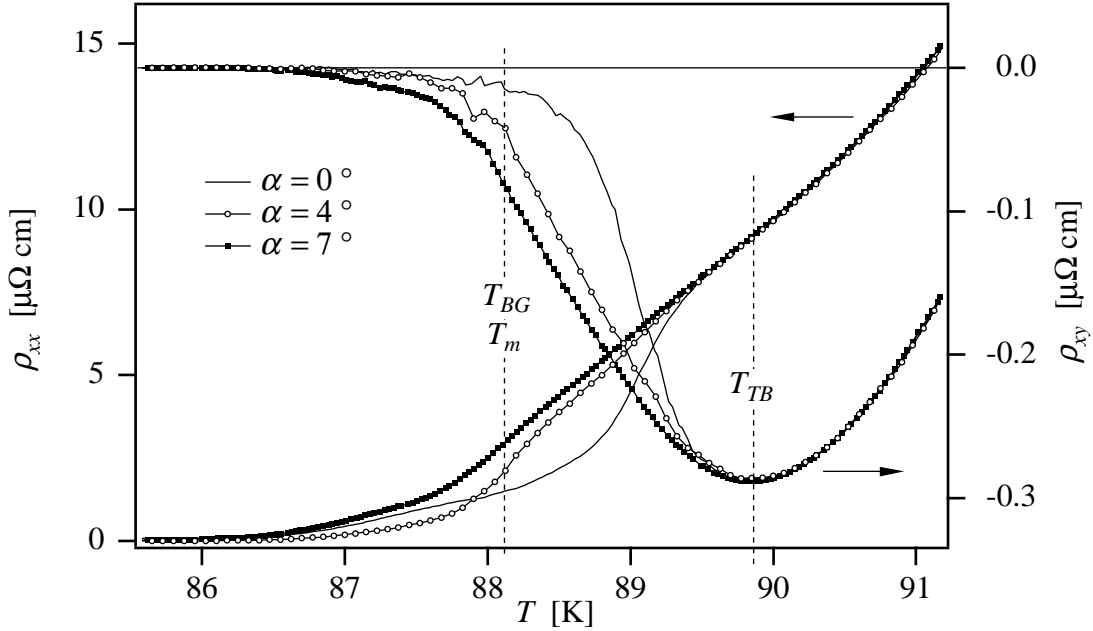

**Figure VI-9 :** Longitudinal resistivity $\rho_{xx}$ and Hall resistivity $\rho_{xy}$ as a function of the temperature, in a magnetic field of $B = 2$ T inclined from the $c$-axis by $\alpha = 0°$, $4°$ and $7°$. The curves for $\alpha = 0°$ and $7°$ are averages between data for both polarities of the current density, whose magnitude is $j = 150$ A/cm$^2$ dc $+ 50$ A/cm$^2$ pp. For $\alpha = 4°$, the current has the same value, but the measurement is done with only one polarity.

The picture formed by the $\rho_{xx}$ curves is similar to Fig. V-2 of the preceding chapter, where we have noted that when $\alpha = 4°$, the resistivity is higher in the liquid phase than in the $\alpha = 0°$ case (field along the twin planes), but is slightly reduced in the solid phase. Now we can add that for larger angles, the longitudinal resistivity is always larger than when $\alpha = 0°$, even in the vortex solid. From Fig. VI-9 we also immediately see that the Hall behavior does not follow the same trend : the relative positions of the different curves observed in the liquid state – which is the same as the corresponding order between the longitudinal resistivities – is preserved in the vortex solid phase : we always have an increasing Hall resistivity for increasing angles $\alpha$.

This difference in the angle dependence of both the longitudinal and Hall resistivities has a very important consequence on the Hall conductivity, shown in Fig. VI-10. At $\alpha = 4°$ and $7°$, the vortex liquid behavior is the same, and the curves spread at the vortex phase transition, with a sharp slope change for $\alpha = 4°$, and a much smoother deviation from the liquid trend at $\alpha = 7°$. This can be seen as the same pinning effect as in Fig. VI-7 : at large angles $\alpha$, the pinning of twin boundaries is reduced, such that the usual current density ($j = 150$ A/cm$^2$ dc $+ 50$ A/cm$^2$ pp) is more "efficient" than at lower angles, or in other words it is bigger with respect to the critical current, leading to a behavior close to the data for larger currents measured at $\alpha = 4°$ (see the curve at $j_{dc} = 200$ A/cm$^2$ in Fig. VI-7).

For $\alpha = 0°$, on the other hand, the conductivity starts to deviate from the other curves in the vortex liquid already, below $T_{TB}$ but clearly above $T_{BG}$. After a small bump below the usual $\sigma_{xy}$ curve (*i.e.*, at larger negative values), the Hall conductivity seems to reach a constant value in the vortex solid, at approximately $\sigma_{xy} = 6 \times 10^{-3}$ $\mu\Omega^{-1}$ cm$^{-1}$. This can in no way be interpreted following the same pinning argument as above : at $\alpha = 0°$, the twin planes influence should be the largest, and the conductivity should sink event faster than for $\alpha = 4°$.

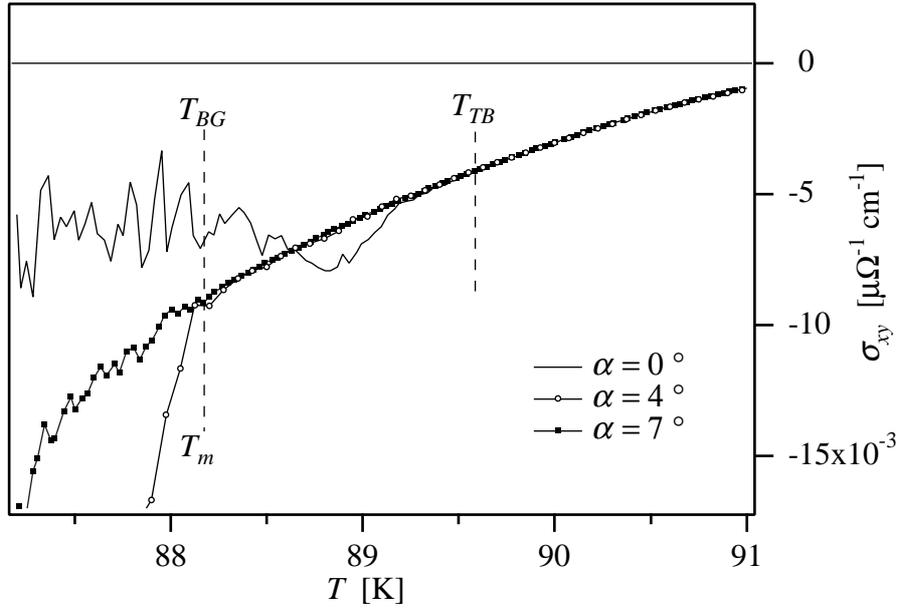

**Figure VI-10 :** Hall conductivity $\sigma_{xy}$ as a function of the temperature calculated from the data of Fig. VI-9. $B = 2$ T, $\alpha = 0°$, $4°$ and $7°$. The current is $j = 150$ A/cm$^2$ dc $+ 50$ A/cm$^2$ pp (both polarities for $\alpha = 0°$ and $7°$, single polarity for $\alpha = 4°$).

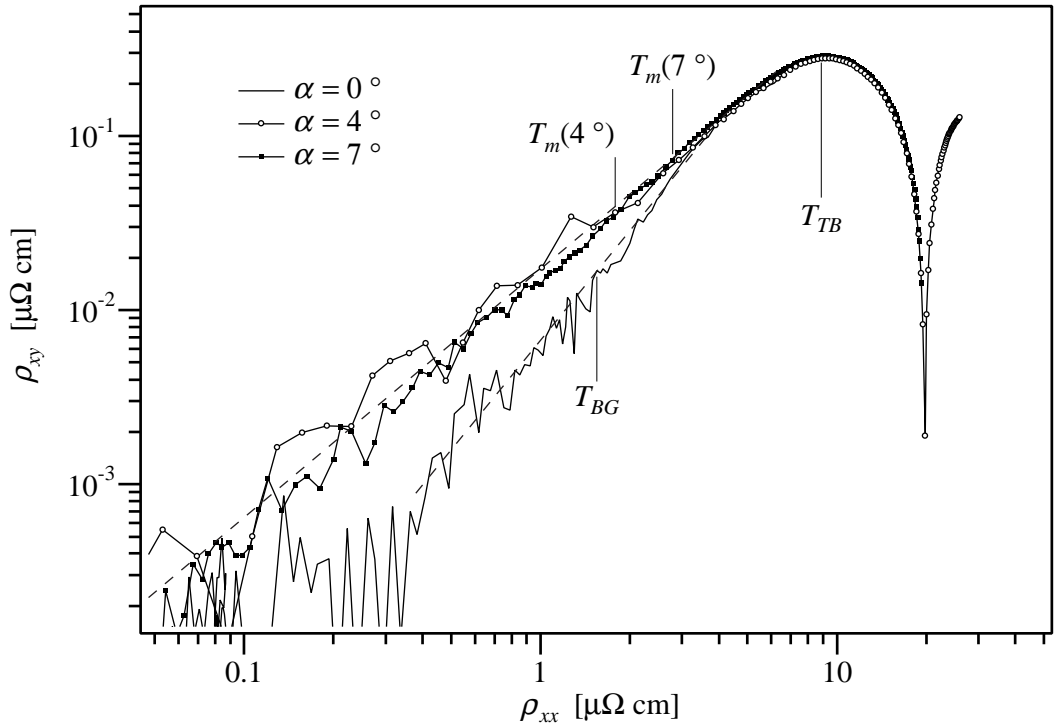

**Figure VI-11 :** Log-log plot of the absolute value of the Hall conductivity $\rho_{xy}$ as a function of the longitudinal resistivity $\rho_{xx}$, for a magnetic field of $B = 2$ T tilted away from the $c$-axis by different angles $\alpha$. The current is $j = 150$ A/cm$^2$ dc $+ 50$ A/cm$^2$ pp (both polarities for $\alpha = 0°$ and $7°$, single polarity for $\alpha = 4°$). The temperature is the implicit running parameter for each of the curves. The power-law scaling relation $|\rho_{xy}| \propto \rho_{xx}^{\beta}$ is verified with $\beta = 1.4$ for $\alpha = 4°$ and $7°$ and $\beta = 2$ for $\alpha = 0°$ (dashed lines).

# VI

The repercussion of this different Hall behavior for the Hall scaling law is illustrated in Fig. VI-11 : whereas the data for $\alpha = 7°$ still match the same scaling relation as all the measurements shown before, namely with an exponent $\beta = 1.4$, the curve corresponding to $\alpha = 0°$ clearly displays something different : it still scales down to very low values of the resistivity and *through the vortex phase transition*, but with a different slope on the log-log plot. Practically, the new value of the exponent is $\beta \approx 2.0$ (the curve fit more precisely leads to $\beta = 2.06$, meaning that $\sigma_{xy}$ possibly could even vanish at lower temperatures; however, this small nuance is beyond the noise level). Actually, this approximate value of $\beta$ could already be expected from the behavior of the Hall conductivity $\sigma_{xy}$ shown in Fig. VI-10 : since from the scaling law we can write $\sigma_{xy} \propto \rho_{xx}^{\beta-2}$, a constant conductivity can only correspond to $\beta = 2$ provided $\rho_{xx}$ is far from being constant, as it drops by more than one order of magnitude in the scaling region.

## 2. Discussion

### 2.1 Phenomenology of the results

As a first step in the discussion of our results, we will summarize the observed behavior for the Hall conductivity, and discuss it empirically. The link with the different theories for the Hall anomaly and the Hall scaling law will be given in the next paragraph.

The results of this chapter give a new insight into the low temperature / low magnetic field mixed state Hall behavior. Although they do not provide new information on the sign change of the Hall effect that happens higher in the vortex liquid, they include most of the rich physics characteristic of the mixed state : linear and non-linear conductivity, solid and liquid vortex phases, partial and "tunable" influence of twin boundaries, *etc*. The results around the vortex phase transition can be condensed into two main points :

- when the magnetic field is tilted away from the *c*-axis (and thus from the twin planes) by more than approximately $\alpha^* \approx 2°$ to $3°$, the Hall resistivity obeys the scaling law $|\rho_{xy}| \propto \rho_{xx}^{\beta}$ with $\beta = 1.4$ from about the negative maximum of $\rho_{xy}$ down to the vortex solid, through the vortex phase transition. As a consequence, the negative Hall conductivity diverges as the longitudinal resistivity $\rho_{xx}$ vanishes, since $\sigma_{xy} = \rho_{xy} / (\rho_{xx}^2 + \rho_{xy}^2) \approx \rho_{xy} / \rho_{xx}^2 \propto \rho_{xx}^{\beta-2} \propto \rho_{xx}^{-0.6}$.
- when the magnetic field is parallel to the twin planes ($\alpha = 0°$), the scaling law still holds in the non-linear region, namely slightly above and all the way below the vortex phase transition, but with $\beta \approx 2$. The Hall conductivity $\sigma_{xy}$ is then basically constant.

Since in the considered regime the Hall behavior is completely determined by the longitudinal resistivity through the scaling relation, we start by commenting on the scaling law. We first note that our value of $\beta = 1.4$ is very constant, in the sense that it is independent of the temperature, magnetic field intensity and orientation (for $\alpha > \alpha^*$), as well as current intensity, contrary to some measurements reported in the literature. Actually, as we have seen in chapter III, the few available experimental values of the exponent are quite scattered, and the general dependences on the above parameters is far from being clearly established. We believe that these disagreements are partly the consequence of a considerable sensitivity of the scaling relation to backgrounds and offsets in the measured resistivities : for example, a small, non-constant inductive contribution to the signal, which might be negligible for most of the measured data (*e.g.* in the vortex liquid), will have a strong influence on the scaling behavior, which shifts the focus on the low-level data (considered in logarithmic scale). However, note that our value of $\beta = 1.4$, even though very constant and reproducible, is still associated with a small fitting statistical error (of the order of $\pm 0.05$ to $\pm 0.1$) because of the noise level. Therefore, it can also be compatible with the (often reported) value of $\beta = 1.5$, which has the particularity to lead to a Hall conductivity expressed as a function of the longitudinal resistivity as

$$\sigma_{xy} \propto 1/\sqrt{\rho_{xx}}.$$

Another recurrent controversy in the experimental literature that can be addressed with our measurements is the pinning dependence of the Hall conductivity. As we have already noted on page 94 and in the results summary just above, this information is trivially obtained in the scaling regime : since $\rho_{xx}$ is obviously always pinning dependent, $\sigma_{xy}$ is pinning independent if and *only* if $\beta = 2$. In all other cases, $\sigma_{xy}$ depends on pinning through $\rho_{xx}$. In our sample, we therefore have a clear dependence of $\sigma_{xy}$ on pinning for tilted magnetic fields. On the other hand, we can assume from the scaling law that at $\alpha = 0°$ the Hall conductivity is pinning *independent*, even though we do not have directly measured $\sigma_{xy}$ at different currents for this field orientation.

Although the Hall scaling is valid almost up to $T_{TB}$ (or $B_{TB}$ for measurements performed as a function of the magnetic field), the pinning dependence revealed on $\rho_{xx}$ and $\rho_{xy}$ by changing the magnetic field orientation $\alpha$ (see Fig. VI-9), which is well marked far above $T_m$ (respectively $B_m$), is not apparent in the Hall conductivity, probably because the effect of the field rotation on $\sigma_{xy}$ is too small to be observed correctly. The influence on $\sigma_{xy}$ is visible only when $\rho_{xx}$ and $\rho_{xy}$ become non-linear and more temperature (or field) dependent, namely close to $T_m$ (or $B_m$). However, between $T_m$ and $T_{TB}$ (respectively $B_m$ and $B_{TB}$), $\beta$ is in clearly not close to 2, such that $\sigma_{xy}$ is rigorously pinning dependent (and thus $\alpha$ dependent) through $\rho_{xx}$. Therefore, in some sense, the scaling argument is much more sensitive to address the question of pinning dependence of the Hall conductivity than a direct measurement of this conductivity under different pinning conditions. For example, Samoilov [125] have concluded that $\sigma_{xy}$ is independent of pinning on the basis of data performed strictly in the linear regime. However, it is clear from their data that the Hall scaling leads to an exponent $\beta$ smaller than 2.

Aside from these considerations on scaling, the influence of twin boundaries should also be discussed. It might indeed be argued that the Hall effect in general, and notably the change of behavior at the vortex phase transition in our crystal is strongly influenced by the twins. However, a couple of arguments can cancel these doubts. First of all, as already noted several times in the literature, the Hall effect of both twinned and untwinned samples is very similar. This is confirmed by our results, which are in very good quantitative agreement with measurements in an untwinned crystal [127,51], where it was reported that no even Hall effect, namely no guided motion, was observable [127].

Even more convincingly, we can point out that the observed change of behavior of $\sigma_{xy}$ at the vortex phase transition is very sharp, whereas we have experimentally shown in chapter V that the onset of guided motion is smooth, and starts already deep in liquid. From this we can affirm that the (odd) Hall effect is not affected by guided motion in our twinned sample; in this sense, the statement we have made at the beginning of this chapter, namely the fact that extracting the antisymmetric part of the transverse resistivity upon magnetic field reversal allows us to eliminate the influence of guided motion, is justified *a posteriori*. The reason for this is that, as we have seen in chapter V, there is probably no *absolute* guided motion, even in the vortex solid, for the orientation of current we have used here (at 45 ° from the twin planes, namely at $\beta_j = 0$ ° to follow the notation of the preceding chapter). Therefore, as soon as the vortices leave the twin planes to achieve a trajectory not strictly parallel to them, the overall motion can be influenced by the intrinsic bulk Hall angle.

Of course, if the pinning produced by the twin boundaries is stronger (in dirtier samples, for example), the guided motion can have a much stronger influence, and even dominate the antisymmetric transverse response. To illustrate this, we recall that Morgoon *et al.* [49] have measured the Hall effect in three different unidirectionally twinned Y:123 films, each of them having a different direction of the current with respect to the twin planes. It is very interesting to first note that in one of their samples (labelled S#1 in Ref. [49]), in which the current is at −60 ° from the twin planes (i.e. $\beta_j = 75$ ° in our notation), they actually measure a sharp change of slope in the Hall conductivity, just like we report in the present chapter for $\beta_j = 0$ °. However, their data for the longitudinal resistivity do not show a corresponding sharp step indicating the simultaneous occurrence of a vortex phase transition (the first oder lattice melting is not observed in film samples).

Moreover, the interpretation of this behavior is all the more delicate since the two other samples (S#2 and S#3) have a very different Hall conductivity. Although the sample S#3 has an acceptable behavior, since its conductivity is qualitatively[1] similar to that of S#1, the last sample (S#2) presents rather puzzling data. Even though the current orientation is not fundamentally different from that of S#1 (it is at 45 ° from the twin planes, namely at $\beta_j = 0$ °), the Hall conductivity indeed shows two additional sign reversals in the low field regime, separated by a sharp and very large positive peak. Since this behavior has never been reported by other authors for similar samples (and for the same current orientation, which is most often used), and since this difference between S#1 and S#2 is too important to be explained by the minor change of current orientation with respect to the twin planes, we conclude that these observations are rather the consequence of a sample dependence than the result of the different current orientations. In some sense, this shows how interesting it might be to improve our eight terminal technique to extract the Hall effect for any current orientation *in the same sample* (see discussion on page 85).

Finally, we turn to the Hall behavior for the magnetic field oriented along the twin planes ($\alpha = 0$ °). Whereas we have noted that the Hall scaling exponent $\beta$ is independent on the temperature and the magnetic field, we have seen that it depends strongly on the field orientation $\alpha$ (Fig. VI-11). This indicates that the scaling law is disorder-type dependent, with $\beta = 1.4$ for point disorder and $\beta = 2$ for correlated disorder (see discussion in paragraph 2.2), when vortices are localized inside the twin boundaries.

We have noted in chapter IV (see Fig. IV-18 and discussion on page 68) that the vortex solid phase is presumably a vortex lattice (or a Bragg glass) for $\alpha > \alpha^* \approx 2$ ° to 3 °, and a Bose glass when $\alpha < \alpha^*$. Therefore, the corresponding change in the scaling exponent $\beta$ might be related to this change of the nature of the solid phase. It is worth pointing out that in the original description of the Bose glass [23], the Bose glass transition is indeed expected to have a different universality class of critical exponents than the vortex glass from Fisher *et al.* [22]. However, it is still not yet clear whether there is a direct link between the critical exponents of the transition from the vortex liquid to solid and the exponent of the Hall scaling law (see discussion on page 40), as was proposed by Dorsey *et al.* [86]. Another similar development in terms of critical exponents, this time in the frame of percolative processes, will also be given further on (see next paragraph, page 104).

---

1. the particular current orientation, which is just orthogonal to the twin boundaries, $\beta_j = 45$ °, can justify the quantitative difference

In any case, the very different behavior of the Hall conductivity for $\alpha > \alpha^*$ and $\alpha < \alpha^*$ allows us to confirm that the Bose glass and the vortex lattice phases are not only different from a configurational (static) point of view, but also have very different dynamic properties. Further, the value of $\beta = 2$ for $\alpha = 0°$ have led us to state that the Hall conductivity is independent of disorder. Moreover, it is important to stress that *this is the first time that an exponent of $\beta = 2$ is precisely shown to correspond to a constant Hall conductivity $\sigma_{xy}$*. Although most of the reports of the same value of the exponent do not show the corresponding conductivity, a few works report both a diverging conductivity and a contradicting exponent of $\beta = 2$. In this case, the explanation can probably be found in a small though significant error in the numerical estimation of $\beta$ (see the comment about Ref. [120] on page 45). In this sense, our simultaneous report of $\beta = 2$ and a constant Hall conductivity in the same regime shows that in our case the exponent is *accurately* equal to 2. It would be very interesting to know if these properties are intrinsic attributes of the Bose glass in general. This is of course an open question.

As a last comment, we also want to emphasize that the strong dependence of the exponent on the spatial correlation of the defects together with the magnetic field orientation can explain at least a part of the scattering in the experimental determinations of $\beta$. This also leads us to a remark on Ref. [124], in which twinned Y:123 crystals are irradiated to different doses, and the scaling exponent is measured at different magnetic fields for each irradiation level. The authors then report that $\beta$ is found to be 1.5 for all the irradiated samples, but is 2 for almost all the data of the unirradiated crystal (except the measurement at $B = 1$ T, for which they also find $\beta = 1.5$). This is at first sight very surprising, since this behavior is exactly opposite to our own findings : we have $\beta = 2$ when correlated disorder (twins in our sample) is relevant, and $\beta = 1.4$ in the other case. We can tentatively explain this striking difference by first noting that the samples in Ref. [124] are indeed also twinned, and that, as mentioned in the text, the irradiation beam "was aligned approximately parallel to the *c*-axis" [124]. Therefore, it is quite probable that the disorder is actually *more* "correlated" in the *non*-irradiated sample, where they have performed the measurements with the field well aligned with the twin planes, leading to $\beta = 2$, just like in our sample, than in the irradiated samples which have columnar defects at an angle from the twin planes, leading to a so-called "splayed defects" configuration, in which the vortices are much more entangled and can have very different properties [160]. It is possible that the measurement at $B = 1$ T has been done in a slightly different orientation, at $\alpha > \alpha^*$, which is of the order of the sample alignment accuracy for the usual geometry of cryostats, as we have noted at the very beginning of chapter IV. This hypothesis, if verified, is very interesting, since it would then suggest that a vortex glass in the case of splayed defects has the same Hall scaling exponent as the vortex solid in a twinned crystal in inclined magnetic fields. The corresponding theories (Ref. [160] and [155], respectively) have to be completed before they can bring an answer to this question. The two families of twin boundaries could event actually play the same role for inclined field as the two directions of columnar defects considered in splayed disorder models. Even though these comments are rather speculative, a systematic experimental investigation of these types of samples would be informative.

## 2.2 Theoretical implications

We will now discuss the results reported in this chapter more specifically from the theoretical point of view. However, we first have to note that, as we have already mentioned at the beginning of the preceding paragraph, our data are more concerned with the relation between the vortex phases and the Hall effect, focusing on the vortex liquid to solid phase transition, than on the Hall anomaly itself, namely the Hall sign reversal. As a consequence, the theories presented for the Hall anomaly in chapter III are not really adequate, since they do not take into account the vortex phase, nor even vortex-vortex interactions in its full complexity. The difficulty is even increased by the fact that even the vortex lattice melting itself has no satisfying microscopic explanation at the present time. Moreover, the new behaviors of the Hall conductivity that we have seen here, namely a sharp change in its slope at the vortex lattice melting transition, as well as a constant conductivity in (and slightly above) the Bose glass phase, occur in the regime of the Hall resistivity scaling, which is itself not convincingly predicted either.

To start with, we want to point out that in a recent comment on the present work [161], Ao sees the vortex solid Hall behavior as the verification of the vortex lattice defects model presented in chapter II, paragraph 3.2 (see also Ref. [64,65]). Of course, we agree with some of his arguments. First, the pinning certainly plays a dominant role on the Hall conductivity in the vortex solid. Second, since the collective pinning, in close relation with vortex-vortex interactions, is well known to be necessary to successfully describe the longitudinal flux flow resistivity in presence of weak disorder, it is reasonable to expect the same interactions to be required for a correct description of the flux flow Hall component of the resistivity under the same conditions. However, we are not absolutely convinced that our experimental data (and more generally all the phenomenology of the Hall anomaly) is completely described by this theory. Since it is based on the independent motion of vacancies in an otherwise rigidly pinned vortex lattice as the source of the negative Hall effect, it is very difficult to see how it might explain, for example, the presence of a sign reversal extremely deep in the vortex liquid, very close to $T_c$, as it is most often observed. This theory is, indeed, a purely mean field hydrodynamic model, based on a single vortex equation of motion.

On the other hand, the consideration of this model in the vortex solid phase, where the mean field description is probably closer to the reality (the fluctuations are then greatly reduced) should be done more carefully. There could be a sudden enhanced contribution of the vortex lattice vacancies, when reaching the vortex solid phase, increasing the absolute value of the Hall conductivity. Unfortunately, since the magnitude of the Hall effect is fundamentally determined by the lattice defect density, it is difficult to imagine why the vortex solid would actually have more defects than the liquid disordered phase. Even the transition between the vortex liquid and the vortex solid brings forward a major problem. To explain the role of vacancies in the disordered solid, Ao assumes the existence of "local crystalline structures, like lattice domains" in the vortex assembly already above the vortex phase transition. But, if this transition then becomes the progressive ordering, or growth, of these crystallites, it is doubtful that the resulting phase transition can be of first order.

Moreover, note that this theory also fails to account for the doping dependence of the flux flow Hall effect, which apparently determines the Hall effect sign, independent of the vortex phase. Finally, the picture of motion of independent vacancies in the immobile lattice is in any case in disagreement with the well accepted[1] notion of a vortex solid moving by plastic motion, more resembling avalanches along specific channels.

We now quickly address the other models for the Hall anomaly presented in chapter II. For example, the vortex charge model is not compatible with the observed influence of current in our data : since in this model the extrapolation of the zero temperature Hall conductivity $\sigma_{xy}(T \to 0)$ should provide information on the charge carrier density [43], the strong current dependence of the Hall conductivity slope in the vortex solid would lead to a carrier density varying over a couple of orders of magnitude in a moderate current range, which is not acceptable, confirming that the observed effect is rather related to the pinning.

---

1. and observed, *e.g.* through noise measurements

Therefore, the pinning related model of Wang *et al.* [57] (page 28) is *a priori* an interesting theory. For example, the change of behavior at the vortex phase transition might be explained through this theory by a sudden increase of the critical current (namely of the effective pinning force), in very good agreement with experimental facts. However, it is not clear whether this model can account for the observed disorder-type dependence of the Hall scaling exponent. Similar to Ao's scenario of lattice defects motion, this model is not compatible with the electronic doping dependence, nor with critical fluctuations close to $T_c$, which apparently also play a role in the Hall effect. We shall come back to this model in the conclusion (next chapter), but we already can see that a really complete description of the mixed state Hall effect necessarily has to include *both* fluctuations and pinning interactions, as well as presumably take into account vortex-vortex interactions, to explain the influence of the vortex phase observed in the present work.

For this reason, the most promising rigorous model probably lies in a time dependent Ginzburg-Landau (TDGL) approach including pinning interactions, as proposed by Ikeda. In a first model in some way in the continuation of the vortex glass theory from Fisher *et al.* [22], he considers the effect of point-like disorder on the Hall conductivity [72]. Thanks to his perturbative approach of fluctuations in the lowest Landau level (LLL) approximation, he can then account for the vortex phase transition (called the vortex glass transition). Note that, even though this model is a very complete and rigorous approach to the Ginzburg-Landau equations, it is not in itself a truly microscopic theory, since the complex part of the relaxation time, notably determining the Hall effect, is a parameter of the model that still has to be calculated on a microscopic basis.

Although the formalism of this theory is very exacting and demanding, we can summarize some important points concerning the Hall effect in a few words. The idea is to decompose the superconducting part of the Hall conductivity $\sigma_{xy}^s$ (Eq. (I.3) on page 10) into three terms :

$$\sigma_{xy}^s = \sigma_{xy}^{mf} + \sigma_{xy}^{vg} + \delta\sigma_{xy}$$

where $\sigma_{xy}^{mf}$ is the mean field contribution, reducing to the vortex flow deep in the liquid regime, $\sigma_{xy}^{vg}$ is precisely the term of Gaussian fluctuations related to the vortex glass order and the associated transition, and $\delta\sigma_{xy}$ includes other pinning-related perturbative terms due to superconducting amplitude fluctuations (also called critical fluctuations).

The first contribution $\sigma_{xy}^{mf}$ has a form similar to the previous TDGL derivations (see chapter III). However, in two dimensions (in layered systems), the vortex glass fluctuation part $\sigma_{xy}^{vg}$ is shown to be *opposite* to $\sigma_{xy}^{mf}$, and depend on the pinning strength [72]. Therefore, if the imaginary part of the relaxation time is such that $\sigma_{xy}^{mf}$ is itself opposite to the normal state Hall effect, the total Hall conductivity can undergo two successive sign reversals in these systems, in agreement with the experimental observations. Similarly, $\delta\sigma_{xy}$ is also proportional to the pinning strength, though with the same sign as the mean field contribution $\sigma_{xy}^{mf}$. With this result, Ikeda can qualitatively explain why some authors experimentally report a pinning independent Hall conductivity : it is, according to him, the result of the competition between $\sigma_{xy}^{vg}$ and $\delta\sigma_{xy}$. In three dimensions, the calculation seems to be more delicate, but apparently all three terms have the same sign in this model for point-like defects, therefore explaining the presence of only one sign change in the less anisotropic systems. Note that the pinning strength dependence of both $\sigma_{xy}^{vg}$ and $\delta\sigma_{xy}$ also explains why $|\sigma_{xy}|$ is increased as the pinning starts to dominate, yielding a Hall conductivity which diverges faster than $1/B$, as is often observed experimentally [107,108,119,159].

# VI

Very recently, in response to the experimental results presented in this chapter, Ikeda proposed a new, more complete version of his theory, including both point-like and line-like disorder [73,74]. The main conclusion is that the line-like defects play, in some sense, the same role as the two dimensional fluctuations, again giving a Hall effect contribution opposite to the mean field term, whereas the point-like disorder (in the tridimensional case that we consider for our Y:123 sample) still leads to a strictly negative superconducting part of the Hall conductivity.

As a consequence, our observation of a constant Hall conductivity when the magnetic field is aligned with the twin boundaries would reflect the competition between line-like and point-like disorder. However, the behavior deeper in the vortex solid is not yet clear, since it is apparently not straightforward whether or not the Gaussian fluctuation part of the conductivity $\sigma_{xy}^{vg}$ eventually diverges near the phase transition.

Unfortunately, due to its complexity, Ikeda's theory is still essentially qualitative, and many more analytical developments are required. For example, a precise prediction of the Hall scaling exponent in different conditions is still lacking.

Finally, we would like to present an alternative, much more empirical and phenomenological explanation for the scaling behavior that we observe. In the case of a mixed metallic/insulating system, the conductivity is governed by percolation processes. The corresponding theory leads to a longitudinal conductivity expressed as $\sigma_{xx} \propto \delta p^t$, where $\delta p = p - p_c$ is the difference between the conducting metallic phase density $p$ and the critical percolation threshold $p_c$ of this density. The critical exponent is reported to be $t \approx 1.3$ in two dimensions, and $t \approx 1.6$ in three dimensions [162]. Similarly, the Hall number $R_H$ diverges as $R_H \propto \delta p^{-g}$, where $g = v(d-2)$ [163], where $d$ is the dimension. As a consequence, we see that $g$ is obviously zero in two dimensions, such that the Hall number is unchanged by the percolation process. Therefore, the Hall conductivity $\sigma_{xy} \approx H R_H \sigma_{xx}^2$ is exactly proportional to $\sigma_{xx}^2$ (at constant magnetic field). In three dimensions, $g = v \approx 0.9$ [162], such that from the above relations $\sigma_{xy} \propto \sigma_{xx}^{2-g/t} \propto \sigma_{xx}^{1.44}$.

From this point, Geshkenbein [164] has proposed that if one views the vortex freezing as an inhomogeneous, non-simultaneous process, with regions where vortices are pinned (thus with vanishing resistivity), and others where they can still move, inducing a non-zero electric resistivity, the behavior should have many analogies with the usual percolative transition in inhomogeneous conductors [162,163]. Nonetheless, we have to be aware that the above model for the percolation of charge carriers across an inhomogeneous conductor can apply to moving vortices in inhomogeneous pinning, provided we finally interpret the vortex *conductivity* (the values for $\sigma$ above) as the electric *resistivity*, since a high vortex mobility means large electric dissipations. Therefore, the relations for $\sigma_{xy}$ above would become scaling relations for the Hall resistivity.

In the latter case, the dimensionality would be determined either by the intrinsic anisotropy of the material (like the usual assignation of a two dimensional nature to the most strongly layered compounds), or by the vortex localization along correlated defects, suppressing one vortex degree of freedom, resulting in a geometry therefore two dimensional-like, explaining our change of scaling exponent when the vortices are oriented along the twin planes. It is then impressive to see how well the exponents from this percolation picture are in agreement with our data : one indeed finds $\beta \approx 2$ in two dimensions (correlated disorder), and $\beta \approx 1.4$–$1.5$ in three dimensions (point-like disorder). We also see how the introduction of splayed defects, allowing for a strong entanglement of the vortex phase, therefore bringing back the third degree of freedom, can explain the reduction of the exponent from 2 to 1.5 (see comment about Ref. [124] on page 101).

## 3. Conclusion

We have studied here the Hall effect for a current oriented at 45 ° from the twin boundaries. Thanks to the very low noise level, we have measured the Hall conductivity down to large negative values not yet reported for $YBa_2Cu_3O_{7-\delta}$, about ten times larger than the level corresponding to the vortex lattice melting. We can therefore compare the vortex liquid and solid Hall behaviors. When the magnetic field is tilted away from the twin planes, we see that the Hall conductivity as a function of both the temperature and the magnetic field is much steeper in the vortex solid phase than in the liquid. On the other hand, the scaling power law for the Hall resistivity is absolutely not affected by the vortex phase transition. We exclude an influence of vortex guided motion for the sharp change of slope of the conductivity, since the twin boundaries were shown to affect the electric response only very smoothly and progressively, without any sharp signature of the vortex phase transition (preceding chapter).

The slope of the conductivity is the largest for low currents, revealing the major role played by the pinning dependence in this new effect. More generally, we insist on the fact that the Hall resistivity scaling exponent of $\beta = 1.4$ necessarily implies that the Hall conductivity is anyway pinning dependent. Therefore, even if it is very tempting to avoid the complexity of pinning in the description of vortex dynamics, the disorder has to be taken into account for the mixed state Hall effect. Such a description can be either done in the frame of a complete time dependent Ginzburg-Landau model including fluctuations and pinning, as proposed by Ikeda, or, if a mean-field description is considered as acceptable, with a simpler hydrodynamic single vortex equation of motion, of course also including pinning, as developed by Wang *et al*.

When the magnetic field is applied parallel to the twin planes, the scaling exponent is then $\beta = 2$. We also show for the first time that this particular value corresponds to a constant Hall conductivity, as should indeed follow from its definition. We also insist on the fact that only this value can lead to a pinning *independent* Hall conductivity.

Considering the difficulty to accurately align the magnetic field (or a particle beam for sample irradiation) with the crystallographic *c*-axis, and consequently to the very sensitive dependence of the Hall behavior to rotations by angles as low as 2 °, we believe that alignment errors can explain an important part of the discrepancy between the various existing data for irradiated and twinned samples.

Finally, we recall that we propose here a new phenomenological model for the Hall resistivity scaling law, suggested by Geshkenbein (page 104). Based on an analogy with the percolative description of a conducting to insulating transition, it is meant to be used very generally in presence of pinning. Describing the progressive transition to a frozen vortex assembly, it considers the system as a random distribution of domains in which the vortices are pinned, providing a vanishing electric resistivity, the remaining volume being dissipative. Using the universal critical exponents for the percolation processes, we find a Hall scaling exponent of $\beta = 2$ at constant magnetic field in two dimension, and $\beta \approx 1.4 - 1.5$ in three dimensions, in good agreement with our data. Note that the dimensionality has, here, to be understood in terms of degrees of freedom for the vortices : in the presence of unidirectionally strong correlated disorder, the vortices are straight and behave two-dimensionally; for point-like disorder or splayed correlated defects, the system is three-dimensional.

This model is very interesting, since it is independent of the vortex phase : it simply describes the scaling behavior as the resistivity vanishes (that is as the vortices become pinned). This is again in good agreement with the fact that the scaling regime starts roughly where the pinning of twin boundaries is seen to first have an influence on the resistivity, that is at $T = T_{TB}$ (see Fig. VI-9 and VI-11). Note that another explanation of the scaling law in terms of universal critical exponents had been given before (see page 39). However, two problems were emerging in this case. First, this model was predicting only one single universal exponent, in disagreement with the rather scattered data of the literature. In this sense, the percolation picture is more acceptable, since the two predicted values indeed correspond to the most often reported experimental estimations. Secondly, it is related to the vortex phase transition, such that the shift of the conductivity divergence to lower temperatures at higher currents is not well accounted for.

# CHAPTER VII  *CONCLUSIONS*

In the present work, we have experimentally investigated vortex dynamics in a twinned high quality YBa$_2$Cu$_3$O$_{7-\delta}$ single crystal by way of resistivity measurements. The main objective was to study the mixed state Hall effect in the vortex solid phase. Thanks to the nine contact pattern deposited on the sample surface, the guided motion of vortices along twin boundaries has also been studied as a function of the current direction. Since the voltage signal is much smaller in the vortex solid than in the liquid phase, very special attention has been paid to the data processing in order to avoid any undesired, non-intrinsic contribution. For example, the resistivity is obtained from the measured total impedance after the subtraction of a complex inductive background. Moreover, the orientation of the magnetic field with respect to the sample *c*-axis is controlled with a high accuracy. All these steps are absolutely determinant for the study of the above mentioned effects.

In the considered sample, the vortex phase transition is clearly observable through the resistivity measurements. It is apparently of first order when the magnetic field is sufficiently inclined from the twin planes (by more than $\alpha^* \approx 2°$ to $3°$), revealing the transition from a vortex liquid to a vortex lattice or a Bragg glass. When the vortices are parallel to the twin boundaries, the solid phase is instead shown to be a Bose glass, as already reported by other authors [25].

The investigation of guided motion has shown us that one twin family manifestly dominates the vortex dynamics. We have also seen that when the vortices are aligned to the twin planes, the longitudinal resistivity is not reduced mainly because of a reduction of the vortex mobility, but rather as a consequence of the rotation of the electric field away from the current direction, lowering its projection in the direction of measurement.

Another important result with respect to Hall measurements is that, as soon as the Lorentz force is not along the twin boundaries, the vortex motion is apparently never strictly channeled within the twin planes, a lateral component of the vortex velocity always remains. Therefore, thanks to this lateral motion, the Hall effect is not inhibited by guided motion. Before we further specifically comment on our results for the Hall effect, we first give a brief general synthesis of the different theories that we believe might bring conclusive explanations for the mixed state Hall effect.

Considering the various dependences of the Hall anomaly, which indeed depends on the nature of the compound and its doping, or on the orientation of the current with respect to the crystallographic axis, for example, its explanation should be sought in microscopic processes, such as the localized vortex core states, the influence of the *d* symmetry of pairing, the existence of preformed pairs, *etc*. In order to rigorously reveal the role of vortex-vortex interactions, as well as the interplay be-tween the pinning and the fluctuations, the approach the most likely to succeed is the use of the time-dependent Ginzburg-Landau equations, including all of these effects, as has been proposed by Ikeda [72-74]. In this theory, the various microscopic processes can by included through a key parameter, like the imaginary part of the relaxation time.

However, since this complete formalism is rather demanding, a more phenomenological simplified model would be welcome, for example, in a regime in which the mean field description is a good approximation, that is when fluctuations can be reasonably neglected (at low enough fields or temperatures). In this case, a hydrodynamic approach of the vortex equation of motion, of course taking the pinning into account (which will always be dominant in this regime) can be very useful. The question is then to know whether a single vortex equation of motion may contain all of the relevant physics, or if a many-body theory should necessarily be considered. If we assume that the pinning force already reflects the collective behavior of the vortex assembly (through the elastic constants of the vortex lattice, for example), it is then possible that a single vortex model including this pinning force as a parameter, as proposed by Wang *et al.* [57], would be satisfying in the appropriate regime.

On the other hand, the Hall scaling relation, which is much more general, is hence probably not related to such specific microscopic quantities. It is indeed observed in any compound, for any doping (that is, even if the Hall anomaly is not present). Since the scaling is verified in the regime where the resistivity vanishes, the pinning is certainly a key factor in this behavior. The best model, in this sense, is the percolation description of the vortex freezing that is suggested here, following an original idea from Geshkenbein [164], since it correctly predicts two different possible exponents ($\beta = 2$ and $\beta \approx 1.4 - 1.5$) depending on the dimensionality, independent of the vortex phase, current density or any material specific parameter. Such a very phenomenological and general model is certainly required considering the omnipresence of scaling.

Turning back to our results, we note that the scaling behavior that we report for the vanishing part of the Hall resistivity is in excellent agreement with the percolation picture : we find $\beta \approx 1.4$ when the point-like defects are dominant (that is when the magnetic field is inclined away from the twin planes), for any current density, magnetic field, temperature, and in the vortex liquid and solid alike. When the magnetic field is exactly parallel to the twin boundaries, presumably localizing the vortices along them, the exponent becomes $\beta = 2$ (as confirmed by the constant Hall conductivity).

Aside from that, the sharp change of the Hall conductivity at the vortex phase transition in inclined fields, together with its magnetic field direction and current dependences, show that the Hall conductivity is unambiguously and strongly pinning dependent, in contrast to often reported contrary assertions. Note that, since the Hall conductivity can be expressed from the scaling relation approximately as $\sigma_{xy} \propto 1/\sqrt{\rho_{xx}}$, the sharp change of $\sigma_{xy}$ at the vortex lattice melting can be phenomenologically explained by the corresponding change of the longitudinal resistivity. However, if we microscopically interpret the Hall conductivity in terms of the transverse force acting on vortices, its vortex phase dependence certainly is a challenge to the theories mentioned above.

In the direct continuation of this work, many points still have to be investigated. For example, we have seen that the Hall conductivity becomes steeper and steeper in the vortex solid as the magnetic field is progressively (but not completely) aligned with the twin boundaries. However, when the alignment is perfect, the Hall conductivity is then constant, a behavior that is not at all a continuation of the above mentioned trend. Therefore, it would be very interesting to further explore this angle dependence (for $0° < \alpha < 4°$) to see how the transition between these two regimes occurs.

Moreover, the fact that the Hall conductivity is independent of the temperature, when the vortices are aligned with the twin planes, is perfectly explained by the percolation theory for a two dimensional case. We have however noted that in this model the Hall conductivity, although indeed independent of the temperature, is proportional to the magnetic field. It would be very nice to confirm this prediction by complementary measurements as a function of the magnetic field, or as a function of the temperature but for different magnetic fields.

More generally, we recall that the main originality of this work, from the experimental point of view, is the use a multicontact configuration allowing us to apply a current in any direction in the sample plane, and to measure the magnitude as well as the direction of the resulting electric field. We would like to point out that this technique can also be used to address several other major open questions in vortex dynamics.

Obviously, the method is particularly adapted to the investigation of extended defects providing anisotropic pinning, such as twin boundaries. First, the effects of these twins on the vortex dynamics can be studied as a function of the current direction in much more detail. The advantage of this experimental setup compared to other studies is that all of the investigations can be done in the same sample, avoiding sample dependent effects, quite unavoidable when dealing with twinned samples. One can then, for example, address more accurately the magnetic field orientation dependence of the vortex mobility, by skirting the problems of the projection of the electric field on the current direction, which are possibly not parallel to each other as a results of guided motion. Another very important possibility provided by this method is the check of the still debatable efficiency of sample detwinning [156,157]. A residual anisotropy of the resistivity in a direction corresponding to twin planes would indeed mean that the twin boundaries are still present, even if the twins are not visible anymore, for example because one twin family is largely dominating, the domains of the other being too small to be optically observed.

For other types of intrinsic effects that are expected to have a much smaller amplitude, naturally untwinned samples are required. As proposed by Vicente Alvarez *et al*. [165], a similar setup can, for example, be used to study the intrinsic current orientation dependence of the Hall effect that they have predicted for *d*-wave superconductors. The dynamics of the vortex solid phase can also be investigated with a rotating current density, to try to reveal the stronger non-linearity expected for the transverse response of a Bragg glass.

# VII

## *REMERCIEMENTS*

After four years of enriching work, I will keep many memories of people to whom I am indebted. First, I have to acknowledge Prof. W. Benoit, for letting me participate in the activities of the vortex matter section of his group. This project was actually composed of two successive periods, the first supervised by M.V. Indenbom and C.J. van der Beek, who were later replaced by G. D'Anna. In order to avoid making anyone jealous, I can affirm without hesitation that it was the same pleasure to work with each of them. Their heartwarming enthusiasm, sincere encouragements and their communicative interest were an unforgettable experience. For a long time I will remember this motivating supervision and the team spirit it has brought...

Apart from that, I have been extremely lucky to get really invaluable help from László Forró. His truly generous contribution to this work was absolutely essential, as he was the one who could connect microwires to the electrical contacts on the sample surface, with a watchmaker's accuracy that only he could achieve. He even repeated this exploit a couple of times, always so obligingly and efficiently, when some of the wires became loose. László, thank you very very much !

The gold contacts were evaporated with kind assistance from Michel Schaer, and Andreas Erb provided the very high quality sample. I also acknowledge Gérard Gremaud and Alessandro Ichino for electronic designs and Bernard Guisolan for the development of mechanical parts.

Finally, I thank Andreas Erb, Ping Ao, Ryusuke Ikeda, Z.D. Wang, Bei-Yi Zhu, Thierry Giamarchi, Pierre Le Doussal and particularly Vadim B. Geshkenbein for useful discussions about various aspects of vortex physics.

\* \* \*

En dehors de ces remerciements professionnels, je souhaite exprimer ma gratitude aux proches que j'ai côtoyés et appréciés tout au long de ce travail. En premier lieu, je dois énormément au grand nombre de collègues et camarades qui m'ont soutenu et encouragé lors des moments difficiles, m'offrant ainsi un encadrement indispensable, si ce n'est scientifique, du moins humain. Il m'est impossible de tous les remercier ici, mais je tiens malgré tout à citer Emmanuelle Giacometti, Sylvie Goncalves-Conto, Stéphane Bolognini, Armand Hirt, Didier Michoud, Frédéric Oulevey et Jean-Paul Salvetat pour leur énorme soutien, ainsi que László Forró, qui a su allier assistance technique et encouragements. Merci aussi a Romuald Houdré pour son aimable appui "officiel" !

D'autre part, je remercie vivement ma famille, particulièrement mes parents et mes soeurs, pour leur compréhension face à ma faible disponibilité pendant ces longues années d'études, ainsi que pour l'aide qu'ils m'ont régulièrement offerte.

Finalement, soit depuis Manchester, soit sur place à Lausanne, le support inconditionnel et quotidien de Nicole, sa patience aussi, ont été salutaires dans l'accomplissement de ce travail. MERCI !